\documentclass[12pt,a4paper]{article}
\pdfoutput=1
\usepackage[utf8]{inputenc}
\usepackage[T1]{fontenc}
\usepackage{style}
\usepackage[english]{babel}
\usepackage{url}
\usepackage{footnote}
\usepackage{float}

\usepackage{amsmath}
\usepackage{graphicx} 
\usepackage{verbatim} 

\newcommand{\Mpc}{\text{Mpc}}
\newcommand{\Gpc}{\text{Gpc}}
\newcommand{\fsky}{f_{\rm sky}}
\newcommand{\fid}{{\rm fid}}

\renewcommand{\L}{\mathcal{L}}

\renewcommand{\l}{\lambda}
\renewcommand{\d}{\partial}

\newcommand{\be}{\begin{equation}}
\newcommand{\ee}{\end{equation}}
\newcommand{\beqa}{\begin{eqnarray}}
\newcommand{\eeqa}{\end{eqnarray}}
\newcommand{\bsm}{\begin{smallmatrix}}
\newcommand{\esm}{\end{smallmatrix}}

\newcommand\m{\mu}

\newcommand\G{\Gamma}

\renewcommand\r{\rho}

\renewcommand\t{\theta}

\renewcommand\l{\lambda}

\newcommand{\HH}{{\cal H}}

\newcommand\x{{\bf x}}
\renewcommand\k{{\bf k}}
\newcommand\q{{\bf q}}

\newcommand{\e}{\eta}

\def\e{{\rm e}}
\def\d{\partial}
\newcommand{\bseq}{\begin{subequations}}
\newcommand{\eseq}{\end{subequations}}

\renewcommand{\ln}{\mathop{\rm ln}\nolimits}

\def\half{\frac{1}{2}}

\renewcommand{\L}{\Lambda}

\renewcommand{\k}{{\bf k}}

\newcommand{\z}{{\bf z}}

\newcommand{\eV}{\,\text{eV}}
\newcommand{\meV}{\,\text{meV}}

\newcommand{\disp}{\displaystyle}
\renewcommand{\d}{\partial}

\newcommand{\lin}{\mathrm{lin}}
\newcommand{\tree}{\mathrm{tree}}
\newcommand{\lp}{\mathrm{1\text{-}loop}}

\newcommand{\tot}{{\rm tot}}
\newcommand{\obs}{{\rm obs}}
\newcommand{\minn}{{\rm min}}
\newcommand{\maxx}{{\rm max}}
\newcommand{\true}{{\rm true}}
\newcommand{\eff}{{\rm eff}}
\def\l{\left(}
\def\r{\right)}

\usepackage[textwidth=3cm,textsize=footnotesize]{todonotes}

\title{
Measuring neutrino masses with large-scale structure: Euclid forecast with controlled theoretical error
}

\author[a,b]{Anton Chudaykin\footnote{\texttt{chudy@ms2.inr.ac.ru}}}
\author[c,a]{Mikhail M. Ivanov\footnote{\texttt{mi1271@nyu.edu}}}
\affiliation[a]{Institute for Nuclear Research of the
Russian Academy of Sciences, \\ 
\normalsize \it  60th October Anniversary Prospect, 7a, 117312
Moscow, Russia}
\affiliation[b]{Moscow Institute of Physics and Technology,\\
	Institutsky lane 9, Dolgoprudny, Moscow region, 141700, Russia}
\affiliation[c]{Center for Cosmology and Particle Physics, Department of Physics,
New York University,\\
New York, NY 10003, USA}

\abstract{
We present a Markov-Chain Monte-Carlo (MCMC) forecast for the 
precision of neutrino mass 
and cosmological parameter measurements with a Euclid-like galaxy clustering survey. 
We use a complete perturbation theory model 
for the galaxy one-loop power spectrum and tree-level bispectrum, which includes
bias, redshift space distortions, IR resummation for baryon acoustic oscillations and UV counterterms. 
The latter encapsulate various effects of short-scale dynamics which cannot be modeled within perturbation theory. 
Our MCMC procedure consistently computes the non-linear power spectra and bispectra
as we scan over different cosmologies.
The second ingredient of our approach is the theoretical error covariance which captures uncertainties due to higher-order non-linear
corrections omitted in our model. 
Having specified characteristics of a Euclid-like 
spectroscopic survey,
we generate and fit mock galaxy power spectrum and bispectrum likelihoods.
Our results suggest that even under very agnostic assumptions about non-linearities and short-scale physics
a future Euclid-like survey will be able to measure
the sum of neutrino masses with a standard deviation of 28 meV.
When combined with the Planck cosmic microwave background likelihood, this uncertainty
decreases to 13 meV. Over-optimistically reducing the theoretical error on the bispectrum 
down to the two-loop level marginally tightens this bound to 11 meV.
Moreover, we show that the future large-scale structure (LSS) spectroscopic data will 
greatly
improve constraints on the other cosmological parameters, 
e.g. reaching a percent (per mille) error on the Hubble constant with LSS alone (LSS + Planck).
}

\begin{document}

\begin{flushright}
	INR-TH-2019-014
\end{flushright}

\maketitle
\flushbottom

\section{Introduction}
\label{sec:intro}

The detection of neutrino flavor
oscillations has revealed that neutrinos have at least three individual mass states. 
The neutrino masses are described by the dimension-5 Weinberg operator 
in the Standard model treated as an effective field theory.
Thus, the neutrino mass scale hints on the energy cutoff above which the Standard model
must be completed, and constrains its possible extensions. 
The target value relevant for particle physics model building is 
$100$ meV, which would allow one to discriminate between two different hierarchies: 
normal (two light states and one more massive one) or inverted (one light state and two massive ones)
\cite{Qian:2015waa,Tanabashi:2018oca}.

Laboratory oscillation experiments are only 
sensitive to the mass gap between various neutrino species and  
bound the total neutrino mass $\sum m_\nu $ to be above $60$~meV.
An upper limit can be obtained by measuring the edge of the electron spectrum in $\rm ^3H$ $\beta$-decay experiments \cite{Lobashev:2003kt,Kraus:2004zw,Eitel:2005hg,Aseev:2011dq}. 
Using this method, 
the `Troitsk nu-mass' experiment obtained
the bound on the neutrino mass 
for the electron antineutrino $\bar{\nu}_e$, $m_{\bar{\nu}_e}<2.05\eV\,(2\sigma)$ \cite{Lobashev:2003kt,Aseev:2011dq}. In principle, the sensitivity to $m_{\bar{\nu}_e}$ can be improved down to $\sim 0.2\eV$ with the ongoing KATRIN facility \cite{Eitel:2005hg}, 
which recently obtained the most stringent laboratory bound to date $\sum m_{\nu}<1.1\eV\,(90\%\text{CL}.)$ \cite{Aker:2019uuj}.
However, further progress toward a more accurate measurement of the absolute neutrino mass 
may be challenging
with current technologies.

As of now, the most stringent upper bound on the total neutrino mass comes from cosmology. The Planck cosmic microwave background (CMB) observations have set a $2\sigma$ upper limit $240$ meV from the CMB data alone \cite{Aghanim:2018eyx}.
Combining CMB with measurements of the scale of baryon acoustic 
oscillations (BAO) tightens this constraint down to $120$ meV. 
Similar bounds have been obtained by combining the CMB data with the Ly$\alpha$-flux measurements \cite{Palanque-Delabrouille:2015pga}, the Sunyaev-Zeldovich power spectrum and cluster counts \cite{Bolliet:2019zuz}, and the full-shape galaxy clustering measurements, e.g.~\cite{Gil-Marin:2014baa,Beutler:2014yhv,Cuesta:2015iho,Alam:2016hwk,Grieb:2016uuo,Upadhye:2017hdl}.
In general, the combination of large-scale structure (LSS) and CMB data is essential for the 
neutrino mass measurements \cite{Hu:1997mj,Lesgourgues:2006nd,Audren:2012vy,Archidiacono:2016lnv,Sprenger:2018tdb,Brinckmann:2018owf}.
However, in principle, some constraints can be derived from the LSS data alone,
even though currently the LSS data \textit{per se} are not competitive with the CMB measurements \cite{Ivanov:2019pdj}.

The situation will likely change in the near future as we enter the era of high precision LSS data. Next-generation surveys (e.g. SKA\footnote{\href{https://www.skatelescope.org}{
\textcolor{blue}{https://www.skatelescope.org}}
}, LSST\footnote{\href{https://www.lsst.org}{
\textcolor{blue}{https://www.lsst.org}}
}, DESI\footnote{\href{https://www.desi.lbl.gov}{
\textcolor{blue}{https://www.desi.lbl.gov}}
}, Euclid\footnote{\href{https://www.euclid-ec.org}{\textcolor{blue}{https://www.euclid-ec.org}}}) 
will map a large volume of the Universe 
and generate a highly detailed three-dimensional galaxy distribution.
For example, the Euclid satellite
is expected to measure 
more than 50 million redshifts of distant galaxies over a large fraction of the sky \cite{Laureijs:2011gra} and thus harvest a huge amount of information about galaxy clustering at different scales and redshifts. 
This offers a unique opportunity to improve measurements of  cosmological parameters including the total neutrino mass. 
However, the LSS data analysis is complicated by effects 
of non-linear clustering,
galaxy bias and redshift space distortions. 
Our ability to fully exploit the potential of upcoming 
surveys will strongly depend on 
the understanding on these effects, which is not yet complete.

Fortunately, the bulk of information about the neutrino free-streaming is encoded in mildly non-linear scales which can be robustly and systematically 
described within perturbation theory. 
One of the most popular approaches is Eulerian standard cosmological perturbation
theory (SPT) \cite{Bernardeau:2001qr}.
The basic formulation of SPT, however, does not correctly capture 
the non-linear evolution of baryon acoustic oscillations (BAO) \cite{Crocce:2007dt}
and short-scale physics beyond the 
single-stream pressureless perfect fluid hydrodynamics \cite{Pueblas:2008uv,Baumann:2010tm,Blas:2015tla}. 
These problems have been intensely studied in the recent years.

First, it has been shown that the non-linear suppression and distortion 
of the BAO can be captured by 
a resummation of contributions describing the tidal effects of 
large-scale bulk flows. 
This procedure, called infrared (IR) resummation,
is essential for an accurate description of the BAO and has been formulated within various 
theoretical frameworks \cite{Senatore:2014via,Vlah:2015zda,Blas:2016sfa,Baldauf:2015xfa,Ivanov:2018gjr,Noda:2017tfh}.
We will adopt a systematic approach of~\cite{Blas:2016sfa,Ivanov:2018gjr} streamlined in
the context of time-sliced perturbation theory~\cite{Blas:2015qsi}.

Second, we will use the effective field theory of large-scale structure (EFT) to account for the back-reaction of small scale nonlinearities on larger scales \cite{Baumann:2010tm,Carrasco:2012cv,Lewandowski:2014rca}.
This approach removes the unphysical UV sensitivity of perturbation theory loop integrals 
and parameterizes the ignorance about short scale-dynamics by various effective operators in the equations of motion for the dark matter fluid. 
The EFT addresses rather general short-scale phenomena including 
halo virialization,
baryonic feedback, 
and the fingers-of-God effect due to virialized motions. These phenomena are encapsulated in a number of free parameters, called ``counterterms,'' whose values and 
time-dependences are not known \textit{a priori}.

Third, we will employ the full non-linear bias model.
Galaxies are biased tracers of the underlying matter field which consists of baryons and dark matter.
The relation between matter and galaxies on large scales 
is encapsulated in a perturbative expansion built out of all possible
operators allowed by rotational symmetry and the equivalence principle \cite{McDonald:2009dh,Senatore:2014eva,Mirbabayi:2014zca,Desjacques:2016bnm}. 
The relevance of these operators is controlled by the corresponding non-linear bias coefficients. 
One could try to derive bias parameters analytically or extract them from 
N-body simulations \cite{Lazeyras:2015lgp,Lazeyras:2019dcx,Lazeyras:2017hxw,Abidi:2018eyd}.
In this work we adopt an agnostic approach for the bias expansion.
We will treat both the counterterms and bias coefficients as nuisance parameters and marginalize over their values and time-dependence.

The galaxy distribution is commonly characterized by the two-point 
correlation function or its Fourier space counterpart, power spectrum.
The power spectrum analysis, however, suffers from degeneracies
between cosmological and bias parameters. 
This problem can be alleviated by adding the information from the tree-point correlation function, or its Fourier space counterpart called ``bispectrum.'' 
The bispectrum introduces new shape dependencies that break parameter degeneracies and yield 
more robust constraints on cosmological parameters \cite{Gil-Marin:2014pva,Kitaura:2014mja,Yankelevich:2018uaz,Karagiannis:2018jdt,Gil-Marin:2014sta,Gil-Marin:2016wya}.

It is desirable to use the Markov Chain Monte Carlo (MCMC) technique 
for parameter inference from LSS surveys. 
This is a common practice for the CMB, weak lensing and photometric galaxy clustering data, but not 
in the full-shape (FS) Fourier-space power spectrum analysis \cite{Beutler:2016arn}\footnote{
With a few notable exceptions, e.g.~Refs.~\cite{Upadhye:2017hdl,Ivanov:2019pdj}.
Note that the situation is different for the position space full-shape 
correlation function and redshift-space wedges analyses, which do vary relevant cosmological 
parameters \cite{Satpathy:2016tct,Grieb:2016uuo}, but only in combination with the Planck likelihood.
}. 
The main goal of the recent FS analyses is the measurement of the radial and angular diameter distances 
from the Alcock-Paczynski (AP) effect \cite{Alcock:1979mp}, and the 
amplitude of velocity fluctuations $f\sigma_8$ probed by redshift-space distortions.
In that case one usually 
keeps the shape of the power spectrum fixed,
which is correct if one does not vary 
the physical densities of massive neutrinos, dark matter and baryons,  
$\Omega_{\nu} h^2$, $\Omega_{cdm} h^2$ and $\Omega_{b} h^2$, respectively. 
The latter two have been measured by the CMB data significantly more
precisely than by any other observation.
This justifies the standard practice 
for the baseline BOSS power spectrum model with the minimal neutrino mass. 
This practice clearly becomes 
inaccurate if 
one varies the neutrino masses, 
which generates peculiar scale-dependent shape distortions 
of the power spectrum. 
Besides, the precision of the upcoming LSS surveys can surpass the CMB precision,
in which case the standard approach will also be inadequate.
In these situations it is 
imperative to vary all cosmological parameters in MCMC chains
and consistently recompute the shape of the 
non-linear power spectrum during the sampling.

Performing a full MCMC analysis for the Fourier space power spectrum was
unfeasible for a long time 
because the calculation of 
perturbation theory convolution integrals was not fast enough.
Recently, there have been a number of attempts to 
boost the computational efficiency by using some advanced numerical algorithms  
\cite{McEwen:2016fjn,Schmittfull:2016yqx,Fang:2016wcf,Simonovic:2017mhp}.
In our work we will employ the FFTLog method originally proposed in Ref.~\cite{Hamilton:1999uv} and recently revisited in the context of perturbation 
theory loop integrals in \cite{Simonovic:2017mhp}.
In this approach the linear matter power spectrum is decomposed over a basis of 
power-law functions, whose convolution integrals can be done analytically. 
We implement this algorithm in the publicly available \texttt{CLASS} code \cite{Blas:2011rf}
and show that its performance is fast enough for MCMC
parameter estimation. 
To our knowledge, the present analysis is the first MCMC 
forecast of an LSS survey that features a consistent perturbative 
treatment of non-linearities and other short-scale phenomena for galaxies in redshift space.

Another important ingredient of our study is the theoretical error covariance. 
Perturbative calculations of a given order are valid only for a limited range of scales where the next-order corrections are small. 
These corrections can be estimated and 
added to the covariance matrix as a correlated error. 
This approach was pioneered in Ref.~\cite{Audren:2012vy} and was recently revisited in 
the context of perturbation theory in Ref.~\cite{Baldauf:2016sjb}.
The theoretical error method is different from a usually employed approach of trusting the theory completely until 
a certain wavenumber $k_{\text{max}}$ \cite{Yankelevich:2018uaz,Sprenger:2018tdb,Boyle:2017lzt,Boyle:2018rva}. 
Using such a cut-off means that all information coming from wavenumbers higher than $k_{\text{max}}$ is thrown away. 
However, the theoretical error grows gradually as a function of the wavenumber
and is correlated across different $k$-bins. 
This implies that even the short scales dominated by the theoretical error can still yield some cosmological information. 
For example, the BAO wiggles in the matter power spectrum can be accurately described
even at the scales where the broadband part can have a large theoretical uncertainty, see e.g. Ref.~\cite{Beutler:2019ojk}
for a related study. 
If the coherence frequency of this uncertainty is bigger than the BAO frequency, the BAO 
wiggles will still have a significant signal-to-noise even if they are superimposed 
on top of a poorly known broadband signal. 
The situation here is similar to the common BAO scale measurements upon 
marginalizing over the broadband shape \cite{Beutler:2016ixs}.

Our paper has several objectives. 
First, we demonstrate that cosmological parameters can be extracted from 
a spectroscopic galaxy survey by means of an MCMC analysis of the full shape power
spectrum and bispectrum data. 
This includes a rigorous computation of non-linear loop corrections 
for each sampled set of cosmological parameters.
Second, we show that even under the most agnostic assumptions 
about the short-scale physics and galaxy bias one is still able to obtain decent 
constraints on the total neutrino mass and cosmological parameters of the minimal $\Lambda$CDM.
Third, we scrutinize various effects forming the neutrino mass constraints: redshift space
distortions, the Alcock-Paczynski effect, baryon acoustic oscillations, 
inclusion of the bispectrum, and combination of 
LSS measurements with the CMB data.

The sensitivity of upcoming LSS surveys to neutrino masses has been studied in a number 
of works \cite{Takada:2005si,Audren:2012vy,Baldauf:2016sjb,LoVerde:2016ahu,Raccanelli:2017kht,Boyle:2017lzt,Brinckmann:2018owf,Vagnozzi:2018pwo,Boyle:2018rva,Parimbelli:2018yzv,Mishra-Sharma:2018ykh,Copeland:2019bho}. 
In this paper we combine, for the first time, all important ingredients of the analysis.
First, we use a full MCMC approach which accurately captures relevant parameter correlations
and is free from inaccuracies of the Fisher matrix calculus.
Second, we use a complete one-loop perturbation theory model for the redshift-space galaxy power spectrum,
and adopt a conservative approach to nuisance parameters describing 
non-linear short-scale dynamics, baryonic effects, bias and redshift space distortions. 
Third, we add the tree-level bispectrum likelihood to the analysis.
Our baseline theoretical model includes the one-loop power spectrum 
with all relevant nuisance parameters, and the tree-level bispectrum, which
includes quadratic bias parameters.\footnote{Technically, the tree-level bispectrum formula does not include counterterms and other short-scale corrections,
which appear only at the one-loop order \cite{Baldauf:2014qfa,Angulo:2014tfa}. 
If one is working at this order, consistency demands the
power spectrum to be computed at the two-loop order.}
Thus, we consistently take into account all perturbative corrections up to the third order in the 
linear density field. 
Fourth, we employ the theoretical error covariance, 
which is based on perturbation theory
arguments and does not require any input from N-body simulations.

The paper is organized as follows. Section \ref{sec:theor} discusses our theoretical model.
Section \ref{sec:euclid} specifies the Euclid galaxy 
survey and fixes fiducial parameters for our mock
datasets. In Section \ref{sec:binning} we give a detailed description of our method 
and the covariance matrix treatment, including the theoretical error. 
The readers who are not interested in the technical aspects of our analysis can jump directly to Section \ref{sec:results},
where we present results of our MCMC parameter extraction and 
discuss how various effects contribute to the constraints on the neutrino masses 
and cosmological parameters. 
We conclude in Section \ref{sec:concl}.
Some supplementary material is collected in the Appendices. 
We give explicit expressions for the 
one-loop redshift space galaxy power spectrum in Appendix \ref{app:mult}
and the corresponding covariance matrix in Appendix \ref{app:Cov}.
Appendix \ref{sec:BAO} scrutinizes the information 
content of the BAO wiggles. 
Appendix \ref{app:fog} presents results for the worst-case scenario
of the theoretical error with a strong fingers-of-God effect. 
Appendix \ref{app:kmax} shows that using a sharp momentum cutoff or the linear power 
spectrum model can significantly degrade the forecasted constraints. 
Finally, Appendix \ref{app:gauss_planck} contains the results obtained by using the 
Planck Gaussian prior on the cosmological parameters instead of the full likelihood.

\section{Theoretical Model}
\label{sec:theor}

We will use cosmological perturbation theory to compute   
templates for the galaxy power spectrum and bispectrum.
Let us briefly discuss some general 
approximations which we will make. 
These approximation are 
common in the LSS literature.
Our theoretical model will be an extension of standard Eulerian
perturbation theory \cite{Bernardeau:2001qr}, which is based 
on several core assumptions.
First, SPT assumes that the dynamics of matter 
on large scales is governed by the Eulerian pressureless
perfect fluid hydrodynamics. 
Second, SPT assumes that the initial 
density field is a Gaussian stochastic variable and its rms deviations are small on large scales. 
Third, SPT uses the so-called Einstein de-Sitter (EdS) approximation, 
i.e. the time-dependence of loop corrections is 
factored out and approximated by powers of the linear growth factor
just like in a matter-dominated (Einstein de-Sitter) universe. This approximation 
was checked to be sub-percent accurate on mildly non-linear scales \cite{Pietroni:2008jx,Fasiello:2016qpn}.
In this paper we will go beyond the first approximation 
and take into account corrections to the pressureless perfect fluid dynamics within effective field theory (EFT). 
As we argue now, the latter two SPT assumptions mentioned above need not be revisited for the purposes of our study.

The presence of massive neutrinos requires several modifications to the standard formalism.
Most importantly, the linear growth rate of matter fluctuations becomes wavenumber-dependent.
This happens because matter clustering slows down on scales smaller than the 
free-streaming length after the neutrinos become non-relativistic. 
Strictly speaking, the EdS factorization does not take place in this case.
A proper perturbative treatment requires 
an accurate calculation of the scale-dependent Green's function~\cite{Blas:2014hya,Fuhrer:2014zka,Senatore:2017hyk}.
This approach is quite laborious and has not yet been extended to galaxies in redshift space.
To overcome this difficulty we adopt the following approximation. 
For a given redshift we will evaluate non-linear predictions with the standard perturbation theory  kernels, but using the exact linear matter
power spectrum computed in the presence of massive neutrinos \cite{Saito:2008bp}.
This approximation was shown to agree with the full calculation within a two-fluid extension of standard perturbation theory with a few percent precision \cite{Blas:2014hya}. Recently, this result was confirmed within the EFT \cite{Senatore:2017hyk}, which showed that the residual difference between the full calculation and the EdS
can be absorbed into the $k^2$-counterterm (to be discussed shortly). 
This suggests that the EdS approximation with the appropriate countertems will be sufficient for the precision of upcoming LSS surveys. 
We emphasize that the EdS approximation takes into account the leading 
non-trivial time- and scale-dependence
of the perturbation theory Green's functions in the presence of massive neutrinos.



As far as the bispectrum is concerned, 
recent analytic and numerical analyses \cite{Ruggeri:2017dda,deBelsunce:2018xtd} have 
shown that on mildly non-linear scales the neutrino effect is also dominated by the 
free-streaming damping of the linear matter power spectrum. 
In other words, the power spectrum picture presented above applies to the bispectrum as well.

To sum up, for the purposes of this paper the leading effect of massive neutrinos on matter clustering can be approximated as a time- and scale-dependent suppression of the linear density field.
All more complicated phenomenology beyond this approximation, e.g.~\cite{VillaescusaNavarro:2012ag,
LoVerde:2013lta,Fuhrer:2014zka,Dupuy:2015ega,Chiang:2017vuk,Chiang:2018laa}
is captured 
by the EFT corrections (counterterms)~\cite{Senatore:2017hyk}.
Now let us discuss in detail various ingredients of our theoretical approach.




\subsection{Non-linear galaxy bias}
\label{subsec:traces}

Galaxies are biased tracers of the underlying matter density field. 
Galaxy bias has been a subject of various studies carried out over last years, e.g.
see \cite{McDonald:2009dh,Senatore:2014eva,Mirbabayi:2014zca} and \cite{Desjacques:2016bnm}
for a comprehensive review.
In our base bias model we assume that the clustering properties of galaxies are determined by the dark matter and baryons, and \textit{not} by the total matter including massive neutrinos. 
This approach, called the ``cb'' prescription, has a simple physical interpretation: halos form 
on short scales 
where the neutrinos have significant 
velocity dispersion and hence do not participate in clustering. 
The ``cb'' prescription was first advocated on the basis of N-body simulations \cite{Villaescusa-Navarro:2013pva,Castorina:2013wga,Costanzi:2013bha,Castorina:2015bma} and then explained theoretically in Ref.~\cite{Biagetti:2014pha}.
Moreover, Refs.~\cite{Raccanelli:2017kht,Vagnozzi:2018pwo} pointed out its importance for parameter
inference.
Note that within our prescription all the matter statistics, e.g. the linear power spectrum
 $P_{\text{lin}}(k)$
 will refer to cold dark matter and baryons without massive neutrinos.

In perturbation theory the galaxy density contrast $\delta_g$ 
is generally expressed as a series of operators that are constructed 
out of the Newtonian gravitational potential $\Phi$ and the velocity potential $\Phi_v$,
and satisfy rotational symmetry and 
the equivalence principle. 
For the purposes of this paper the bias expansion will be built upon the 
cold dark matter + baryon (cb) density fields $\delta_{\text{cb}}$, 
hence $\Phi$ and $\Phi_v$ will refer to the effective potentials sourced by the `cb' fluid.
The expansion sufficient for the
one-loop matter power spectrum is
\begin{equation}\label{bias}
\begin{split}
\delta_g=&~b_1\delta_{\text{cb}}+\varepsilon-R^2_* k^2 \delta_{\text{cb}}+\frac{\disp b_2}{\disp 2}\delta^2_{\text{cb}}+b_{\mathcal{G}_2}\mathcal{G}_2\\
&~+\frac{b_3}{6}\delta^3_{\text{cb}}+b_{\mathcal{G}_3}\mathcal{G}_3+b_{(\mathcal{G}_2\delta)}\mathcal{G}_2\delta_{\text{cb}}+b_{\Gamma_3}\Gamma_3
\,,
\end{split}
\end{equation}
where $\varepsilon$ is the stochastic part which is not correlated with the 
large-scale density field, and we introduced the following operators:
\be{}
\begin{split}{}
&\mathcal{G}_2(\Phi)\equiv(\d_i\d_j\Phi)^2-(\d^2\Phi)^2\,,\\
&\mathcal{G}_3(\Phi)\equiv-\d_i\d_j\Phi\d_j\d_k\Phi\d_k\d_i\Phi-\half(\d^2\Phi)^3+\frac{3}{2}(\d_i\d_j\Phi)^2\d^2\Phi\,,\\
&\Gamma_3\equiv\mathcal{G}_2(\Phi)-\mathcal{G}_2(\Phi_v)\,.
\end{split}
\ee

The $k^2R_*^2 \delta_{\text{cb}}$ contribution in \eqref{bias} is the so-called higher-derivative bias, 
which is characterized by a length scale $R_*$. For the higher derivative 
expansion to make sense one demands that $R^2_* k^2 \ll 1$. 
Higher-derivative terms are expected to originate from an effective viscous tensor in the Euler equation, which accounts for the effects of shell-crossing and virialization.
On dimensional grounds one may expect that $R_*$ should be given by the Lagrangian radius of typical halos that host the galaxies of interest. 
However, the higher-derivative operator also encapsulates other effects
of short-scale dynamics beyond the pressureless perfect-fluid approximation. 
First, it corrects for the error introduced by integrating over infinite momenta in standard perturbation theory loop integrals. 
Indeed, the shape of the higher-derivative bias term,
\be
P_{\nabla^2 \delta}(k)= -2 R^2_* k^2 P_{\text{lin}}(k)\,,
\ee
coincides with the UV part of the one-loop power spectrum. 
When summed together, the higher-derivative contribution reduces the overall amplitude of 
the one-loop correction (for positive $R_*^2$), and because of that it is often referred to as `the counterterm'. 
From this argument we see that the measured value of $R_*$ depends on the UV-cutoff (smoothing scale) of one-loop integrals. Second, the $k^2 P_{\text{lin}}(k)$-contributions naturally appear as a long-wavelength limit of other physical effects 
relevant for galaxy statistics, e.g. 
velocity bias,
baryonic feedback \cite{Lewandowski:2014rca} and non-linearity in the neutrino fluid \cite{Senatore:2017hyk}.
Hence, by including the higher-derivative contribution we effectively take all these effects into account, and the 
parameter $R^2_*$ need not be precisely equal to the Lagrangian radius of the halo. 
We will treat $R^2_*$ as a free nuisance parameter in what follows.

It is known that at the level of the one-loop power spectrum the cubic bias parameters 
$b_3$, $b_{\mathcal{G}_3}$ and $b_{(\mathcal{G}_2\delta)}$ do not form independent shapes, 
or, in other words, renormalize the other bias parameters. 
Hence, there are only five independent free parameters
relevant for the deterministic part of the one-loop power spectrum and tree-level bispectrum: 
$b_1$, $b_2$, $b_{\mathcal{G}_2},b_{\Gamma_3}$ and $R^2_*$. 
We will treat them as nuisance parameters and marginalize over their values and time-dependence.

The two-point function of the galaxy density field in real space \eqref{bias}
is given by\footnote{By default, we will assume that all the power spectra and bispectra
are functions of redshift, and suppress the
explicit time-dependence in the relevant expressions in what follows.} 
\be
\begin{split}
 \label{Pg}
P_{g}(k)=&~b^2_1 (P_{\text{lin}}(k)+P_{\text{1-loop}}(k))+b_1b_2\mathcal{I}_{\delta^2}(k)+2b_1b_{\mathcal{G}_2}\mathcal{I}_{\mathcal{G}_2}(k)\\
&+(2b_1b_{\mathcal{G}_2}+\frac{4}{5}b_1b_{\Gamma_3})\mathcal{F}_{\mathcal{G}_2}(k)+\frac{1}{4}b^2_2\mathcal{I}_{\delta^2\delta^2}(k)+b^2_{\mathcal{G}_2}\mathcal{I}_{\mathcal{G}_2\mathcal{G}_2}(k)\\
&+b_2b_{\mathcal{G}_2}\mathcal{I}_{\delta^2\mathcal{G}_2}(k)+P_{\nabla^2 \delta}(k)+P_{\text{shot}}\,,
\end{split}
\ee
where the term $P_{\nabla^2 \delta}$ originates from the higher derivative bias contribution. 
The term $P_{\text{shot}}$ 
denotes the shot noise. 
This contribution is produced by the stochastic part $\varepsilon$
and reflects the discrete nature of galaxies observed in a finite volume.
For the purposes of this paper we assume that 
the stochastic noise is scale-independent. 
This is supported by the results of Ref.~\cite{Schmittfull:2018yuk}. 
This study also showed
that the amplitude of the shot noise can be super- or sub-Poissonian for very light or massive halos, respectively.
To capture this effect we sample 
$P_{\text{shot}}$ in our MCMC chains along with other nuisance parameters.

\subsection{Redshift space distortions}
\label{subsec:power2}

The distance to a galaxy is inferred from its observed redshift, which 
gets contaminated by the peculiar velocity field. 
This effect, known as redshift space distortions (RSD), 
generates an anisotropy in the galaxy distribution due to the mixture of the 
velocity field $\textbf{v}$ (and its divergence $\t$) and the real-space galaxy density $\delta_g$ \cite{Bernardeau:2001qr}.

We will use the flat-sky approximation in which the redshift-space power spectrum 
depends only on the module of the
wavevector $\k$ and the cosine 
of the angle between this vector and the line-of-sight $\textbf{z}$,
\be 
\mu=\frac{(\textbf{k}\cdot \textbf{z})}{k}\,.
\ee
The galaxy density field in redshift space can be obtained by mapping the real space
bias expansion~\eqref{bias}
onto redshift space,
\be 
\label{eq:dens}
\delta^{(s)}_g(\k)=\delta_g(\k)+\int
d^3x\,\e^{-i\k\cdot
  \x}\Big(\exp\left\{-i (\k\cdot \z) (\z\cdot {\bf v}(\x))/\HH\right\}-1\Big)\big(1+\delta_g(\x)\big)\,, 
\ee
where ${\bf v}$ is the peculiar velocity field and $\HH$ is the conformal Hubble parameter.
Squaring Eq.~\eqref{eq:dens} and averaging over the statistical ensemble  
one can obtain the anisotropic galaxy power spectrum in redshift space. Its explicit expression 
can be found in Appendix~\ref{app:mult}.
In practice, one usually expands it 
over Legendre polynomials $L_\ell(\mu)$,
\begin{equation}
P^{(s)}_g(k,\mu)=\sum_{\ell=0} P_{\ell,g}(k)L_\ell(\mu)\,,
\end{equation}
where we introduced redshift-space multipoles of the power spectrum
\begin{equation}\label{mu_integral}
P_{\ell,g}(k)=\frac{2\ell+1}{2}\int_{-1}^1d\mu\, P^{(s)}_g(k,\mu)L_\ell(\mu)\,.
\end{equation}
In linear theory the redshift-space power spectrum is fully 
characterized by the first three non-vanishing moments: the monopole ($\ell=0$), quadrupole ($\ell=2$) and hexadecapole ($\ell=4$).
In principle, loop corrections generate multipoles higher than the hexadecapole, but their
amplitude is suppressed on large scales because they do not have tree-level contributions.
Hence, in accordance with previous studies \cite{Kazin:2011xt,Taruya:2011tz,Beutler:2013yhm},
we expect that the bulk of the information on cosmological parameters are encoded in the first three even moments, and will focus on them in our further analysis.

The velocity field appearing in Eq.~\eqref{eq:dens} has a stochastic short-scale component
which does not correlate with the large-scale modes. 
This component is responsible for the so-called fingers-of-God effect \cite{Jackson:2008yv},
which is
caused by virialized motions of galaxies and 
cannot be captured within standard perturbation theory.
From Eq.~\eqref{eq:dens} one observes that each power of the 
stochastic velocity field is accompanied by a power of $k\mu$, 
hence the fingers-of-God effect can be 
described by appropriate higher-derivative operators 
similarly to the discussed backreaction of short scale modes \cite{delaBella:2017qjy,Senatore:2014vja,Perko:2016puo}.
Following \cite{Senatore:2014vja}, we will refer to these operators as `counterterms'.
There are several phenomenological models widely used in the literature 
to model the fingers-of-God e.g. \cite{Taruya:2010mx,Hand:2017ilm}. In our paper we adopt an agnostic effective field 
theory point of view and assume that each redshift space multipole requires its own $\sim k^2P_{\lin}$ counterterm with a free normalization.
For simplicity, we also ignore selection bias effects \cite{Desjacques:2018pfv}.

Upon doing the integral \eqref{mu_integral} one obtains the following expressions for the
multipoles of the galaxy power spectrum,
\begin{subequations}
\label{Pell}
\begin{align}
\notag
 P_{0,g}(k)= & P^{\text{tree}}_{0,\theta \theta}(k)+ 
P^{\text{1-loop}}_{0,\theta \theta}(k)
 +b_1 (P^{\text{tree}}_{0,\theta \delta}(k)
 + P^{\text{1-loop}}_{0,\theta \delta}(k))
 +b_1^2(P^{\text{tree}}_{0,\delta\delta}(k)
  +P^{\text{1-loop}}_{0,\delta\delta}(k))\\
  \notag
 &+0.25b_2^2\mathcal{I}_{\delta^2\delta^2}(k)+b_1b_2\mathcal{I}_{0,\delta \delta^2}(k)
+b_2\mathcal{I}_{0,\theta \delta^2}(k)
+b_1b_{\mathcal{G}_2}\mathcal{I}_{0,\delta \mathcal{G}_2}(k)
+b_{\mathcal{G}_2}\mathcal{I}_{0,\theta \mathcal{G}_2}(k)
 \\
 \notag
& +b_2b_{\mathcal{G}_2}\mathcal{I}_{\delta^2\mathcal{G}_2}(k)
+b_{\mathcal{G}_2}^2\mathcal{I}_{\mathcal{G}_2\mathcal{G}_2}(k)
+(2b_{\mathcal{G}_2}+0.8b_{\Gamma_3})
(b_1\mathcal{F}_{0,\delta \mathcal{G}_2}(k)+\mathcal{F}_{0,\theta \mathcal{G}_2}(k))\\
&+c_0 P_{0,\nabla^2\delta}(k)+P_{\text{shot}}\,,\\
\notag
P_{2,g}(k)= &
P^{\text{tree}}_{2,\theta \theta}(k)
+P^{\text{1-loop}}_{2,\theta \theta}(k)
+b_1(P^{\text{tree}}_{2,\theta \delta}(k)
+P^{\text{1-loop}}_{2,\theta \delta}(k))
+b_1^2P^{\text{1-loop}}_{2,\delta \delta}(k)\\
\notag
& +b_1b_2\mathcal{I}_{2,\delta \delta^2}(k)
+b_2\mathcal{I}_{2,\theta \delta^2}(k)
+b_1b_{\mathcal{G}_2}\mathcal{I}_{2,\delta \mathcal{G}_2}(k)
+b_{\mathcal{G}_2}\mathcal{I}_{2,\theta \mathcal{G}_2}(k)
\\
&+(2b_{\mathcal{G}_2}+0.8b_{\Gamma_3})\mathcal{F}_{2,\theta\mathcal{G}_2}(k)+c_2 P_{2,\nabla^2\delta}(k)\\
P_{4,g}(k)= & 
P^{\text{tree}}_{4,\theta \theta}(k)+
P^{\text{1-loop}}_{4,\theta \theta}(k)
+b_1P^{\text{1-loop}}_{4,\theta \delta}(k)+b_1^2P^{\text{1-loop}}_{4,\delta\delta}(k)\\
\notag 
&+b_2\mathcal{I}_{4,\theta \delta^2}(k)+b_{\mathcal{G}_2}\mathcal{I}_{4,\theta \mathcal{G}_2}(k)+c_4 P_{4,\nabla^2\delta}(k)\,,
\end{align}
\end{subequations}
where $P_{\delta \delta}$, $P_{\theta \delta}$, $P_{\theta \theta }$ are density, cross and velocity power spectra respectively, as computed in SPT. 
The different contributions ${\cal I}_{\ell,n}$ and ${\cal F}_{\ell,n}$ 
are redshift-space generalizations of the real space bias loop integrals  
which can be computed using the explicit formulas given in Appendix~\ref{app:mult}. 
We keep them in the expression above to illustrate the sensitivity of different multipoles  
to bias parameters.
The new contributions $P_{0,\nabla^2\delta}$, $P_{2,\nabla^2\delta}$ and $P_{4,\nabla^2\delta}$ are counterterms in redshift 
space, which are characterized by free coefficients $c_0$,$c_2$,$c_4$. 

The presence of massive neutrinos requires a modification of the logarithmic growth rate 
\be
\label{eq:f}
f=\frac{d\ln D_+}{d\ln a}\,,
\ee
where $D_+$ is the linear growth factor and $a$ is the scale factor. 
Refs.~\cite{Castorina:2015bma,Villaescusa-Navarro:2017mfx} argued on the basis of N-body simulations that the logarithmic growth rate, similarly to bias, should be computed for dark matter and baryons only. 
In this case $f$ (whose definition \eqref{eq:f} is meaningful only in linear theory) 
is scale-independent with precision better than $0.1\%$ for $m_\nu \leq 100$ meV.
The negligible residual scale-dependence of $f$ can be easily taken into account at the linear level,
but introduces ambiguity in the loop calculations.
We have explicitly checked that the difference is too small to affect our results,
and hence prefer to use the scale-independent approximation in what follows.
The ``cb'' prescription ensures that the redshift-space linear power spectrum (given by the Kaiser formula \cite{Kaiser:1987qv})
evaluated with the ``cb'' linear bias and logarithmic growth rate 
matches the result of N-body simulations on large scales.
Motivated by this prescription, we evaluate the redshift-space loop integrals using the standard 
perturbation theory kernels but with the logarithmic growth rate computed for the ``cb'' component.


\subsection{IR resummation}
\label{subsec:IR}

Another ingredient required to accurately predict the clustering of galaxies on large scales is IR resummation. IR resummation takes into account tidal effects of large-scale bulk flows 
that suppress and distort the pattern of BAO. 
This effect was pointed out many years ago \cite{Crocce:2007dt} and 
since then there were a number of studies aimed at capturing it, see Refs.~\cite{Senatore:2014via,
Baldauf:2015xfa,
Vlah:2015zda,
Senatore:2017pbn,
delaBella:2017qjy,
Lewandowski:2018ywf}.
We implement the IR resummation procedure developed in the context of 
time-sliced perturbation theory \cite{Blas:2015qsi,Blas:2016sfa,Ivanov:2018gjr}. 
This procedure is based on rigorous power counting rules and gives an
accurate estimate of the theoretical error at each order of IR resummation.
Besides, it is numerically stable and fast, which is crucial 
for the MCMC analysis.

IR resummation in real space 
requires splitting the matter linear power spectrum
into the smooth and the wiggly parts,
\be
P_{\text{lin}}= P_{\text{nw}}(k)+P_{\text{w}}(k)\,,
\ee
where $P_{\text{nw}}$ is a power-law function, 
and $P_{\text{w}}$ contains the BAO wiggles.
At leading order one `dresses' the wiggly part with the damping exponent,
\be\label{PO_real}
P_{\text{LO}}(k) \equiv P_{\text{nw}}(k)+e^{-k^2\Sigma^2}P_{\text{w}}(k)\,,
\ee
where 
\be
\Sigma^2\equiv\frac{4\pi}{3}\int_0^{k_S}dqP_{\text{nw}}(q)\left[1-j_0\l\frac{q}{k_{osc}}\r+2j_2\l\frac{q}{k_{osc}}\r\right]\,,
\ee
$k_{osc}$ is the BAO wavelength $\sim 110\,h/$Mpc, $k_S$ is the separation
scale controlling the modes to be resummed,
and $j_n$ are the spherical Bessel function of order $n$. 
In principle, $k_S$ is arbitrary and any dependence on it should be
treated as a theoretical error. Following \cite{Blas:2016sfa} we define it 
to be $k_S=0.2\,h$/Mpc, which gives the same result as an alternative choice $k_S=k/2$,
adopted in \cite{Baldauf:2016sjb}.

At next-to-leading order one uses the expression \eqref{PO_real} as an input in the one-loop power spectrum,
\be
\begin{split}
P_{g}(k) \to &  \quad P_{\text{nw}}(k) +e^{-k^2\Sigma^2}P_{\text{w}}(k)(1+k^2\Sigma^2) + P_{\text{1-loop}}[P_{\text{nw}}+e^{-k^2\Sigma^2}P_{\text{w}}]  \,,
\end{split}
\ee
where $P_{\text{1-loop}}$ should be considered a functional of the 
linear power spectrum. 

In redshift space the damping factor becomes $\mu$-dependent since peculiar velocities 
additionally wash out the BAO wiggles along the line-of-sight. The IR resummed
anisotropic power spectrum at leading order takes the following form,
\be
\label{Plo}
P_{\text{LO}}(k,\mu) \equiv P_{\text{nw}}(k)+e^{-k^2\Sigma^2_{\text{tot}}(\mu)}P_{\text{w}}(k)\,.
\ee
where we introduced the anisotropic damping factor,
\be
\label{eq:sigmatot}
\Sigma^2_\tot(\mu)=(1+f\mu^2(2+f))\Sigma^2+f^2\mu^2(\mu^2-1)\delta\Sigma^2\,,
\ee
and a new contribution $\delta\Sigma^2$ given by,
\be
\delta\Sigma^2\equiv4\pi\int_0^{k_S}dqP_{\text{nw}}(q)j_2\l\frac{q}{k_{osc}}\r \,.
\ee
In general, IR resummation in redshift space at next-to-leading (one-loop) order requires a computation of anisotropic loop integrals which cannot be reduced to one-dimensional ones. 
One can simplify these integrals by splitting the one-loop contribution itself into a smooth 
and wiggly parts. More precisely, one first computes the one-loop integrals with a smooth 
part only. At a second step one evaluates these integrals with one insertion of the wiggly power spectrum and suppresses the output with a direction-dependent damping factor \eqref{eq:sigmatot} to get
\be
\begin{split}
	P_{g}(k,\mu) \to &  \quad P_{\text{nw}}(k,\mu) + P_{\text{nw},\,\text{1-loop}}(k,\mu) \\
	& \quad +e^{-k^2\Sigma^2_{\text{tot}}}P_{\text{w}}(k,\mu)(1+k^2\Sigma^2_{\text{tot}}(\mu))+e^{-k^2\Sigma^2_{\text{tot}}(\mu)} P_{\text{w},\,\text{1-loop}}(k,\mu)\,.
\end{split}
\ee
At a last step the eventual IR-resummed anisotropic power spectrum should be used to compute the multipoles 
in Eq.~\eqref{mu_integral}.

\subsection{Alcock-Paczynski effect}
\label{subsec:AP}

Another source of anisotropy in the density distribution is introduced at the 
galaxy catalog level through the so-called Alcock-Paczynski (AP) effect \cite{Alcock:1979mp}. Galaxy surveys probe a volume of the universe spanning over some range in angles and redshifts. 
To convert them into physical distances one has to assume some fiducial cosmology.
If the fiducial cosmology is different from the true one, the inferred distribution of galaxies will be anisotropically deformed. 
The distortions perpendicular and parallel to the line-of-sight are proportional to the angular diameter distance $D_A(z)$ and the Hubble parameter $H(z)$, respectively, thus providing us with an additional probe of underlying cosmology which is insensitive to the evolution of matter perturbations. 

To account for the AP effect one has to compute the observable galaxy power spectrum,
\be
P_\obs(k_\obs,\mu_\obs) = P_{g}(k_\true[k_\obs,\mu_\obs],\mu_\true[k_\obs,\mu_{\obs}]) \cdot \frac{D^2_{A,\text{fid}}H_\true}{D^2_{A,\true}H_\text{fid}},
\ee
where $k_{\text{true}}$ and $\mu_{\text{true}}$ are wavevectors and angles in the true cosmology, whereas $k_{\text{obs}}$ and $\mu_{\text{obs}}$ refer to quantities obtained using the values of $D_{A,\text{fid}}$ and $H_{\text{fid}}$
computed in the fiducial cosmology. 
The relation between the true and observed wavevector modules and directions
is given by
\be
\begin{split}\label{AP_k_mu}
	k^2_\true&=k^2_\obs\left[\l\frac{H_\true}{H_\text{fid}}\r^2\mu_\obs^2+\l\frac{D_{A,\text{fid}}}{D_{A,\true}}\r^2(1-\mu_\obs^2)\right]\\
	\mu^2_\true&=\l\frac{H_\true}{H_\text{fid}}\r^2\mu^2_\obs\left[\l\frac{H_\true}{H_\text{fid}}\r^2\mu_\obs^2+\l\frac{D_{A,\text{fid}}}{D_{A,\true}}\r^2(1-\mu_\obs^2)\right]^{-1}\,.
\end{split}
\ee
Note that $H_\true$ and $D_{A,\text{true}}$ are sampled during the MCMC analysis while 
$H_\text{fid}$ and $D_{A,\text{fid}}$ are always fixed.
The galaxy multipoles with the AP effect are given by
\be 
P_{\ell,\text{AP}}(k) = \frac{2\ell+1}{2} \int_{-1}^1 d\mu_{\text{obs}}\, P_{\text{obs}}(k_{\text{obs}},\mu_{\text{obs}}) \cdot L_\ell(\mu_{\text{obs}})\,.
\ee

\subsection{Bispectum}
\label{subsec:bispectrum1}

The bispectrum is a function of three 
wavevector norms (sides) $k_1$, $k_2$, $k_3$, 
which satisfy momentum conservation and 
form a certain triangle configuration in Fourier space. 
In real space the tree-level galaxy bispectrum takes the following form \cite{Baldauf:2016sjb,Yankelevich:2018uaz}:
\be 
\begin{split}
\label{Bg}
B_g(k_1,k_2,k_3)=&[F^{(b)}_2({\bf k}_1,{\bf k}_2)b_1^2P_\lin(k_1)P_\lin(k_2)+{\rm cycl.}]\\
&+P_{\text{shot}}\sum_{a=1}^{3}b_1^2P_\lin(k_a)+B_{\text{shot}}\,,
\end{split}
\ee
where the non-linear kernel $F^{(b)}_2({\bf k}_1,{\bf k}_2)$ is given by 
\be
F^{(b)}_2(\k_1,\k_2)  =\frac{b_2}{2}+b_{\mathcal{G}_2}\left(\frac{(\k_1\cdot \k_2)^2}{k_1^2k_2^2}-1\right)
+b_1 F_2(\k_1,\k_2)\,. 
\ee
To account for the effect of IR resummation on the bispectrum one has to simply substitute $P_\lin$ by the leading order IR resummed 
spectrum \cite{Blas:2016sfa}, see Eq.~\eqref{PO_real}.

Note that in Eq.~\eqref{Bg}
there are two different noise contributions $P_{\text{shot}}$ and $B_{\text{shot}}$.
Following \cite{Yankelevich:2018uaz}, we assume that the contribution $P_{\text{shot}}$
is the same for the power spectrum and bispectrum.
Similarly to the power spectrum case, we also assume that the stochastic contributions 
to the bispectrum are scale-independent, but their amplitudes need not be exactly
Poissonian.

The bispectrum analysis is more intricate in redshift space. We are going to focus on the monopole part of the 
tree-level bispectrum ($\ell=0,m=0$ in the notation of \cite{Scoccimarro:1999ed}), obtained upon averaging over all directions,
\be
\label{Bg0}
B_{0,g}(k_1,k_2,k_3)=\frac{1}{4\pi}\int_0^{2\pi}d\phi\int_{-1}^{1}d\cos\t\, B^{(s)}_g(\k_1,\k_2,\k_3)\,,
\ee 
where $\t$ is the angle between $\k_1$ and $\z$, and $\phi$ is the azimuthal angle around $\k_1$. The redshift space bispectrum
$B^{(s)}_g(\k_1,\k_2,\k_3)$ is a simple generalization of the real space expression \eqref{Bg} \cite{Scoccimarro:1999ed,Gil-Marin:2014sta,Karagiannis:2018jdt},
\be 
\begin{split}\label{Bg_rsd}
B^{(s)}_g(\k_1,\k_2,\k_3)=&[Z_2({\bf k}_1,{\bf k}_2)Z_1(\k_1)Z_1(\k_2)P_\lin(k_1)P_\lin(k_2)+{\rm cycl.}]\\
&+P_{\text{shot}}\sum_{a=1}^{3} Z^2_1(\k_a)P_\lin(k_a)+B_{\text{shot}}\,,
\end{split}
\ee
where the kernels $Z_1$ and $Z_2$ 
can be found in Appendix~\ref{app:mult}.
For simplicity, we do not consider the AP effect in the bispectrum 
\cite{Song:2015gca}.

\section{Survey Characteristics}
\label{sec:euclid}

The main effect of neutrinos on the matter power spectrum is the suppression of its 
amplitude at high wavenumbers due to free-streaming. 
Observing the onset of this suppression is hard because of large 
sample variance on large scales.  
On mildly non-linear scales the suppression can be mimicked 
by decreasing the amplitude of the primordial power spectrum. 
However, the suppression effect has a non-trivial redshift-dependence,
which helps to break the degeneracy between the neutrino mass and the primordial fluctuation amplitude 
by measuring the power spectrum at different redshifts. 
From this argument it is clear that in order to robustly detect the neutrino mass,
we need a deep survey with a wide redshift range and good redshift resolution. 
Given this reason we will 
focus on a Euclid-like spectroscopic survey. 

\subsection{Euclid survey specification}

The next generation of space-based galaxy redshift surveys such as Euclid \cite{Laureijs:2011gra} and  WFIRST-AFTA \cite{Spergel:2015sza} will use near-IR (NIR) slitless spectroscopy to collect large samples of emission-line galaxies. These surveys will target luminous star-forming galaxies containing H$\alpha$ emitters 
at near-infrared wavelengths (around $z>0.5$). 
The obtained maps of LSS will be used to study the BAO, the matter power spectrum and other statistics, the structure growth rate and RSD.
In this context the space density of H$\alpha$ emitters (i.e. their luminosity function) is a key ingredient essential to forecast the sensitivity of future space missions.

The rate of cosmic star formation is believed to peak near $z\sim 2$ \cite{Madau:2014bja}, providing us with a large number of H$\alpha$ sources which can be detected by future galaxy redshift surveys. 
Thanks to a unique combination of high-resolution optical and multi-band NIR imaging,
the Euclid survey will be able to identify H$\alpha$ emitters out as far as $z\sim 2$. 
However, the abundance of H$\alpha$ emitters is poorly known. It has been firmly established 
only at low redshifts by means of the ground-based optical spectroscopic surveys \cite{Gallego:1995ib}. 
The ground-base NIR single-slit spectroscopic observations are contaminated by the intense airglow, which is the major sources of uncertainty in forecasting H$\alpha$ luminosity function at high redshifts. 
The narrow-band ground-based NIR imaging surveys and space-based redshift observations 
suffer from significant contamination and do not provide us with the unambiguous H$\alpha$ luminosity function at $z>0.7$ \cite{Pozzetti:2016cch}. Given a limited knowledge of the population of emission-line galaxies at high redshifts, we will use empirical approaches based on available ground- and space-based data to model the evolution of the H$\alpha$ luminosity function out to $z\sim2$. Specifically, we will adopt the approach of \cite{Pozzetti:2016cch}, which is based on the largest actual dataset of H$\alpha$ luminosity functions from low- to high-redshifts. 
In particular, we use the empirical `model 1' from \cite{Pozzetti:2016cch} and assume a redshift distribution of H$\alpha$ emitters per square degree ($dN/dz$) based on a limiting flux $F_{H\alpha}>3\times 10^{-16}$ erg cm$^{-2}$ s$^{-1}$. The total number of detected galaxies in a given redshift bin centered at $\bar{z}$ and of width $\Delta z$ can be inferred from the given values of $dN/dz$ as 
\begin{equation}\label{Ntot}
N(\bar{z})=41253\,\, {\rm deg}^2\times \fsky\int_{\bar{z}-\frac{\Delta z}{2}}^{\bar{z}+\frac{\Delta z}{2}} \frac{dN/dz}{1\,{\rm deg}^2}dz
\end{equation}
where $\fsky$ is the fraction of the sky covered by the survey  
\begin{equation}
\fsky=0.3636
\end{equation}
The comoving volume related to a specific redshift bin observed by Euclid can be computed via 
\begin{equation}\label{V}
V(\bar{z})=\frac{4\pi}{3}\fsky\cdot\left[r^3\l\bar{z}+\frac{\Delta z}{2}\r-r^3\l\bar{z}-\frac{\Delta z}{2}\r\right]\,,
\end{equation}
where $r(z)$ denotes the comoving distance up to a object with redshift $z$.

\begin{table}[h]
	\centering
	\begin{tabular}{|c|cc|c|}
		\hline
		$\bar{z}$ & $V(\bar{z})$ & $n_g(\bar{z})$ & $V_\eff(\bar{z})$ \\
		\hline
		0.6	&	4.58	&	3.83	&	4	\\
		0.8	&	6.44	&	2.08	&	4.98	\\
		1.0	&	8.01	&	1.18	&	5.09		\\
		1.2	&	9.23	&	0.7		&	4.37	\\
		1.4	&	10.15	&	0.39	&	2.98	\\
		1.6	&	10.81	&	0.21	&	1.55	\\
		1.8	&	11.25	&	0.12	&	0.68	\\
		2.0	&	11.53	&	0.07	&	0.28		\\
		\hline
	\end{tabular}
	\caption{\label{tab:survey} Specification of the Euclid mission in 8 non-overlapping redshift bins of width $\Delta z=0.2$ centered at $\bar{z}$. The comoving ($V$) and effective ($V_\eff$) volume values are expressed in units of $h^{-3}\,\Gpc^3$, whereas the galaxy number density $n_g$ is quoted in units of $10^{-3}h^3\,\Mpc^{-3}$.}
\end{table}

The Euclid space telescope is expected to measure $\approx5\cdot10^7$ galaxy redshifts in the approximate redshift interval $0.5<z<2.1$ \cite{Laureijs:2011gra}. We divide this spectroscopic volume into 8 non-overlapping redshift bins ($N_z=8$), whose central values $\bar{z}$ are linearly spaced between $0.6$ and $2.0$. The width of each bin is $\Delta z=0.2$.

The mean galaxy number density $n_g(\bar{z})$ in each bin can be obtained from \eqref{Ntot}, \eqref{V} as
\begin{equation}\label{n_g}
n_g(\bar{z})=\frac{N(\bar{z})}{V(\bar{z})}
\end{equation}
We report the comoving volumes covered by the survey and the galaxy number densities as functions of redshift bins centered at $\bar{z}$ in Table \ref{tab:survey}. Note that our specification gives a total number of galaxies to be covered by Euclid $N\approx5.5\cdot10^7$ in full agreement with the previous Euclid forecasts \cite{Laureijs:2011gra,Amendola:2016saw}. Besides, our estimates for the number density and sample volume match the recent study \cite{Yankelevich:2018uaz}
(after an appropriate rescaling of redshift bins).

High redshift observations are subject to 
substantial shot noise, which increases the statistical error 
even for large comoving volumes. 
This effect can be illustrated by
means of the so-called effective volume \cite{Alam:2016hwk},
\begin{equation}\label{Veff}
V_\eff(\bar{z})\approx \left.V(\bar{z})\left[\frac{\bar{n}_g(\bar{z}) b_1^2(\bar{z})P_\lin(k,\bar{z})}{1+\bar{n}_g(\bar{z}) b_1^2(\bar{z})P_\lin(k,\bar{z})}\right]^2\right|_{k=0.1\,h\, \text{Mpc}^{-1}}\,.
\end{equation}
The linear bias parameters $b_1(z)$ for the Euclid galaxy sample will be discussed momentarily.
We list the effective volumes for the chosen redshift bins in the rightmost column of Tab.~\ref{tab:survey}. 
The effective volumes almost coincide with the corresponding comoving volumes at low redshifts, 
but significantly reduce at high redshifts due to small galaxy number densities.
The effective survey volume maximizes near $z\sim 1$. 
Note that our specification yields the cumulative effective volume 24 $h^{-3}$Gpc$^3$.

\subsection{Fiducial cosmology and nuisance parameters}
\label{subsec:fid}

Let us discuss now the fiducial cosmology and nuisance parameters. 
We adopt the baseline Planck $\Lambda$CDM model that corresponds to $\rm TT, TE, EE + lowE + lensing$ dataset \cite{Aghanim:2018eyx}.
We approximate the neutrino sector by one state with mass $m_\nu=0.1\eV$ and two massless states. 
We chose this mass because $m_\nu\equiv \sum_{i}m_{\nu_i}=0.1\eV$ is a border line value defining the neutrino mass hierarchy.
Fiducial values of other cosmological parameters are listed in Table \ref{tab:fid},
where for convenience we used a normalized amplitude of primordial fluctuations defined as $A\equiv{A_s}/{ A_{s,\,\fid}}$ with $A_{s,\text{fid}}=2.1\cdot 10^{-9}$.

\begin{table}[h]
	\centering
	\begin{tabular}{llr}
		\hline
		Parameter & Definition & Fiducial value \\
		\hline
		$h$	& Hubble parameter $H_0/100\,\text{km/s/Mpc}$ & $0.6736$\\
		$\omega_{cdm}$& Cold dark matter density $\Omega_{cdm}h^2$ & $0.12$\\
		$\omega_{b}$ & Baryon density $\Omega_bh^2$ & $0.02237$\\
		$A$ & Amplitude of the primordial power spectrum & $1$\\
		$n_s$ & Spectral index of the primordial power spectrum & $0.9649$\\
		$m_{\nu}$ & Total neutrino mass & $0.1\eV$\\
		\hline
	\end{tabular}
	\caption{\label{tab:fid} Summary of the fiducial cosmological model parameters.}
\end{table}

Since the population of Euclid H$\alpha$ targets is very poorly constrained, we will adopt a semi-analytic model of galaxies formation and N-body simulations to fix the fiducial bias parameters in order to generate mock data. 
Specifically, we use the following model for the Euclid-type galaxies linear bias as a function of redshift \cite{Orsi:2009mj},
\begin{equation}\label{b1}
b_1(z)=0.9+0.4z
\end{equation}
The determination of realistic fiducial values for 
other bias parameters requires additional assumptions, for details see \cite{Desjacques:2016bnm}. 
We will use the following fitting formula for $b_2$,
\be
\label{eq:b2}
b_2(z)= -0.704-0.208z+0.183z^2-0.00771z^3\,,
\ee
which was obtained from a combination of N-body simulations 
and halo occupation distribution modeling, see Refs.~\cite{Yankelevich:2018uaz,DiDio:2018unb}. 
In order to set the other bias parameters we use the co-evolution model \cite{Desjacques:2016bnm}, which gives
\be 
\label{eq:G2}
\begin{split}
 b_{\mathcal{G}_2}(z)=-\frac{2}{7}(b_1(z)-1)\,,\quad  b_{\Gamma_3}(z)= \frac{23}{42}(b_1(z)-1)\,.{}
  \end{split}
\ee
These values agree well with N-body simulations \cite{Abidi:2018eyd}. 
As the parameter $b_{\Gamma_3}$ does not multiply a separate shape, 
it happens to be very degenerate with $b_{\mathcal{G}_2}, c_0$ and $c_2$.
These degeneracies cannot be broken within the errorbars of the Euclid-like survey 
that we consider.
The way to handle this problem is to impose some prior on $b_{\Gamma_3}$. 
We have found that the chains with varied $b_{\Gamma_3}$ yield the same results on cosmological parameters 
as the chains with the fixed $b_{\Gamma_3}$, but the convergence time was significantly longer.
This suggests to fix $b_{\Gamma_3}$ to the theoretically expected value,
which corresponds to the limit of an infinitely narrow prior.
Since we vary the coefficient $b_{\mathcal{G}_2}$ that multiplies the same shape,
this approach will still allow for a sufficient freedom in the fitting procedure.
We believe that fixing $b_{\Gamma_3}$ is
a good compromise between computational efficiency and model generality 
for the parameter extraction from future LSS surveys.

\begin{table}[h]
	\centering
	\begin{tabular}{|c|cccc|cccc|}
		\hline
		 $\bar{z}$&$b_1(\bar{z})$ & $b_2(\bar{z})$ & $b_{\mathcal{G}_2}(\bar{z})$ & $b_{\Gamma_3}(\bar{z})$& $R_*^2(\bar{z})$ & $c_0(\bar{z})$ & $c_2(\bar{z})$ & $c_4(\bar{z})$\\
		\hline
		0.6 & 1.14	&	-0.765	&	-0.04	& 0.077&	0.536	&	13.398	&	13.398	&	0.536\\
		0.8 & 1.22	&	-0.757	&	-0.063	&0.121 &	0.442	&	11.060	&	11.06	&	0.442\\
		1.0 & 1.30	&	-0.737	&	-0.086	& 0.164&	0.369	&	9.236	&	9.236	&	0.369\\
		1.2 & 1.38	&	-0.703	&	-0.109	& 0.208&	0.312	&	7.799	&	7.799	&	0.312\\
		1.4 & 1.46	&	-0.658	&	-0.131	& 0.252&	0.266	&	6.658	&	6.658	&	0.266\\
		1.6 & 1.54	&	-0.600	&	-0.154	& 0.296&	0.230	&	5.740	&	5.740	&	0.230\\
		1.8 & 1.62	&	-0.531	&	-0.177	& 0.340&		0.200		&	4.993	&	4.993	&	0.200\\
		2.0	&1.70	&	-0.450	&	-0.200	&0.383 &	0.175	&	4.380	&	4.380	&	0.175\\
		\hline
	\end{tabular}
	\caption{\label{tab:bias} Fiducial values for bias parameters and free normalizations of conterterms.
	}
\end{table}

As for the higher-derivative bias coefficient, the measurements from N-body simulations \cite{Baldauf:2015aha,Schmittfull:2018yuk,Lazeyras:2019dcx}
suggest that it is an order-one quantity in units of $[\text{Mpc}/h]^2$. 
Thus, we adopt the following fiducial value along with the time-dependence that corresponds
to the one-loop contribution,
\be
R_*^2=1\times D^2_+(z)\,[\text{Mpc}/h]^2\,.
\ee
Note that a recent study \cite{Lazeyras:2019dcx} has found some deviations from this scaling. 

Now let us discuss the redshift space counterterms, which are dominated by the fingers-of-God effect.
At leading order in $(k\mu)^2$ this effect is characterized by a short-scale rms velocity,\footnote{
	Note that we used an additional factor $f$ in the definition \eqref{fog} in order to match the 
	convention of the BOSS DR12 analysis papers \cite{Gil-Marin:2016wya,Beutler:2016arn}.
} 
\be
\label{fog}
\begin{split}
& P_{g,\,\text{FoG}}(k,\mu)\approx -f^2 \sigma^2_v k^2\m^2 (b_1+f\mu^2)^2 P_\lin (k) \,,\\
&\text{where}\quad f^2 \sigma^2_v \equiv z^iz^j\HH^{-2}\langle v_i v_j\rangle \,,
\end{split}
\ee
and $\langle...\rangle$ denotes the average w.r.t short modes.
The velocity dispersion
depends on the galaxy type, environment and the satellite fraction.\footnote{In this discussion we neglect the error of redshift measurements, whose effect is similar to fingers-of-God.}
Since the fingers-of-God is a non-perturbative short-scale effect, one should try to minimize it in order 
to facilitate theoretical modeling. 
It is thus suggestive to build the multipole moments 
using only the central galaxies, in which case the fingers-of-God effect is minimized \cite{Hand:2017ilm}. 
The numerical simulations of the Euclid-type H$\alpha$ - emitting galaxies done in 
Ref.~\cite{Orsi:2009mj}
show that the fingers-of-God effect is small for them. 
A similar conclusion was reached in Ref.~\cite{Orsi:2017ggf}, which argued that the Euclid-type 
sample will typically have small velocity dispersions.

In redshift space the higher derivative terms are expected to take different
values for each multipole moment \cite{Senatore:2014vja,delaBella:2017qjy}. 
Since these counterterms were not yet measured for the Euclid-like galaxies,
we will the expression \eqref{fog} to set their values.
Note that the generic redshift-space counterterms also correct for the error 
introduced by integrating up to infinite momenta in loop corrections and   
take into account other short-scale effects, 
e.g. galaxy formation details and the baryonic feedback. 
However, the characteristic scale of these effect is expected to be $\sim 1$ Mpc$/h$,
which makes them sub-dominant compared to 
the galaxy velocity dispersion, whose characteristic scale is expected to be bigger.
According to~\cite{Orsi:2017ggf}, one may expect it to be approximately twice smaller than the 
velocity dispersion of the BOSS-type galaxies, which is roughly equal to $5$ Mpc$/h$ at $z=0.6$~\cite{Gil-Marin:2016wya,Beutler:2016arn}.
\footnote{These two references, in fact, quote quite different results for $\sigma_v$, which may be explained by a different
	choice of $k_{\text{max}}$ adopted in these two analyses. 
	Thus, the `mean' value $\sigma_v\sim 5$ Mpc/$h$ should be 
	taken with a grain of salt. 
	Note that the BOSS sample satellite fraction is $\sim 10\%$, see
	Ref.~\cite{Rodriguez-Torres:2015vqa}. 

}
Given this reason, we will use the model \eqref{fog} to generate the 
redshift space space counterterms for the monopole and quadrupole moments, 
which are expressed as
\be\label{FoG_02}
c_\ell P_{\ell,\nabla^2 \delta} =-k^2 c_\ell \frac{(2\ell+1)}{2}\int_{-1}^1 d\mu \,
f^2 \m^2 (b_1+f\mu^2)^2 P_\lin (k) L_\ell(\mu)\,.
\ee
We adopt the following fiducial values:
\begin{equation}\label{cs}
\qquad c_0=c_2=25 D^2_+(z)\,[\text{Mpc}/h]^2
\,,
\end{equation}
which correspond to the velocity dispersion $\sigma_v \simeq 2.5$ Mpc$/h$ at $z=1$ (see App.~\ref{app:mult} for our counterterm convention).
The time-dependence in \eqref{cs} is such that $c_\ell P_{\ell,\nabla^2 \delta}$
scales with time just like the one-loop power spectrum.
Note that we will fit the monopole and quadrupole counterterms independently in each redshift bin.
It is worth pointing out that the redshift errors produce an effect similar to the fingers-of-God, 
and hence marginalizing over the counterterms captures this instrumental effect too.

The expression \eqref{fog} gives unsatisfactory results for the hexadecapole as it does not 
correctly capture the behavior observed in simulations (e.g. Ref.~\cite{Hand:2017ilm}).
Indeed, the fingers-of-God noticeably enhance the hexadecapole amplitude at short scales while Eq.~\eqref{fog}
predicts a strong suppression. 
The power enhancement observed in simulations at large momenta
is driven by higher-order corrections that are not present in our model and hence 
must be included in the theoretical error covariance (to be discussed shortly).
In order to fix the counterterm for the hexadecapole, we 
use the following functional form and adopt a value $1$ Mpc/$h$
expected on the effective field theory grounds:
\be
\begin{split}\label{FoG_04}
& c_4 P_{4,\nabla^2 \delta} =-k^2 c_4 \frac{(2\cdot 4+1)}{2}\int_{-1}^1 d\mu \,
(b_1+f\mu^2)^2 P_\lin (k) L_4(\mu) \,,\quad  c_4=D^2_+(z)\,[\text{Mpc}/h]^2\,.
\end{split}
\ee

As far as the stochastic contributions are concerned, we set their fiducial values to the 
Poisson sampling prediction 
\be 
P_{\text{shot}}=\bar{n}_g^{-1}\,,\quad B_{\text{shot}}=\bar{n}_g^{-2}\,.
\ee 
The values of the mean number density $\bar{n}_g$ 
are listed in Table.~\ref{tab:survey}. 

All in all, the fiducial values for bias parameters and counterterms 
at different redshifts are listed in Table~\ref{tab:bias}.
Since we are interested in the constraints on the cosmological parameters and marginalize 
over the nuisance bias parameters, their precise values are not very important for the purposes of this paper.
What really matters is the correlation between the cosmological and nuisance parameters, 
which is expected to be weakly sensitive to the fiducial values.

\section{Methodology and Likelihoods}
\label{sec:binning}

In this section we present the details of our MCMC analysis. 
We start by outlining the method we are going to use.
Then we describe the mock dataset and
discuss the structure of covariance matrices, which include both statistical 
and theoretical errors.

\subsection{Method}

In order to better understand the role of different effects contributing to the eventual neutrino
mass constraints, we will consider the cases of real and redshift space separately. 
The real space case is purely academic, and will serve us as an example illustrating the amount of information
encoded in the shape of the galaxy power spectrum and bispectrum without RSD. 

Our main analysis is done for the redshift space power spectrum and bispectum. 
We start our exploration with the power spectrum. 
We generate and analyze several mock datasets that are aimed to quantify 
the amount of information coming from various sources:
RSD, BAO and the AP effect.
At a next step we quantify the information gain of combining the LSS and CMB 
experiments, for which we use the most recent Planck data release \cite{Aghanim:2018eyx}.
Then we incorporate the bispectrum
and analyze different combinations of the 
power spectrum, bispectrum and Planck likelihoods. 
Finally, we will estimate the information gain from the one-loop
bispectrum under an over-optimistic (unrealistic) assumption that it does not contain 
new bias parameters compared to those present at the tree-level.

We generate mock data samples by computing the fiducial theoretical
one-loop non-linear galaxy power 
spectrum, its redshift space multipoles and the tree-level bispectrum
using a modified version 
of the \texttt{CLASS} code \cite{Blas:2011rf}.\footnote{In principle, we could also randomly spread the mock datapoints according to the statistical error.
However, Ref.~\cite{Perotto:2006rj} showed that this approach yields the same results as the use of the 
fiducial spectrum without the random spread. }
These mock data are assigned statistical and theoretical errors given the survey 
specification and estimates for the higher loop corrections (to be discussed shortly).
The parameter 
constrains are obtained with the April 2018 version of the MCMC code \texttt{Montepython}  \cite{Audren:2012wb,Brinckmann:2018cvx}.
Marginalized posterior densities, limits and contours are produced with the latest version of the \texttt{getdist}
package, which is part of the \texttt{CosmoMC} code \cite{Lewis:2002ah,Lewis:2013hha}.

Our main MCMC analysis samples 6 cosmological and 8 nuisance parameters in each redshift bin $(i)$:
\be
(\omega_b,\omega_{cdm},n_s,h,A,m_\nu)\times \prod_{i=1}^{N_z}(b^{(i)}_1,b^{(i)}_2,b^{(i)}_{\mathcal{G}_2},c^{(i)}_0,c^{(i)}_2,c^{(i)}_4,P_{\text{shot}}^{(i)},B_{\text{shot}}^{(i)})\,.
\ee
We emphasize that we do not assume any time-dependence for the 
bias parameters and counterterms
and fit them separately in each redshift bin.
We treat both $P_{\text{shot}}^{(i)},B_{\text{shot}}^{(i)}$ as 
nuisance parameters and vary them in our MCMC chains to account for the
non-Poissonian nature of shot noise, e.g. halo excursions.
We fit $B_{\text{shot}}^{(i)}$ only for the bispectrum likelihoods.
In the real space analysis instead of three counterterms $c_0,c_2,c_4$ 
we use only one, $R_*^2$.
When including the Planck likelihood, we will also sample an additional parameter - 
the reionization optical depth
$\tau$, whose correlation with the amplitude of 
primordial density fluctuations is important for our eventual 
neutrino mass constraints.

\subsection{Statistical error}
\label{subsec:obs}

To account for sample variance we will 
use the Gaussian approximation to the covariance matrices 
both for the power spectrum and the bispectrum.
In this approximation there is no cross-covariance between these two statistics, which is justified at one-loop order in perturbation theory\footnote{In principle, there is no 
difficulty to include the cross-covariance in the analysis, see e.g. Refs.~\cite{Sefusatti:2006pa,Yankelevich:2018uaz}. This would be required if we worked at the two-loop order.
Note that at this order,  for consistency, one would also need to consider non-Gaussian contributions to 
the power spectrum and bispectrum covariance matrices.
}. 
The Gaussian approximation is quite accurate on large and mildly non-linear scales \cite{Scoccimarro:1999kp,Howlett:2017vwp,Barreira:2017kxd,Mohammed:2016sre,Li:2018scc,Blot:2018oxk} (see however a recent work \cite{Wadekar:2019rdu}).
Of course, it breaks down at short scales, where higher loop corrections 
and the one-halo term become important. The covariance matrix 
at these scales is dominated by the theoretical error, which we implement in the next section.

The real space Gaussian covariance matrix takes the form \eqref{Pg}
\begin{equation}
C_{kk'}=\frac{\l 2\pi\r^3}{V(z)}\frac{\delta_{kk'}}{2\pi k^3d\ln k}P^2_g(k,z)\,,
\end{equation}
where $\delta_{kk'}$ is the Kronecker delta-symbol.
This formula generalizes to redshift-space moments \eqref{mu_integral} as 
(see Appendix \ref{app:Cov} for more details)
\begin{equation}
C^{(\ell \ell')}_{ k k'}=\frac{(2\pi)^3}{V(z)}\frac{(2\ell+1)(2\ell'+1)}{2\pi k^3d\ln k}\int_{-1}^{1}d\m\,L_\ell(\mu)L_{\ell'}(\mu)P_{\ell,g}(k,z)P_{\ell',g}(k',z)\delta_{kk'}\,.
\end{equation}

The covariance matrix is more complex in the bispectrum case \cite{Sefusatti:2006pa,Scoccimarro:1997st,Baldauf:2016sjb,Song:2015gca}. 
An elementary observable in this case is a triangle configuration of three 
wavevectors.
Assuming $k_1\leq k_2\leq k_3$, the sum over triangles can be written as
\begin{equation}
\sum_T\equiv\sum_{k_1=k_\minn}^{k_\maxx}\sum_{k_2=k_\minn}^{k_1}\sum_{k_3=k_*}^{k_2}
\end{equation} 
where $k_*=\maxx(k_\minn,k_1-k_2)$. 
The Gaussian covariance matrix between two triangular configurations $T$ and $T'$ in momentum space is given by
\begin{equation}
\label{covB}
C_{TT'}=\frac{\l2\pi\r^3}{V(z)}\frac{\pi s_{123}}{dk_1dk_2dk_3}\frac{\delta_{TT'}}{k_1k_2k_3} \prod_{a=1}^{3}\l b_1^2(z)P_\lin(k_a,z)+\frac{1}{\bar{n}_g(z)}\r\,,
\end{equation}
where $s_{123}$ is the symmetry factor that equals 6, 2 or 1 for equilateral, isosceles and general triangles, respectively. 
The case of the isotropic redshift space bispectrum is more complicated 
due to different triangle orientations w.r.t. the line-of-sight. 
At leading order in $f/b_1$, the redshift-space bispectrum 
covariance matrix is obtained by multiplying each $P_\lin(k_a,z)$ in Eq.~\eqref{covB}
by the isotropic Kaiser factor \cite{Kaiser:1987qv}\,,
\be\label{a0}
a_0=1+\frac{2f(z)}{3b_1(z)}+\frac{f^2(z)}{5b_1^2(z)}\,.
\ee
At order $(f/b_1)^2$ the covariance matrix receives some non-trivial
shape-dependence, which can be appropriately taken into account, see Appendix~\ref{app:Cov}.

\subsection{Theoretical error}
\label{subsec:theor}

Future LSS surveys will observe a big number of galaxies, which will allow one to significantly decrease sample variance and shot noise compared to current surveys. 
The eventual statistical errors will be minimized on short scales, 
which, however, are hard to describe analytically.
Perturbative calculations are valid only
for wavenumbers sufficiently smaller than the nonlinear scale $k_{\text{NL}}\sim 0.5\,h$/Mpc.
As we go closer to the non-linear scale, a proper theoretical modeling requires 
more loop corrections to be taken into account, which makes the analysis computationally demanding.\footnote{
If the perturbative expansion is asymptotic, it could be that
at a certain order
computing new loop corrections would not increase the momentum 
reach at all \cite{Blas:2013aba,Konstandin:2019bay}. }

A common practice is to set a sharp cutoff $k_{\text{max}}$ and use the theory below this cutoff.
In this approach the theoretical calculations are trusted completely up to $k_{\text{max}}$ 
and discarded after this scale. 
However, it is clear that higher-loop corrections
become important gradually, and neglecting them introduces biases at any $k_{\text{max}}$.
These corrections are smooth functions, whose amplitude and scale dependence can be 
estimated in perturbation theory. 
From this argument it seems more reasonable to use the whole wavenumber range 
and marginalize over higher-order corrections. This is the core idea of the theoretical error approach introduced in Ref.~\cite{Baldauf:2015aha}. 

The theoretical error is a difference between a true theoretical model 
and an explicitly computed approximation to it. 
This error can be seen as a smooth envelope that varies over a characteristic momentum scale $\Delta k$. This scale cannot be arbitrary small -- in that case the theoretical error would be uncorrelated between different k-bins. 
Ref.~\cite{Baldauf:2015aha} proposed to use $\Delta k=0.05\,h/\Mpc$, which is motivated by the BAO wiggles. 
However, the wiggly part of the power spectrum, which oscillates with a frequency similar to $\Delta k=0.05\,h/\Mpc$,
represents only $5\%$ of the total spectrum. The broadband power spectrum varies over 
a much bigger wavenumber scale, which is close to the non-linear momentum $k_{\text{NL}}$.  
Given this reason, in what follows we will use $\Delta k=0.1\,h/\Mpc$.
Note that the correlation length $\Delta k$ makes the theoretical error independent of binning as long as $k_{\rm bin}\ll\Delta k$.

In Ref.~\cite{Baldauf:2015aha} it was shown that the theoretical error acts as a correlated noise that generates the following covariance contribution for the real space power spectrum:
\begin{equation}
(C_e)_{kk'}=E_p(k,z)E_p(k',z)\exp\left\{-\frac{(k-k')^2}{2\Delta k^2}\right\}\,,
\end{equation}
which has to be added to the statistical covariance matrix.
Note that the theoretical covariance is substantially non-diagonal.
We adopt the following envelope based on an explicit two-loop calculation for dark matter in real space \cite{Baldauf:2016sjb}:
\begin{equation}\label{E_P_real}
E_p(k,z)=D^4_+(z)P_g(k,z)\l\frac{\disp k}{\disp 0.45\,h \Mpc^{-1}}\r^{3.3}\,,
\end{equation}
where $P_g$ is the full one-loop galaxy power spectrum. 
Our final results do not
depend on the particular choice of the theoretical error as long as it is small 
on large scales and blows up at short scales quickly enough. 

Now let us focus on redshift space.
The redshift space power spectrum multipoles are correlated among each other, hence
it is natural to write down the theoretical error covariance as
\begin{equation}
(C_e)^{(\ell \ell')}_{ k k'}=E_{\ell,p}(k,z)E_{\ell',p}(k',z)\exp\left\{-\frac{(k-k')^2}{2\Delta k^2}\right\}\,.
\end{equation}
Since the power spectrum calculation in redshift space was not done beyond one-loop order,
we do not have a reliable expression for the two-loop contribution.
On perturbation theory grounds one might expect it to have the same order 
of magnitude as the two-loop real-space power spectrum (which is supported by 
some popular RSD models, e.g.
\cite{Scoccimarro:2004tg}),
yielding the estimate
\begin{equation}\label{E_P_rsd}
\begin{split}
& E_{\ell,p}(k,z)=D^4_+(z)P^{\text{tree}}_{\ell,g}(k,z)\l\frac{\disp k}{\disp 0.45 \,h\text{Mpc}^{-1}}\r^{3.3}\,.
\end{split}
\end{equation}
However, this estimate does not take into account the fingers-of-God effect, 
which is not captured in perturbation theory. 
In this regard, one can use an alternative estimate motivated by the leading order 
correction capturing fingers-of-God \eqref{fog},
\begin{equation}\label{E_P_rsd_fog}
E_{\ell,p}(k,z)=
(k fD_+(z) \sigma_v )^4
\l\ell +\frac{1}{2}\r
\int_{-1}^1 d\mu\,
\mu^4 \,P^{\text{tree}}_{g}(k,\mu,z) L_\ell(\mu)\,.
\end{equation}
The two estimates give comparable results for 
the monopole and quadrupole for \mbox{$\sigma_v\sim 3$ Mpc$/h$}.
The estimate \eqref{E_P_rsd_fog} is, however, very sensitive to the 
velocity dispersion, which 
can vary by a factor of few depending on 
the satellite fraction and galaxy type \cite{Hand:2017ilm}.
In order to be more model-independent, 
we will stick to the estimate \eqref{E_P_rsd}
in what follows. 
As discussed above, this choice is supported by the numerical simulations 
of the Euclid-type galaxies, which imply that the fingers-of-God effect is small for them \cite{Orsi:2009mj,Orsi:2017ggf}.
Another reason to use Eq.~\eqref{E_P_rsd} and not Eq.~\eqref{E_P_rsd_fog} is the following. 
The estimate~\eqref{E_P_rsd_fog} is very $\mu$-dependent, hence, if we analyzed redshift-space wedges 
\cite{Grieb:2016uuo} instead of the usual multipoles, the theoretical error for the wedges with $|\mu| < 1$ would indeed be dominated by the 
two-loop expression \eqref{E_P_rsd} even for relatively large velocity dispersions $\sigma_v$.
This argument suggests that the Fourier space wedges could be a more robust observable for the future spectroscopic surveys.

As for the hexadecapole, its one-loop correction
exceeds the tree-level contribution
on mildly non-linear scales \footnote{In this regard one might be worried that the perturbative expansion breaks down for the hexadecapole. 
This is, however, at artifact of the multipole expansion. One
can check that the one-loop contribution to the total two-dimensional 
redshift-space power spectrum is still smaller than the tree-level expression.
},  
which makes it reasonable to use the full one-loop hexadecapole spectrum  
instead of the tree-level one in the expression for the theoretical error \eqref{E_P_rsd}.

The final redshift space power spectrum likelihood that includes the theoretical error 
is given by 
\be
\begin{split}
&-2\ln \mathcal{L}_P=\sum_{a=1}^{N_{z}} \sum_{\ell,\ell'=0,2,4}\sum_{i,j=1}^{N_{k}}(P^{\text{theory}}_{\ell}(k_j,z_a)-P^{\text{data}}_{\ell}(k_j,z_a))\\
&
\times  (C_{k_i k_j}^{(\ell \ell')}(z_a)+(C_e)^{(\ell \ell')}_{k_i k_j}(z_a))^{-1}
(P^{\text{theory}}_{\ell'}(k_i,z_a)-P^{\text{data}}_{\ell'}(k_i,z_a)) \,.
\end{split}
\ee

As for the bispectrum, 
following Ref.~\cite{Baldauf:2016sjb}, we adopt a Gaussian correlation that is factorisable 
is wavenumbers,
\begin{equation}
(C_e)_{TT'}=E_b(k_1,k_2,k_3, z)E_b(k_1',k_2',k_3', z)\prod_{a=1}^{3}\exp\left\{-\frac{(k_a-k_a')^2}{2\Delta k^2}\right\}\,,
\end{equation}
where the envelope corresponding to the one-loop ($l=1$) and two-loop ($l=2$) orders 
is given by
\begin{equation}\label{E_Breal}
E_b(k_1,k_2,k_3,z)=3B^{\text{tree}}_g(k_1,k_2,k_3,z)
D^{2l}_+(z)
\begin{cases}\l\frac{\disp k_t/3}{\disp 0.31 \,h\text{Mpc}^{-1}}\r^{1.8}\quad l=1\,,\\
\l\frac{\disp k_t/3}{\disp 0.45 \,h\text{Mpc}^{-1}}\r^{3.3}\quad l=2\,,
\end{cases}
\end{equation}
where $B^{\text{tree}}_g$ refers to the tree-level galaxy bispectrum in real space \eqref{Bg} or the redshift space monopole \eqref{Bg_rsd}, respectively. 
We also introduced $k_t\equiv(k_1+k_2+k_3)$.
The one-loop envelope was checked to agree with $\sim 10\%$ 
with the explicit calculations of the dark matter bispectrum in Ref.~\cite{Baldauf:2016sjb}. 
We also checked that the 2-loop bispectrum envelop matches the real space dark matter 
calculation performed in Ref.~\cite{Lazanu:2018yae} at the same level.
Our final bispectrum likelihood is given by
\be
\begin{split}
-2\ln \mathcal{L}_{B}=&\sum_{a=1}^{N_z}\sum_{\text{triangles}\,T,T'}
(B^{\text{theory}}_T(z_a)-B^{\text{data}}_T(z_a))\\
&\times (C_{TT'}(z_a)+(C_e)_{TT'}(z_a))^{-1}
(B^{\text{theory}}_{T'}(z_a)-B^{\text{data}}_{T'}(z_a))\,.
\end{split}
\ee

A comment is in order here.
A proper inclusion of the one-loop bispectrum likelihood is unfeasible at the moment. 
First, the calculation for galaxies in redshift space with all relevant counterterms and biases has not yet been done in principle. 
Second, the computational speed of the one-loop bispectrum calculation 
is not fast enough to enable an MCMC sampling even for the real-space dark matter case, see \cite{Simonovic:2017mhp}
for a study in this direction. 
Yet, one would desire to estimate what could be a possible improvement.
For this reason we will simply use the tree-level formula and the theoretical error that corresponds to a two-loop
envelope.
It is important to stress that the real space galaxy bispectrum at one-loop order 
has 11 new nuisance parameters \cite{Eggemeier:2018qae}. 
One may expect that redshift-space adds up even more additional fitting parameters due to, e.g. fingers-of-God.
Ideally, all these parameters should be included into the fit. 
Since the shapes corresponding to these parameters are unknown, it is not clear how to take them into account in a systematic way. 
To proceed, we adopt the following strategy.
We will not add extra nuisance parameters whatsoever
and will fit the data only with the bias parameters that appear at the tree level
in our one-loop bispectrum likelihood analysis.
This is a rather unrealistic and even inconsistent simplification, which, however,
should already tell us if any improvement can be expected \footnote{
	The reason why the analysis without new bias coefficients may still be meaningful is the following.
	In this paper we restrict ourselves to the monopole moment of the redshift space bispectrum only.
	It might be that other angular moments may give much more information and eventually 
	help measure the one-loop bias parameters without significantly 
	affecting the bounds on the tree-level ones.
Note that a consistent calculation of the one-loop bispectrum also requires taking into account the two-loop
power spectrum.
	}.

We stress that in the case of strong fingers-of-God one can use the redshift-space wedges 
instead of the power spectrum multipoles
and cut the $\mu$-bins around $|\mu|\sim 1$, in which case there are no fingers-of-God and 
the theoretical error should be close to the dark-matter two-loop envelope \eqref{E_P_rsd}. 
Alternatively, the fingers-of-God could also be suppressed by 
removing satellite galaxies at the catalog level \cite{Hand:2017ilm}.
Given these reasons, the main results of our paper are 
presented for the theoretical error dominated by two-loop corrections, which 
corresponds to a more realistic situation.
In Appendix~\ref{app:fog} for purely 
academic reasons we present results for the theoretical error that includes strong fingers-of-God. 
We show that in this scenario the LSS-only constraints degrade by a factor of two.
Additionally, in Appendix \ref{app:kmax} we show results without the theoretical error covariance whatsoever
and also for a simplified power spectrum model without loop corrections. This study shows that these choices 
lead to a significant degradation of cosmological constraints.

Finally, we emphasize that any interpretation of the constraints presented 
in this paper should take into account the underlying assumptions on the theoretical errors. 
These assumptions are based on effective field theory power counting,
which might underestimate (or overestimate) the actual size of the next-to-leading order corrections.
The validity of our theoretical error treatment has to be tested with realistic mock catalogs of Euclid-type 
galaxies, which will be presented in a separate publication. 
The main goal of our proof-of-concept analysis is to show how to build
likelihoods with theoretical errors 
and how their use can improve the efficiency of parameter estimation from the future LSS data.

\subsection{Planck likelihood}

In order to understand the information gain of combining the Euclid survey and the CMB data we will use 
an approximation to the full Planck likelihood. 
Specifically, we use the likelihood \texttt{fake\_planck\_realistic} \cite{DiValentino:2016foa} included in \texttt{Montepython v3.0} \cite{Audren:2012wb,Brinckmann:2018cvx}. This likelihood consists of the mock temperature, polarization and CMB lensing data along  
with the noise spectra that match those from the full Planck results. 
We chose to use the mock Planck likelihood instead of the real one because of two reasons. 
First, in this case we can use the same fiducial model in all mock likelihoods considered in this paper. 
Second, it allows us to be more conservative 
in light of the so-called ``lensing tension,'' which was extensively discussed in many papers 
\cite{Ade:2015xua,Addison:2015wyg,Aghanim:2016sns,Aghanim:2018eyx}.
The problem is that the actual Planck likelihood favors overly enhanced lensing smoothing of the CMB peaks compared to the $\Lambda$CDM expectation. The amplitude of the gravitational lensing potential extracted from the CMB temperature and polarization spectra at high multipoles disagrees with the base $\Lambda$CDM prediction at 2.8$\sigma$ level \cite{Aghanim:2018eyx}. 
This discrepancy might be a statistical fluctuation, 
or indicate unknown systematics and incorrect foreground modeling, 
or even be a manifestation of new physics. 
The lensing excess significantly tightens constraints on the total neutrino mass, 
thus the use of the actual Planck likelihood would inevitably bias our analysis. 
On the contrary, the use of the mock Planck likelihood 
makes the constraints presented on our paper 
independent of the lensing tension.

We have checked that the mock Planck likelihood reproduces 2d posterior contours 
and 1d marginalized distribution for the baseline Planck cosmological model \cite{Aghanim:2018eyx},
except for $m_\nu$, for which we found a twice bigger error as a result of removing the lensing tension.
Note that when combining with Planck, our MCMC chains sample an addition parameter $\tau$, which is 
important for the neutrino analysis as it is strongly correlated with amplitude $A_s$ in the Planck data.

Note that by the time the future LSS surveys are completed, 
there will likely be data from future CMB missions, e.g. LiteBIRD \cite{Matsumura:2013aja}, CORE-M5 \cite{Delabrouille:2017rct}, Stage-4 projects \cite{Abazajian:2016yjj}, 
which will supersede Planck. The mock likelihoods corresponding to these surveys will be analyzed elsewhere.

\subsection{Mock dataset}
\label{subsec:mock}

We generate four mock data samples for each redshift bin: power spectrum in real space,
tree-level real-space bispectrum, 
power spectrum multipoles ($\ell=0,2,4$) in redshift space,
and the tree-level angle-averaged redshift-space bispectrum.

\begin{figure}[!h]
	\centering
	\includegraphics[keepaspectratio,width=0.49\linewidth]{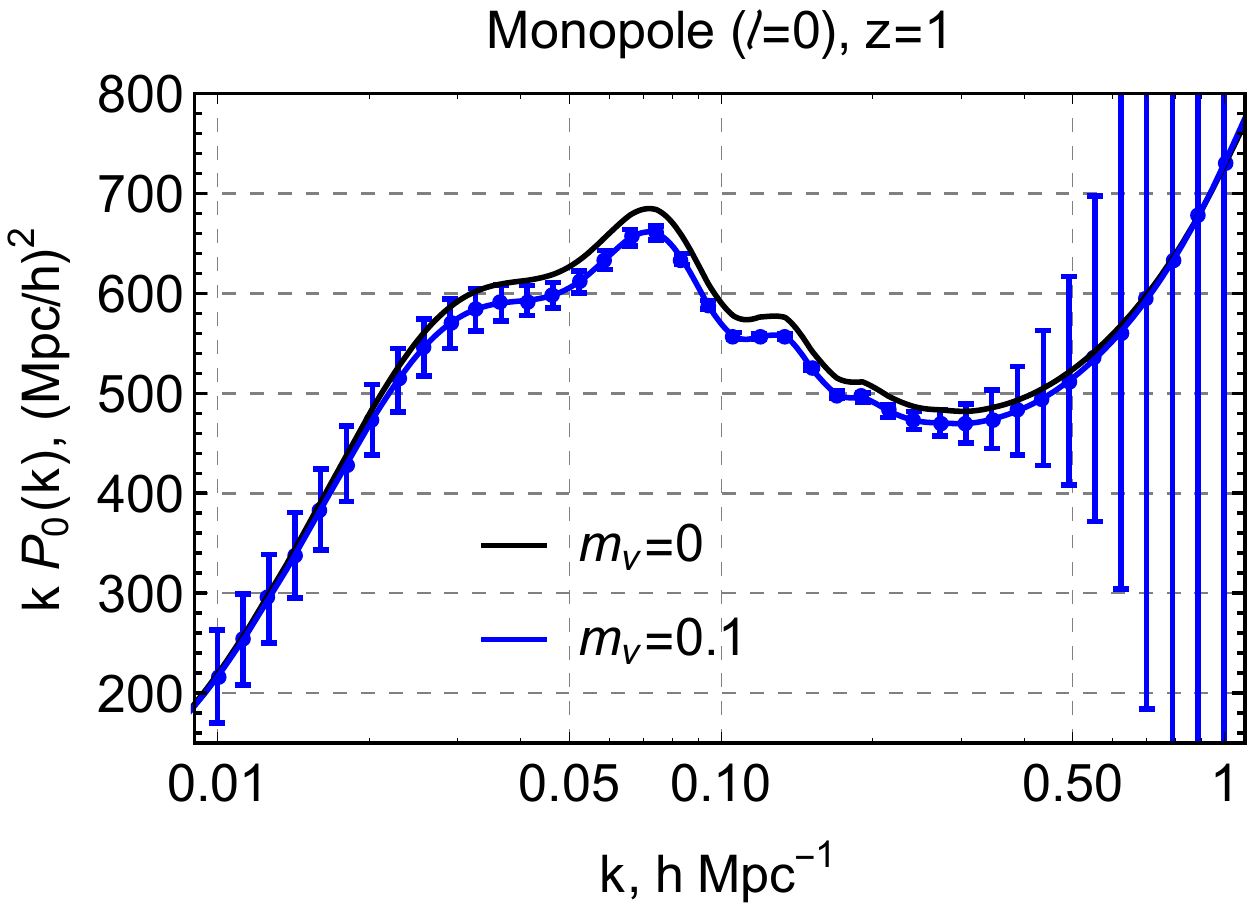}
	\includegraphics[keepaspectratio,width=0.49\linewidth]{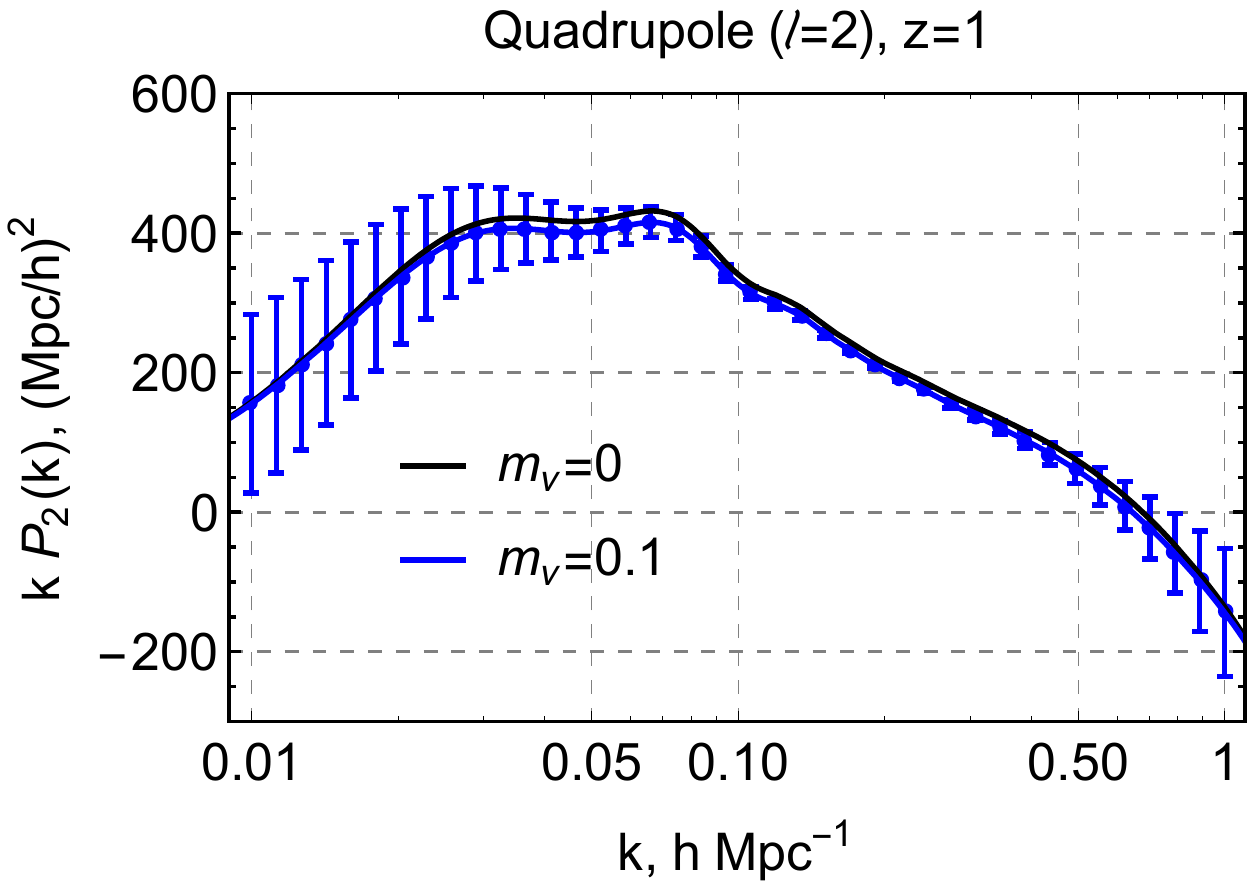} 	
	\includegraphics[keepaspectratio,width=0.49\linewidth]{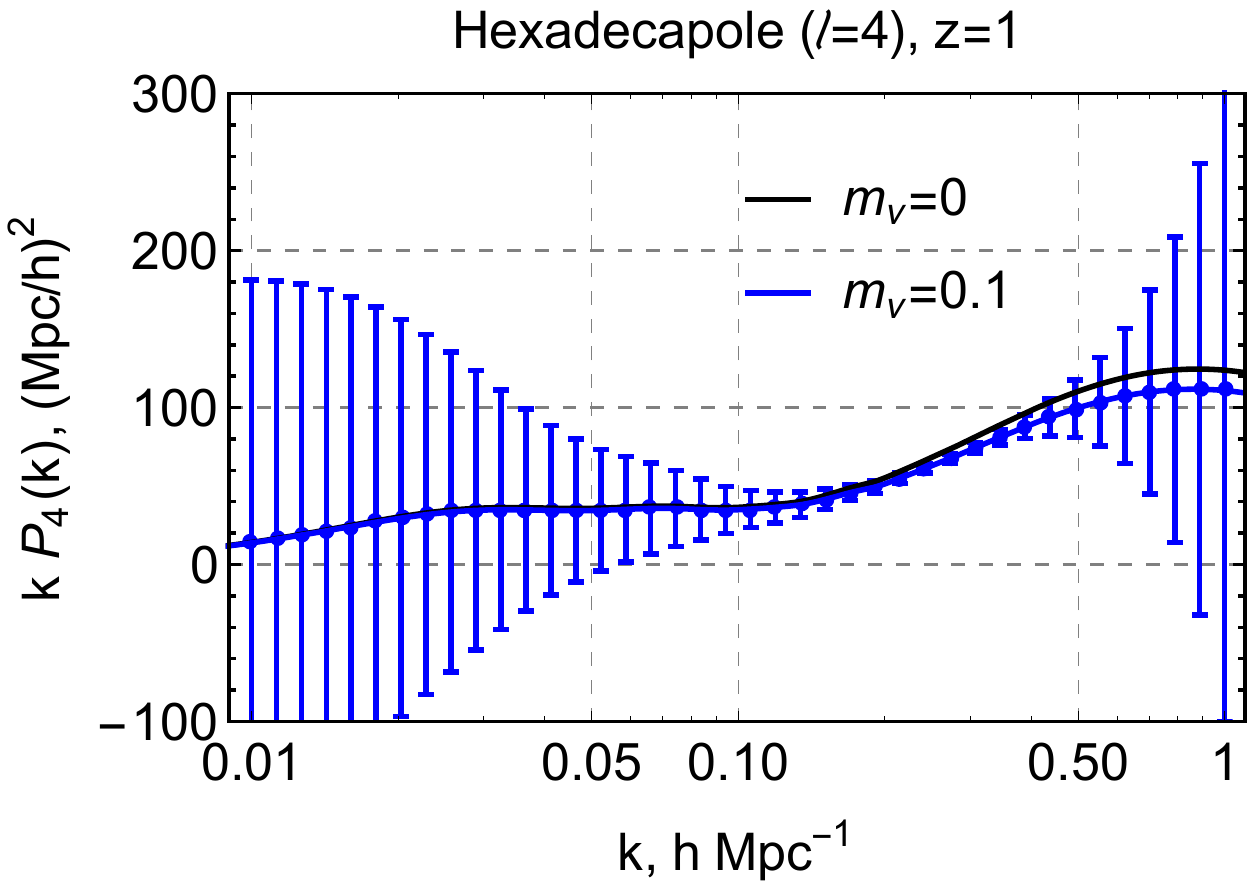}
	\includegraphics[keepaspectratio,width=0.49\linewidth]{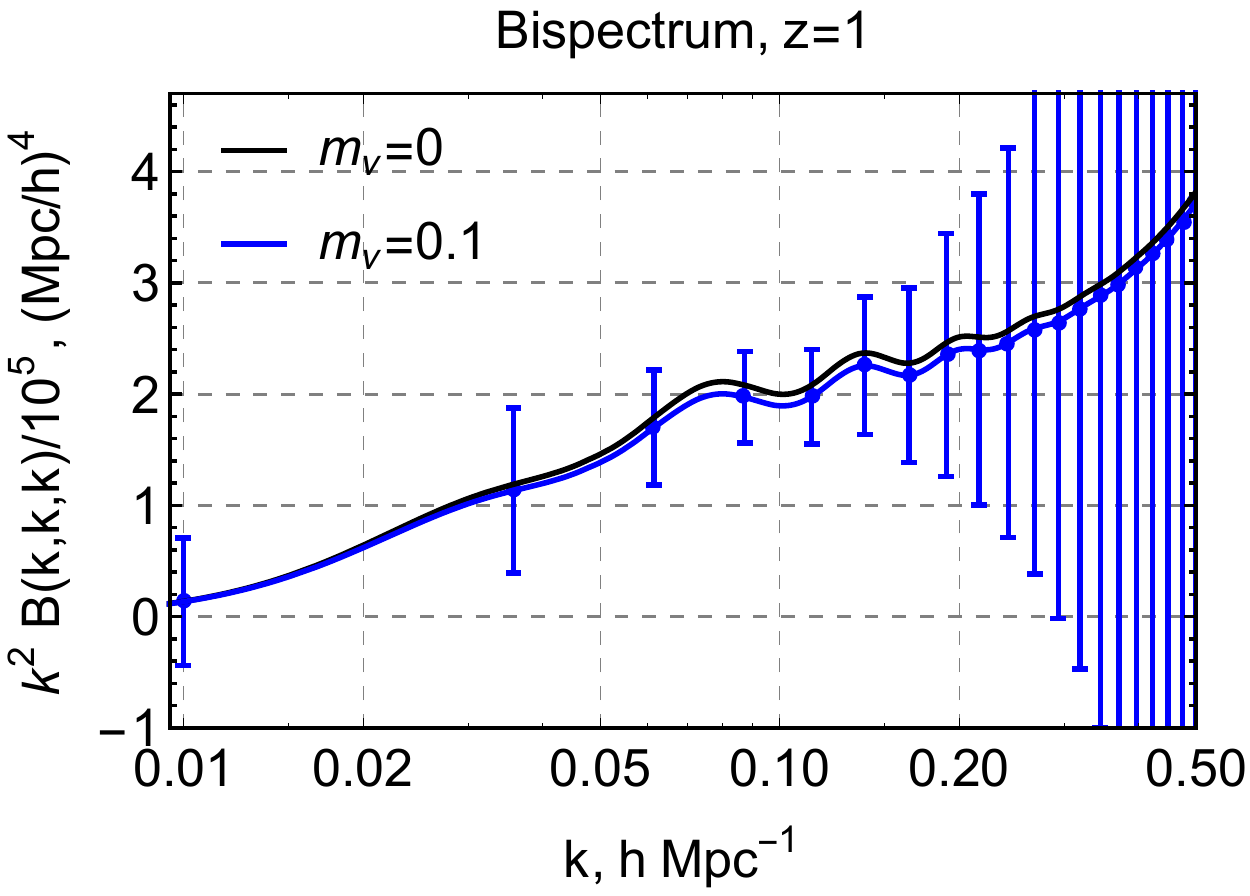}
	\caption{Monopole, quadrupole and hexadecapole of the one-loop galaxy power spectrum and the tree-level monopole bispectrum for the equilateral configuration at $z=1$: theoretical curves for $m_\nu = 0.1$ eV (blue line), $m_\nu = 0$ eV (black line), along 
	with the mock data. 
	The errorbars on short scales blow up because of the theoretical uncertainty added to the 
	covariance. 
	 Note that the monopole power spectrum and bisepectrum
	 increase at high momenta due to the strong shot noise contribution.
	}
	\label{fig:rsd_power} 
\end{figure}

To generate the power spectrum and multipole datasets, 
we evaluate expressions (\ref{Pg},\ref{Pell})
for the fiducial cosmology and bias
parameters in each redshift bin. 
We consider wavenumbers spanning the range $0.01\,h\Mpc^{-1}\leq k\leq1\,h\Mpc^{-1}$ and split them into 40 logarithmically spaced k-bins ($N_k=40$).
Note that the choice of $k_{\text{max}}=1~h$/Mpc is arbitrary, our results do not depend
on it by virtue of the theoretical error.
Recall that the range of fundamental bins covers $k_{\rm f}\equiv2\pi/V^{1/3}=(2.8-3.8)\times10^{-3}\,h\Mpc^{-1}$ depending on a particular redshift bin in the survey. We neglect the effects of the survey window function. 
They are sizable only on large scales (around $k_{\text{f}}$), which contain very little information compared to the mildly non-linear scales.

The mock bispectrum data were generated using the tree-level formulas 
for real and redshift spaces (\ref{Bg},\ref{Bg_rsd}). 
For both analyses we set
$k_\minn=0.01\,h\Mpc^{-1}$, $k_\maxx=0.5\,h\Mpc^{-1}$ and split the corresponding momentum interval into 20 linearly spaced k-bins of width $\Delta k=0.026\,h\Mpc^{-1}$ ($N_k=20$), which produced 825 different triangular configurations. 
For simplicity, we neglect binning effects both for the mock data and theoretical models.

In Fig.~\ref{fig:rsd_power} we display the power spectrum multipoles and the 
equilateral isotropic bispectrum of our mock data at the mean redshift $z=1$ along with the errors representing the diagonal part of the covariance matrix.  
In each panel we show two 
theoretical curves corresponding to zero and fiducial neutrino masses.
The error on large scales is dominated by sample variance, on short scales - by the 
theoretical uncertainty.

\section{Results}
\label{sec:results}

In this section we present our results and discuss various effects 
that contribute to the cosmological parameter measurements from LSS \cite{Lesgourgues:2006nd}. 
We first discuss in detail the neutrino mass limits, and then focus on the parameters of the minimal 
$\Lambda$CDM.

Recall that the main effects of massive neutrinos on the \textit{linear} dark-matter-baryon power spectrum is its uniform 
suppression\footnote{We emphasize that we consider galaxies to be tracers of the baryon+CDM fluid, in which case the massive neutrino suppression will be smaller than the well-known value $1 - 8 f_\nu$ for the \textit{total} matter power spectrum including massive neutrinos. } for $k\gg k_{\text{nr}}$ ($k_{\text{nr}}$ is the comoving free-streaming wavenumber at the time when the neutrinos become non-relativistic),
\be
\frac{P_{\lin}(k,z=0)^{m_\nu \neq 0}}{P_\lin(k,z=0)^{m_\nu = 0}} \approx 1 - 6 f_\nu\,.
\ee
The modes with $k\ll k_{\text{nr}}$ behave as if neutrinos were dark matter,
and the matter power spectrum is the same as in the massless neutrino case.
For small neutrino masses ($m_\nu\lesssim 100$ meV)
the transition between these two asymptotics 
happens at large scales dominated by cosmic variance, 
which makes it hard to observe even with future surveys.
At short scales where the statistical error reduces, 
the massive neutrino effect becomes almost indistinguishable from a simple
reduction of the primordial power spectrum amplitude. 
However, the neutrino suppression has a non-trivial redshift dependence,
which may help to break this degeneracy if several redshifts are combined in the analysis.
The situation becomes more intricate if we consider galaxies, non-linearities and redshift-space distortions. 
To clearly illustrate different effects that impact the neutrino mass measurements, 
we start by the analysis of real space.

\begin{figure}[!t]
	\includegraphics[keepaspectratio,width=0.485\linewidth]{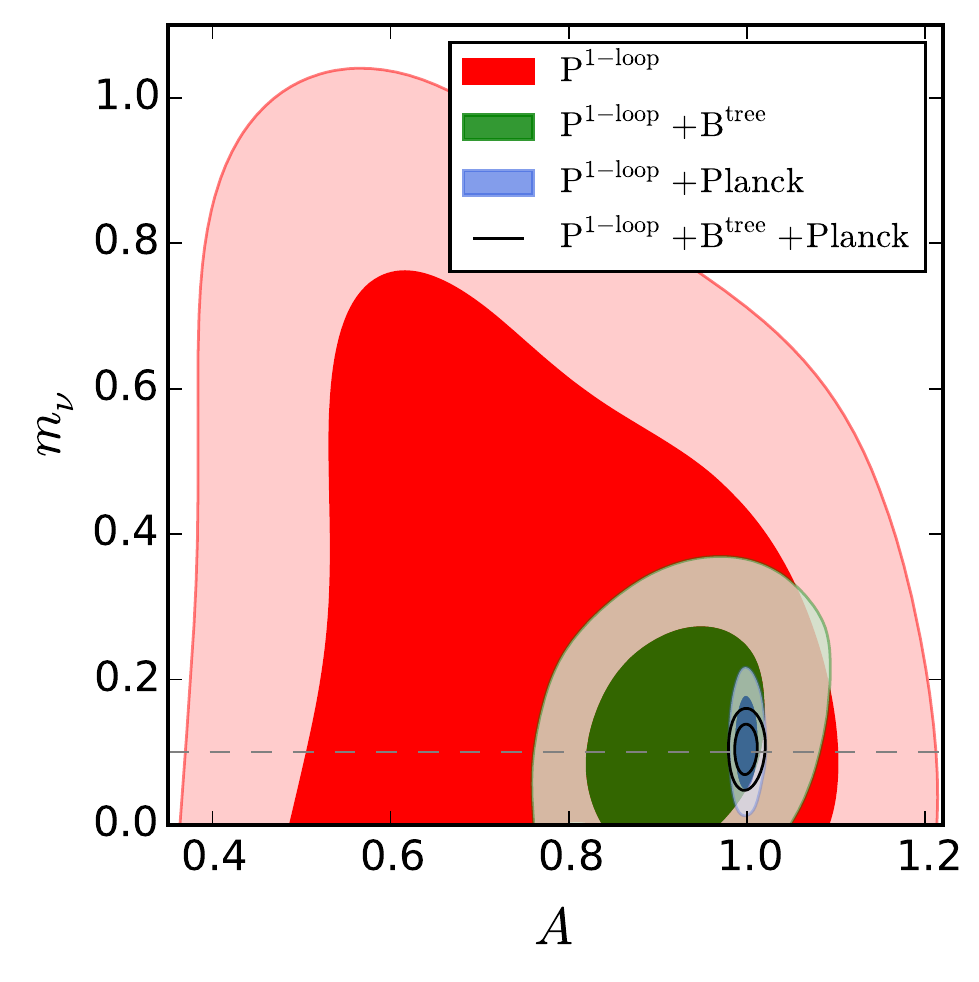}
	\includegraphics[keepaspectratio,width=0.492\linewidth]{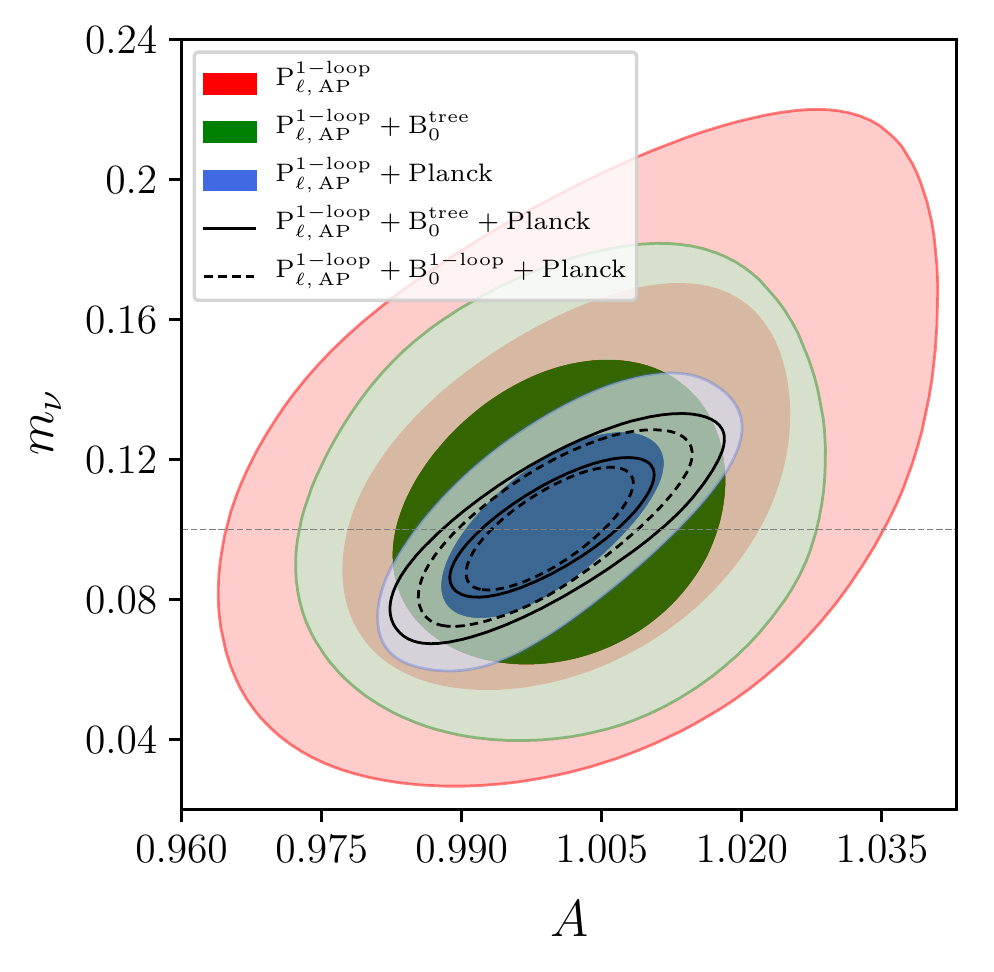}	
	\caption{$1\sigma$ and $2\sigma$ contours in plane $m_\nu-A$ for 
	the real space (left panel) and redshift space (right panel) analyses. $m_\nu$ is quoted in units of $\eV$. See also Tab. \ref{tab:constraints} for the marginalized constraints. 
	}
	\label{fig:real_mnu} 
\end{figure}

\subsection{Real space}

The results of our real space analysis are presented in the left panel of Fig.~\ref{fig:real_mnu}. 
For clarity, we show
only the marginalized 2d distribution for $A$ and $m_{\nu}$.
The marginalized 1-d constraints on neutrino masses and cosmological parameters are listed in the top panel of Tab. \ref{tab:constraints}. Note that the value for the 1$\sigma$ neutrino mass error is approximate 
as the posterior distribution
is quite non-Gaussian in the real space case.

We start with the pure power spectrum case.
For galaxies in real space the situation is complicated by bias, 
which does not allow one to extract the amplitude of primordial fluctuations directly 
from the power spectrum normalization, which goes as $\propto A b_1^2$ on large scales. 
The degeneracy between $A$ and $b_1$ may be partly broken by the loop corrections to the underlying dark matter density, 
which scale as $A^2 b_1^2$. However, at one-loop order there also appear additional loop contributions 
due to the non-linear bias expansion, which are degenerate with the dark matter loops. 

In principle, a non-trivial redshift dependence may help break
the degeneracy between $A$ and $b_1$. However we treat the bias coefficients as nuisance parameters and marginalize over their time-dependence within our conservative analysis. 
Thus, we cannot probe the redshift evolution of the neutrino suppression. Therefore, the degeneracy between $A$ and $b_1$
is very strong, and once we marginalize over $b_1$, the neutrino mass remains largely unconstrained, see the red contour in the left panel of Fig.~\ref{fig:real_mnu}. 
We can rule out only very heavy neutrinos with $m_{\nu}\sim 0.5$ eV,
whose effect on the matter power spectrum is significantly scale-dependent on mildly non-linear scales.

The situation greatly improves upon adding 
the tree-level bispectrum likelihood, see the green contour in the left panel of Fig.~\ref{fig:real_mnu}.
The bispectrum amplitude scales like $A^2 b_1^3$, which helps to
break the notorious degeneracy between $A$ and $b_1$ present in the power spectrum.
Still, the poor accuracy of amplitude and tilt measurements does not allow for a robust detection of the 
neutrino mass.

Once we add the Planck likelihood, it becomes a dominant source of the cosmological information, 
and helps reduce the errorbar on the neutrino masses down to $48$ meV for the joint 
power spectrum + bispectrum + Planck
likelihood. 
This happens mainly because the Planck data fix
the amplitude and tilt of the fluctuation spectrum. 
Since our Planck likelihood is not sensitive to the neutrino mass, 
our results suggest that a joint CMB + LSS analysis yields a large information gain crucial for a robust neutrino mass detection.  
In particular, the joint $\rm P^\lp+\rm B^\tree + \rm Planck$ constraints
on $h$ and $\omega_{cdm}$ are much better than the Planck limits alone.
This happens because the degeneracy direction in $h-\omega_{cdm}$ plane
present in the LSS data is almost orthogonal to the degeneracy direction
in the CMB data. We will return to this question shortly.
However, by comparing the marginalized 
constraints from Planck alone and the joint 
$\rm P^\lp+\rm B^\tree + \rm Planck$ likelihoods one may notice that 
our LSS real-space data tighten the bounds on $n_s$, $\omega_{b}$ and $A$ only 
marginally.

\begin{table}[h]
	\centering
	\begin{tabular}{|c|ccccc|c|}
		\hline
		Set & $10^3\,h$ & $10^2\,A$ & $10^3\,\omega_{cdm}$ & $10^4\,\omega_b$ & $10^3\,n_s$ & $m_\nu, \meV$ \\
		\hline
		Planck	& 15.2& 0.9& 1.4& 1.6& 4.2& $< 260$\\ \hline
$\rm P^\lp$	& 37.4& 18.9& 13.8& 38.1& 62.2& $<510$\\
		$\rm P^\lp\!+\!B^\tree$			& 19.1& 7.2& 7.3& 19.5& 23& $<210$ \\
		$\rm P^\lp\!+\!Planck$			& 1.7& 0.9& 0.7& 1.2& 3& $42$ \\
		$\rm P^\lp\!+\!B^\tree\!+\!Planck$	&1 & 0.8&0.4& 1.1& 2.8& $22$\\
		\hline
		$\rm P_{\ell,nw}^\lp$				&7.9&1.8&2.7 &13.8 &7.6 &$55$ \\
		$\rm P_\ell^\lp$				&7.7&1.7&2.7 &9.3 & 6.5&$48$ \\
		$\rm P_{\rm \ell,\,AP}^\lp$				&7.6&1.6&2.4 &9.1 &6.2 &$38$ \\
		$\rm P_{\rm \ell,\,AP}^\lp\!+\!B_0^\tree$				&5.5 &1.1 &2 &6 &4.6 &$28$ \\
		$\rm P_{\rm \ell,\,AP}^\lp\!+\!Planck$			& 2&0.8 & 0.4& 1.1&2.6 &$17$ \\
		$\rm P_{\rm \ell,\,AP}^\lp\!+\!B_0^\tree\!+\!Planck$			&0.9 &0.7 &0.3 &1.1 &2 & $13$\\
		\hline
		$\rm P_{\rm \ell,\,AP}^\lp\!+\!B_0^\lp$		& 4.8&0.9 &1.8 &5.2 &3.8 & $23$\\
		$\rm P_{\rm \ell,\,AP}^\lp\!+\!B_0^\lp\!+\!Planck$	&0.8 &0.6 &0.3 &1 &1.8 & $11$\\
		\hline
	\end{tabular}
	\caption{ Marginalized $1\sigma$ errors for the cosmological parameters in the base $\Lambda$CDM with one massive neutrino (see Table \ref{tab:fid}) for different combinations of likelihoods. 
If the $1\sigma$ confidence limit on the neutrino mass overlaps with zero, we display a $68\%$ CL upper bound instead.
	We also show the constraints from a realistic Planck mock likelihood,
	which are in good agreement with the Planck 2018 legacy data \cite{Aghanim:2018eyx}.
		}
	\label{tab:constraints}
\end{table}

\subsection{Redshift space}

To track the sources of improvement brought by the redshift space data 
we analyzed several different likelihoods that feature relevant physical effects separately. 
First, 
we study the impact of the BAO by analyzing a mock dataset without the BAO wiggles.
Second, we scrutinize the information content of the one-loop anisotropic power spectrum with and without the AP effect.
Third, we study different combinations of the power spectrum, tree-level bispectrum and Planck likelihoods. 
Finally, we perform a simplified unrealistic analysis of the one-loop power spectrum. Now we present all these case studies separately.

\textit{BAO and IR resummation.} 
Massive neutrinos affect the size of 
the sound horizon at recombination \cite{Lesgourgues:2006nd}.
Roughly speaking, this scale is imprinted in the matter power spectrum in two ways. 
First, it sets the frequency of the BAO wiggles and second, 
it defines the effective Jeans length for baryons, 
which suppress the short-scale part of the matter 
power before recombination.
In this regard it is instructive to quantify how much information 
on the neutrino masses comes directly from the BAO wiggles.\footnote{To avoid confusion, here we mean the non-reconstructed BAO wiggles, i.e. the ones directly extracted from the redshift-space power spectrum.}
To this end we ran an MCMC power spectrum analysis 
for a mock datasample without the BAO wiggles.\footnote{We thank S.~Sibiryakov
for suggesting us to perform this study.} 
The details of this analysis are given in Appendix~\ref{sec:BAO}, 1d marginalized limits 
are shown in the 6th line of Table~\ref{tab:constraints}.
We found that the BAO decrease the error quite weakly,
from $55\meV$ to $48\meV$. 
Our analysis suggests that the BAO impact notably only $\omega_b$ and $n_s$ measurements,
whereas the improvement for other cosmological parameters is marginal\footnote{It should be pointed out that our analysis neglects instrumental systematics, which can affect the power spectrum shape but not the BAO wiggles.}. 
Nevertheless, we stress that an accurate description of the BAO feature 
by means of IR-resummation (see Sec.~\ref{subsec:IR}) is an essential part of any reliable full-shape measurement.

\textit{Redshift space distortions.} RSD help improve the constrains in several ways. 
First, RSD break the degeneracy between $b_1$, $A$ and $n_s$ by allowing one to measure
different multipoles. 
Since massive neutrinos 
produce a similar suppression in all power spectrum multipoles, 
$b_1$, $A$ and $n_s$ are not enough to simultaneously absorb this suppression in all multipoles at different redshifts.
Second, the degeneracies between different bias coefficients in redshift space are 
partly broken because they enter the multipoles 
on different footing (see Eq.~\ref{Pell}).
These effects yield a significant improvement compared to the 
real space case reflected in Table~\ref{tab:constraints}. 

\textit{Alcock-Paczynski effect.} The AP test provides an additional probe of underling cosmology by mapping $H(z)$ and $D_A(z)$. Since all scales in a galaxy survey are measured in units of $\Mpc/h$, the Hubble parameter drops out of the expressions for the AP effect, which turns out to be sensitive only to the total background matter density $\Omega_{m}$.
Thus, the AP effect probes the background neutrino density fraction~$\Omega_{\nu}$, which explains the improvement observed in Table~\ref{tab:constraints}. The AP test helps reduce the errorbar on the neutrino masses down to $38\meV$. The precision gain for other cosmological parameters is more modest, which suggests that the AP 
geometric information is subdominant compared to the power spectrum shape.

\begin{figure}[!t]
	\centerline{
		\includegraphics[keepaspectratio,width=0.43\linewidth]{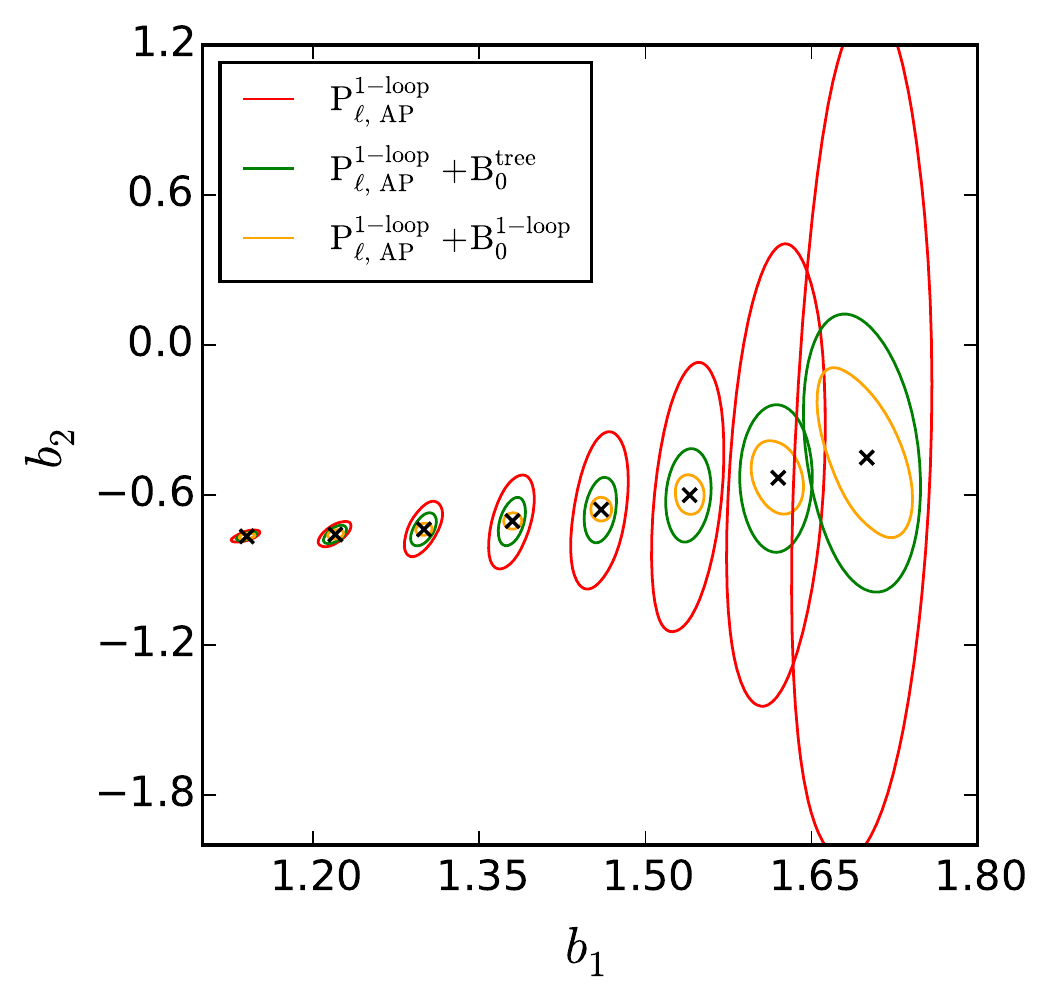}	
		\includegraphics[keepaspectratio,width=0.43\linewidth]{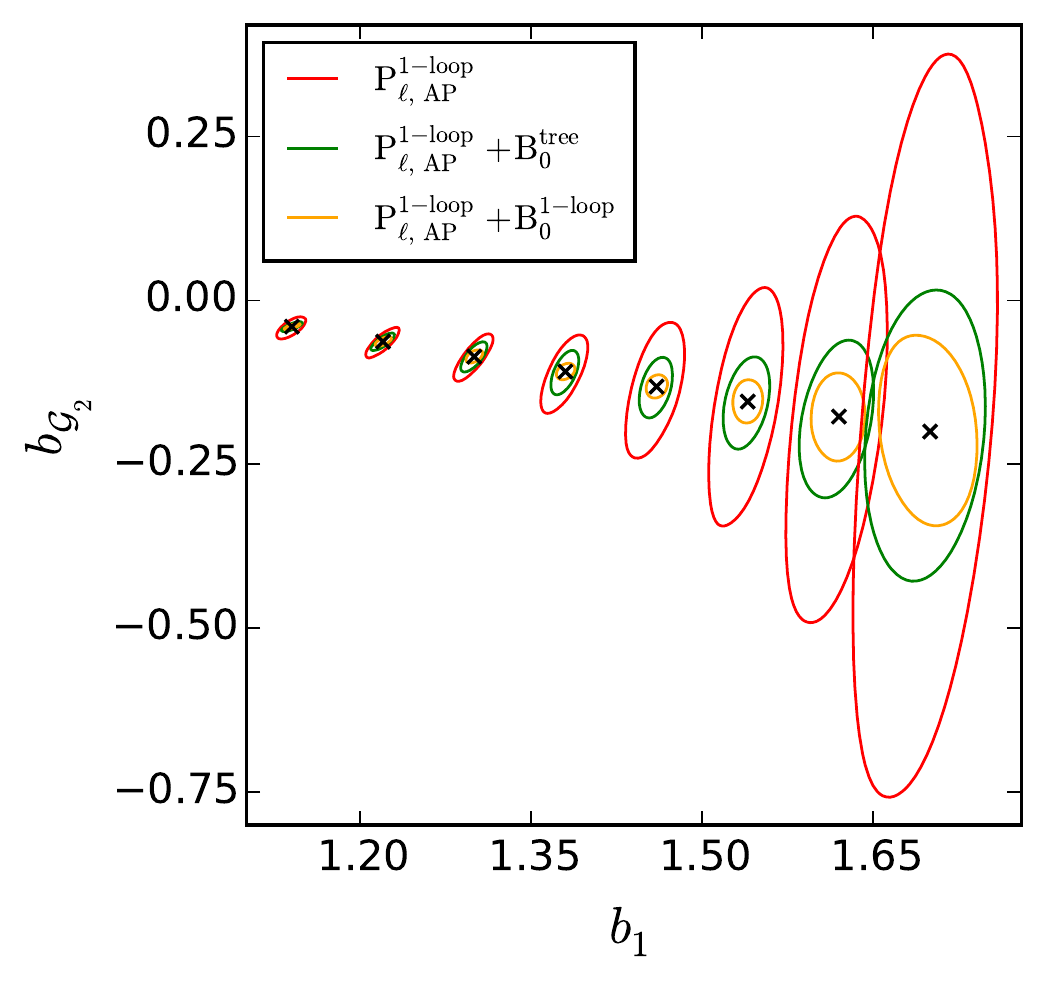} }	
		\includegraphics[keepaspectratio,width=0.43\linewidth]{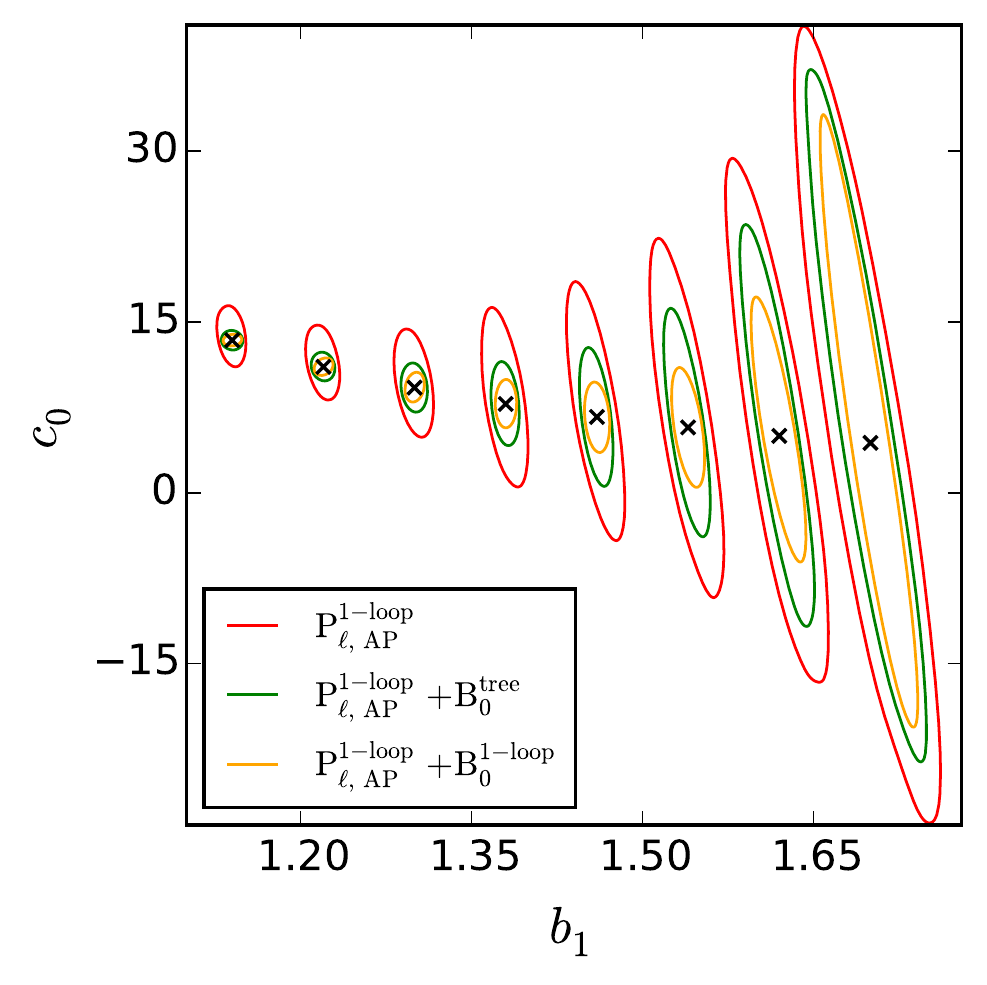}
	\includegraphics[keepaspectratio,width=0.415\linewidth]{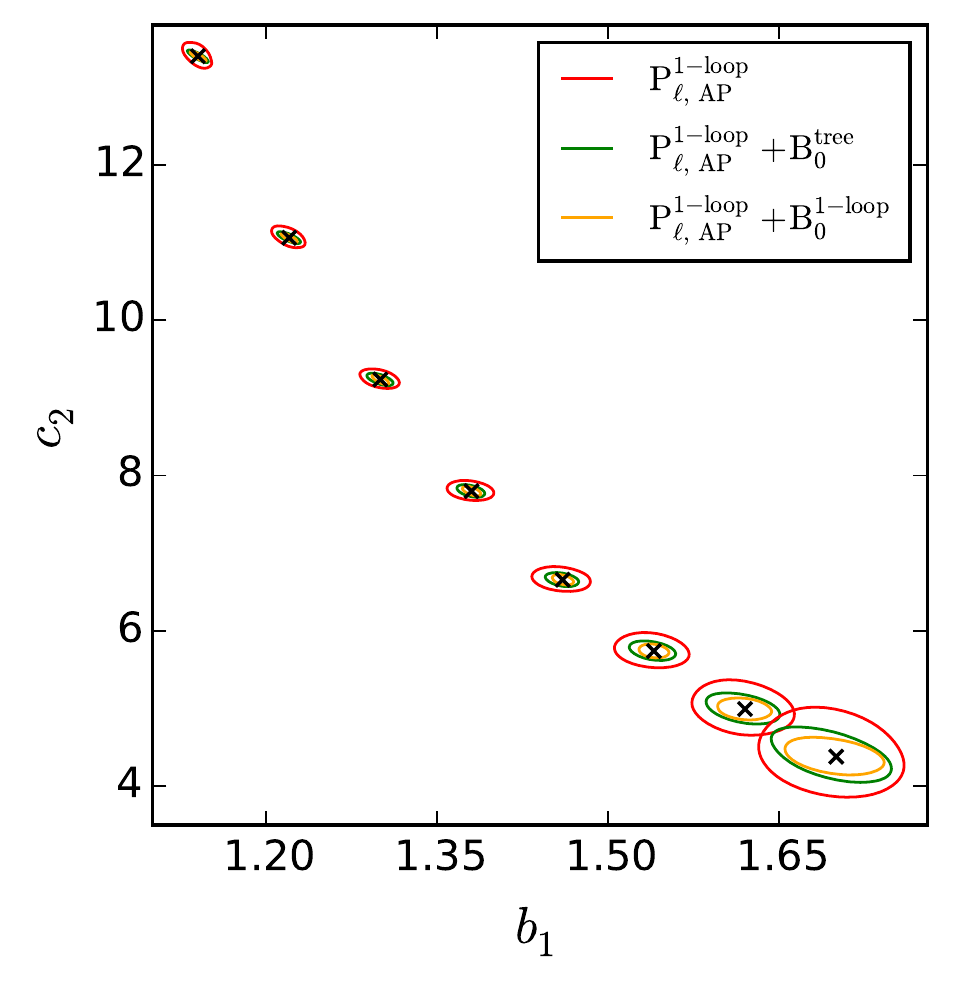}
	\centering{
	\includegraphics[keepaspectratio,width=0.43\linewidth]{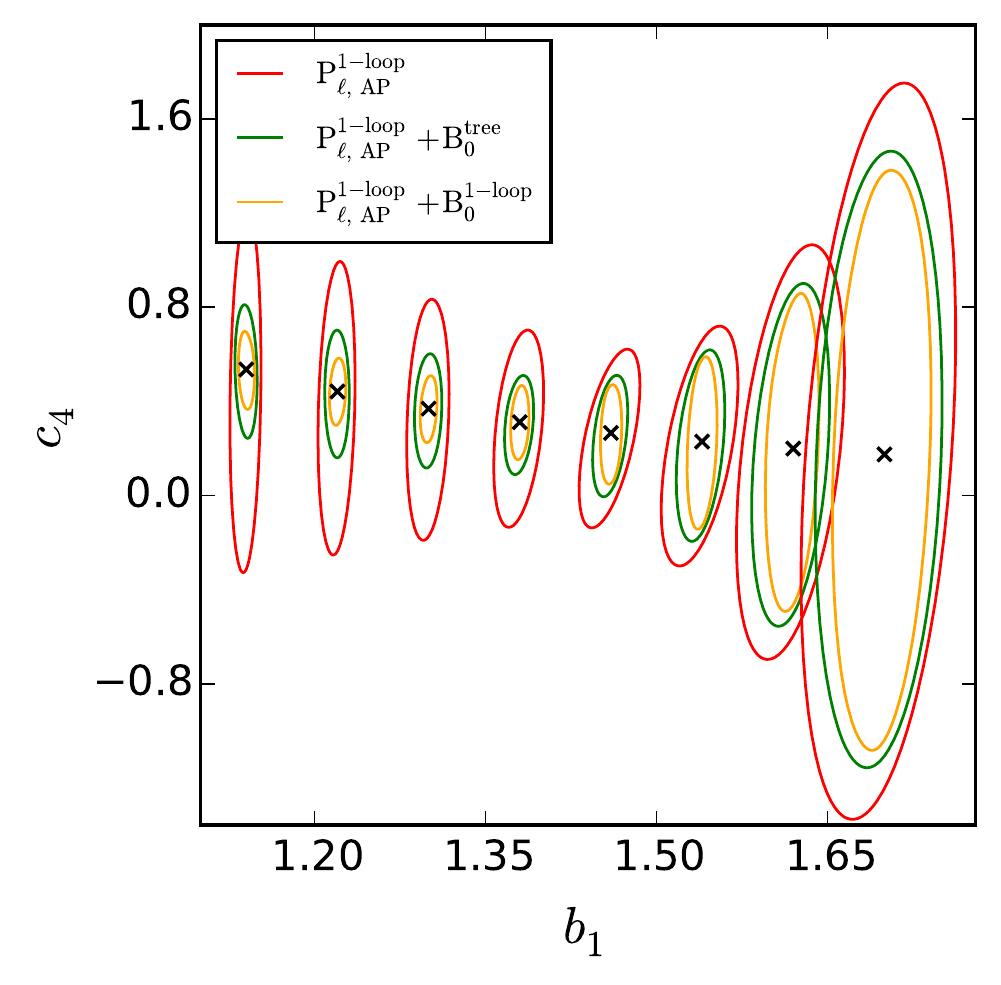}}
	\includegraphics[keepaspectratio,width=0.415\linewidth]{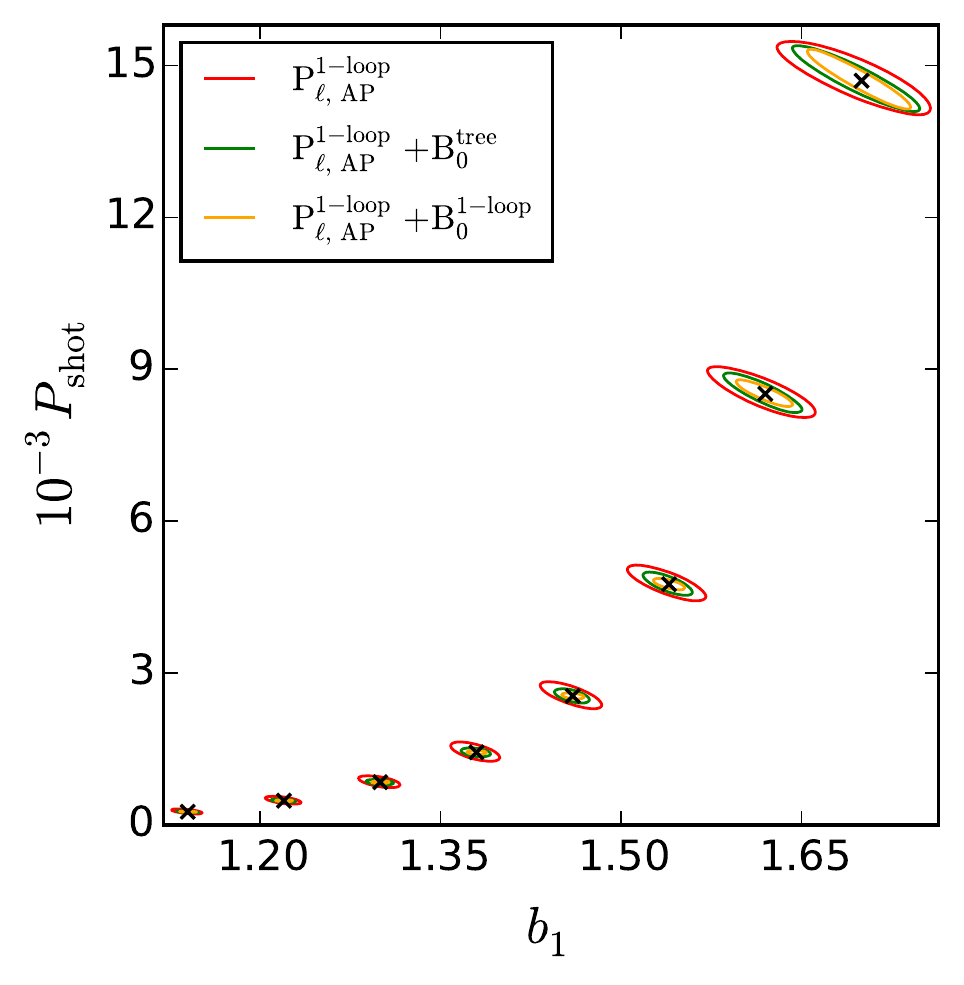}
	\caption{$1\sigma$ contours for bias parameters and RSD counterterm coefficients obtained for different combinations of the power spectrum and bispectrum likelihoods. Black crosses reflect the fiducial values listed in Table~\ref{tab:bias}. The counterterm values $c_i$ are quoted in units [Mpc$/h]^2$, the shot noise 
	$P_{\text{shot}}$ in units [Mpc$/h]^3$.}
	\label{fig:rsd_bias} 
\end{figure}

\textit{Planck likelihood.} As anticipated, the neutrino mass constraints become significantly stronger upon adding the Planck likelihood,
which allows one to define all standard cosmological parameters much better than the Euclid power 
spectrum data alone. 
Remarkably, the joint Euclid $+$ Planck likelihood constrains $\omega_{cdm}$
and $h$ much better than each of these likelihoods separately.
This happens because LSS breaks the corresponding parameter 
degeneracy present in the CMB data. 
This effect will be discussed in more detail in Sec. \ref{subsec:cosmopar}.

As an additional cross-check, in Appendix \ref{app:gauss_planck} we reanalyze the LSS 
likelihoods in combination with 
a multivariate Gaussian approximation to the real Planck likelihood \cite{Aghanim:2018eyx}, 
in which case it effectively acts as a prior on the minimal cosmological parameters. 
We show that the use of the Gaussian approximation 
yields very similar results
for the minimal parameters, but noticeably bigger errors on the neutrino mass
as compared to the realistic mock likelihood.
This implies that breaking the CMB degeneracies 
between $m_\nu$ and the cosmological parameters by the LSS data 
is a very important 
contribution to our eventual constraint on the neutrino mass.


\textit{Bispectrum.} The main effect expected from the bispectrum is a more precise 
measurement of bias parameters. 
This is illustrated in Fig.~\ref{fig:rsd_bias}, where we show the constraints on different bias parameters 
as a function of $b_1$, which can be used as a proxy for redshift.
At large redshifts the contours are very wide because 
the loop corrections are sizable only in the high-$k$ tail, which is dominated by the shot noise. 
At low redshifts the effect of bias parameters and counterterms becomes more pronounced at lower wavenumbers
and dominates over the noise, hence the contours shrink.
One can see that the bispectrum substantially improves the constraints on $b_2$ and $b_{\mathcal{G}_2}$,
while the gain for the counterterms is more modest.

Importantly, including the bispectrum tightens limits on the cosmological parameters.
Regarding the neutrino masses, one can specifically 
emphasize much better measurements of the amplitude and tilt, which are comparable to the current Planck limits. Overall, the Euclid data alone are able to constrain the total neutrino mass with an errorbar of $28\meV$. The main advantage of this constraint is that it entirely comes from Euclid data and does not utilize information from the CMB. Even so, this bound is competitive with that coming from the power spectrum $+$ Planck likelihoods without the bispectrum, see Table~\ref{tab:constraints}.


Including the power spectrum, bispectrum and the Planck likelihoods altogether reduces the errorbar on the neutrino masses down to $13\meV$. The information gain of combining the Euclid survey and the CMB data in this case allows one to detect the fiducial neutrino mass $100$ meV at the $7.7\sigma$ significance. 
This constraint can also be interpreted as 
a forecast for the $4.6\sigma$ detection of the guaranteed minimal
neutrino mass $60$ meV.

As far as the one-loop bispectrum is concerned, our over-optimistic (unrealistic) 
analysis with no new bias parameters 
shows that without the Planck data the constraints improve quite noticeably. 
However, if we add the Planck data, there is almost no difference between 
the tree-level and the one-loop bispectrum likelihoods. 
The interpretation of this result is straightforward: 
adding the tree-level bispectrum data breaks degeneracies between CMB and LSS,
which results in a significant improvement.
But once the degeneracy is broken, the gain from adding more of the bispectrum information is very modest.
It would be interesting to understand to what extent the situation can change after 
taking into account higher-order multipole moments and the AP effect in the bispectrum, omitted in the present analysis.

\begin{figure}[h!]
	\centerline{
		\includegraphics[keepaspectratio,width=1.0\linewidth]{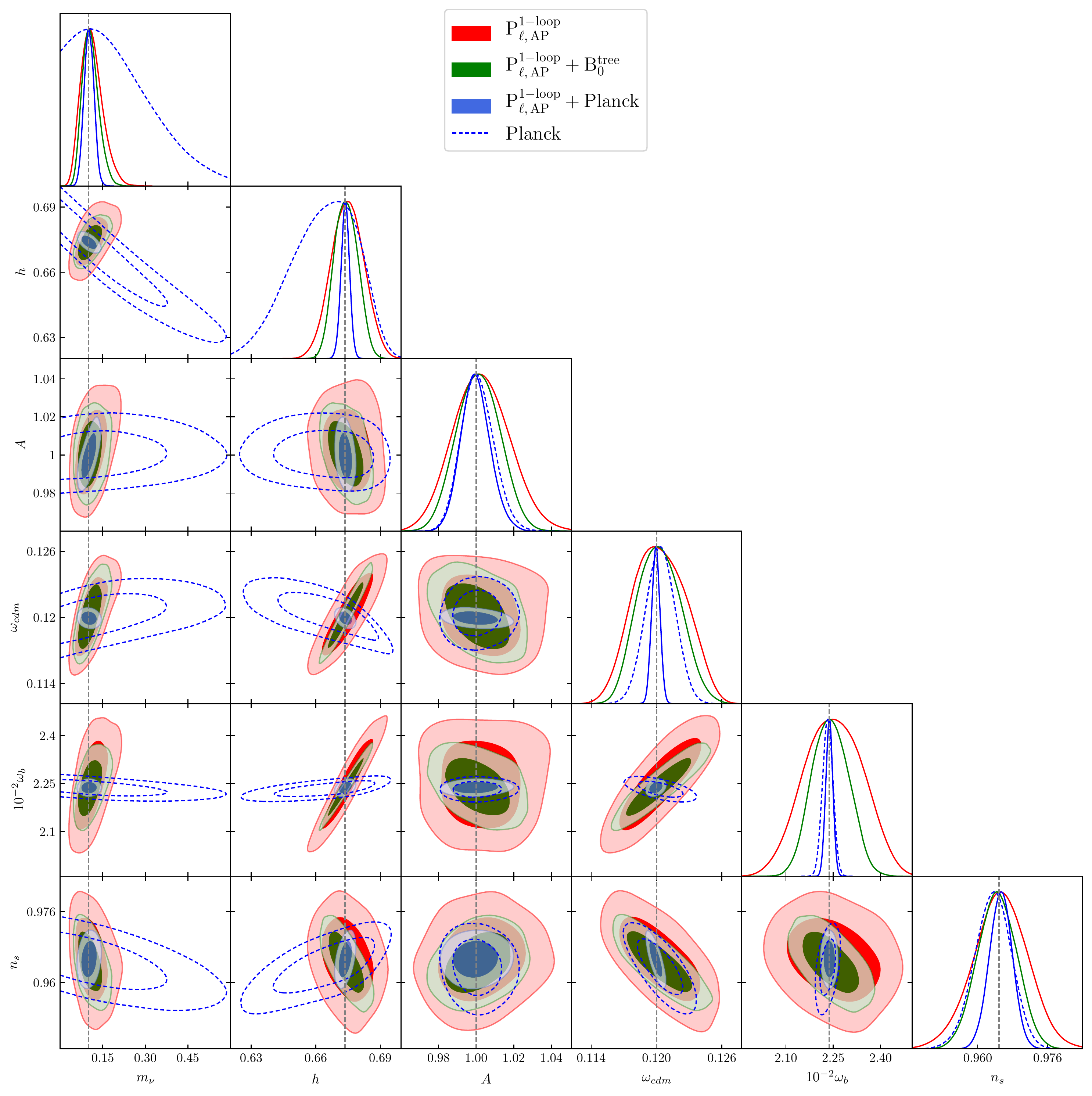}}
	\caption{\label{fig:triang} 
		2d posterior contours and non-normalized 1d  
		marginalized distributions for the total neutrino mass $m_\nu$ in units $[\eV]$ and other parameters of the base $\L$CDM, see also Tab.~\ref{tab:constraints} for the corresponding $1\sigma$ confidence limits.
		The filled and half-filled contours represent
		$68\%$ and $95\%$ confidence limits. The blue dashed lines correspond the 
		Planck 2018 baseline results reproduced with the mock Planck likelihood.
	}
\end{figure}

\begin{figure}[h!]
	\centerline{
		\includegraphics[keepaspectratio,width=1.0\linewidth]{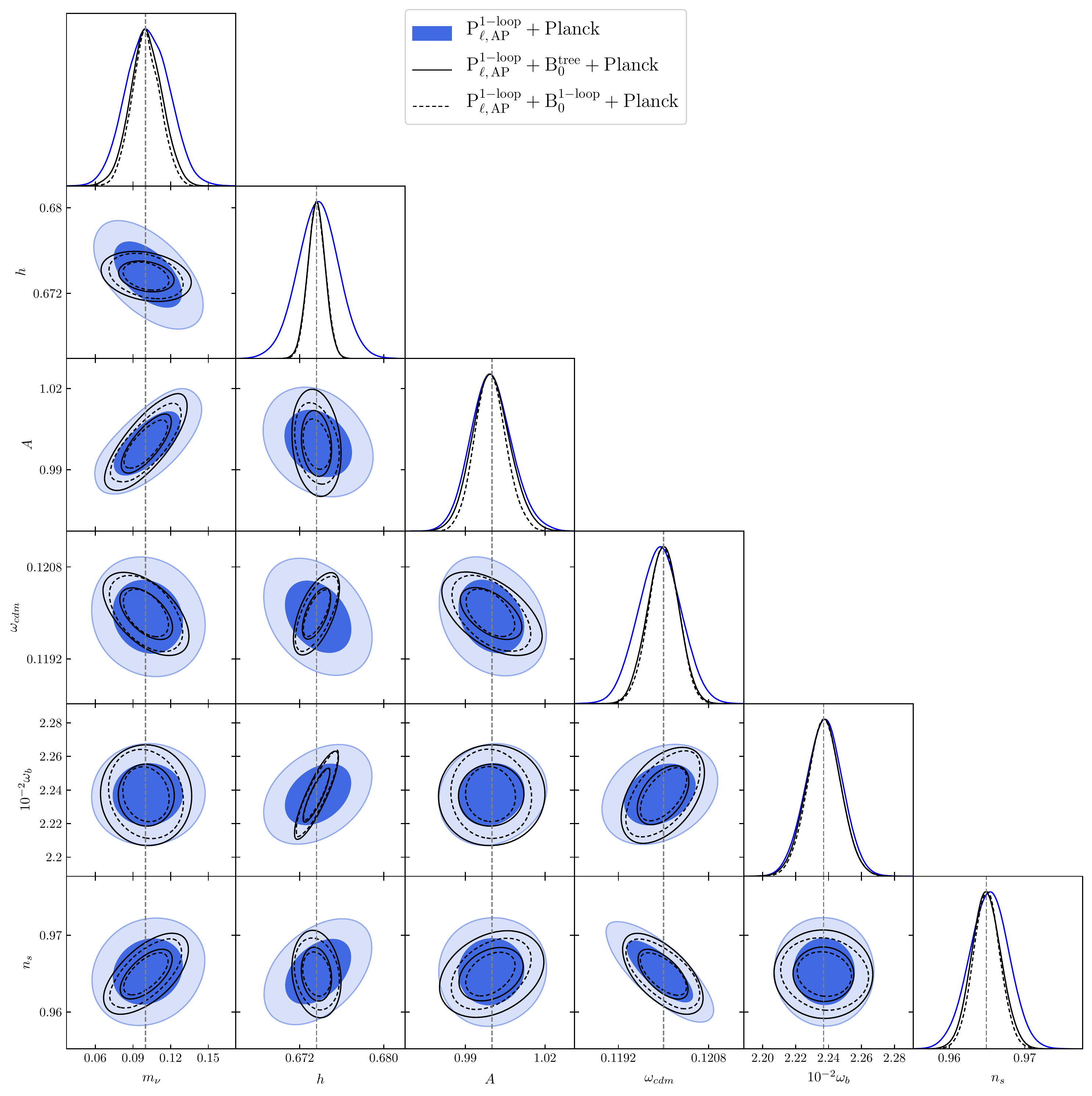}}
	\caption{\label{fig:triang2} 
		The 2d posterior contours and non-normalized 1d  
		2d posterior contours and non-normalized 1d  
		marginalized distributions for the total neutrino mass $m_\nu$ in units $[\eV]$ and other parameters of the base $\L$CDM, see also Tab.~\ref{tab:constraints} for the corresponding $1\sigma$ confidence limits.
		The filled and half-filled contours represent
		$68\%$ and $95\%$ confidence limits. 
	}
\end{figure}

\subsection{Cosmological parameters}
\label{subsec:cosmopar}

Let us discuss now the constraints on the standard cosmological parameters. 
The corresponding triangle plots are displayed in Fig.~\ref{fig:triang} and Fig.~\ref{fig:triang2}. In Fig.~\ref{fig:triang}
we show the posterior distributions for the power spectrum, power spectrum + bispectrum and
power spectrum + Planck likelihoods, respectively in red, green and blue.
The dashed blue line represents the Planck CMB-only constraints on the baseline $\Lambda$CDM with the varied neutrino mass (corresponding values are stated at the very top of Tab. \ref{tab:constraints}).
Fig.~\ref{fig:triang2} shows a similar triangle plot for the combination of Planck and different LSS likelihoods: Planck + power spectrum (also shown in Fig.~\ref{fig:triang}), Planck + power spectrum + tree-level bispectrum, and Planck + power spectrum + one-loop bispectrum (in which we unrealistically
assumed that there are no additional bias parameters compared to the tree-level bispectrum).
For compactness we do not show the contours for the bias parameters and counterterms.

One can observe sizable degeneracies 
between $h$, $\omega_{cdm}$ and $\omega_b$ in the RSD$+$AP power spectrum contours. 
These degeneracies can be readily 
understood from the fitting formulas for the power spectrum \cite{Eisenstein:1997jh,Eisenstein:1997ik,Aubourg:2014yra}.
The shape of the linear matter spectrum expressed in units of Mpc$/h$
depends only on two scales\footnote{We took into account that neutrinos are relativistic
at the matter-radiation equality and do not contribute to the matter density.}:
\begin{subequations}
\label{eq:scales}
\begin{align}
\label{eq:keq}
& k_{\text{eq}}=(7.46\cdot 10^{-2} \omega_{cb}/h)\,\,\,h\text{Mpc}^{-1}\,, 
\\
\label{eq:rd}
& r_d\simeq \frac{55.154\, \e^{-72.3(\omega_\nu+0.0006)^2}h}{\omega_{cb}^{0.25351}\omega_b^{0.12807} }\,\,\,\text{Mpc}/h\,,\quad \text{with} \quad \omega_\nu = \frac{m_\nu}{93.14\,\text{eV}}\,,
\end{align}
\end{subequations}
which define its primary features: 
the equality peak, 
BAO wiggles and the effective Jeans length for baryons.
Similarly to the case of the CMB spectrum, most of the degeneracy directions 
seen in Fig.~\ref{fig:triang}
can be traced back to the locations of these features.
The equality scale $k_{\text{eq}}$ 
controls the position of the power spectrum peak, while the 
sound horizon at the drag epoch sets the frequency of the BAO wiggles along with 
the effective Jean scale for baryons, 
which slow down short-scale clustering before recombination \cite{Lesgourgues:2006nd}.
The amount of this suppression 
is set by the ratio ${\omega_b}/{\omega_{cb}}$.  
This ratio also 
controls the amplitude of the BAO wiggles relative to the broadband.
Note that in units of Mpc both scales $r_d$ and $k_{\text{eq}}$ depend only on
$\omega_{b}$,
$\omega_{cdm}$ and $\omega_{\nu}$. 
If we could measure the power spectrum in units of Mpc,
these physical densities would be the 
only parameters controlling the power spectrum shape.

It is useful to understand the degeneracies seen in the 2d marginalized contours of Fig.~\ref{fig:triang}.
Let us first focus on the pair $\omega_{cdm}-h$. 
Upon marginalyzing over $\omega_{b}$, the constancy of the equality 
and BAO scales in units of Mpc/$h$ fixes the combinations\footnote{More precisely, the principle components extracted from the power spectrum monopole are $r_d/D_V$ and $k_{\text{eq}}D_V$,
where $D_V(z)\equiv (D^2_M(z)z/H(z))^2$, with $D_M=\int_0^z\frac{dz'}{H(z')}$. However, at low redshifts can one approximate $D_V\sim h^{-1}$.} $\omega_{cdm}h^{-1}$
and $\omega_{cdm}h^{-4}$. 
Their geometrical mean roughly 
corresponds to the observed degeneracy direction $\omega_{cdm} h^{-2}$.
The direction $\omega^{0.5}_{b}/\omega_{cdm}$ seen in
the corresponding panel is likely to be a 
combination of $\omega_{b}/\omega_{cdm}$ and the sound horizon~\eqref{eq:rd}.
As for the obtained degeneracy direction $\omega_b h^{-3.3}$, its origin roots in 
the constancy of $r_d$ in units of Mpc/$h$, which leads to $\omega^{0.38}_{b} h^{-1}$
upon marginalization of \eqref{eq:rd} over $\omega_{cdm}$.
Note that $n_s$ has sizable degeneracies with $\omega_{cdm}$, $h$ and $\omega_b$ in the LSS data. 
These degeneracies reflect the fact that the Jeans-like suppression 
of the matter power spectrum by the baryons can be partly compensated by a proper adjustment of the tilt. 

We observe that the Euclid power spectrum data alone (red contours) is competitive with the 
Planck alone (blue dashed contours) in terms of the amplitude, the Hubble parameter and the tilt, whereas $\omega_{cdm}$ and $\omega_{b}$ 
are still measured much worse.
By comparing the green and red contours in Fig.~\ref{fig:triang}, 
one sees that the bispectrum data have almost the same degeneracies
as the power spectrum one, but its inclusion noticeably improves the 
constraints on all cosmological parameters. 
In particular, by combining power spectrum and bispectrum data (green contours) one can notice that the Hubble parameter and tilt measurements reach the precision comparable 
with the recent Planck results whereas the amplitude accuracy even surpasses the Planck limit.
Still, even in this case the Planck CMB data 
remain much better for $\omega_{cdm}$ and $\omega_{b}$. 
All in all, the improvement seen in Fig.~\ref{fig:triang} implies 
that the redshift space bispectrum contains rich information about cosmological parameters.

The bounds on $h$, $\omega_b$ and $\omega_{cdm}$ ameliorate significantly
upon combining the Euclid power spectrum and the Planck likelihoods (blue contours). 
The key observation here is that LSS measures best the combination $\omega_{cdm} h^{-2}$,
which is quite orthogonal to well-known CMB degeneracy related to the angular size of the acoustic horizon $\omega_{cdm}h$. The two degeneracy directions cut each other almost at the right angle 
leaving us with very narrow residual projections onto the $\omega_{cdm}$
and $h$ planes in Fig.~\ref{fig:triang}. 
The resulting errors on $\omega_{cdm}$ and $h$, for instance,
shrink by factors of 4.7 and 18 compared to the current 
Planck limit for the $\L$CDM with the varied neutrino mass. 
A similar situation takes place for the 
$\omega_{cdm}-\omega_{b}$ and $\omega_{b}-h$ pairs,
although the gain is slightly more modest in these cases.
Remarkably, the eventual measurement of the Hubble constant from the combination 
of Planck and future LSS surveys will reach a $0.1\%$ precision, which will be important 
for elucidating the nature of the so-called ``Hubble tension'', see e.g. \cite{Lin:2019htv}
and references therein.

Overall, we conclude that synergy between the CMB and LSS data is essential 
to robustly measure the neutrino masses and cosmological parameters from future surveys.

\subsection{Comparison with previous studies} 

There have been many forecasts on the neutrino mass and cosmological parameter measurements 
with a future galaxy spectroscopic surveys \cite{Takada:2005si,Audren:2012vy,Baldauf:2016sjb,LoVerde:2016ahu,Raccanelli:2017kht,Boyle:2017lzt,Brinckmann:2018owf,Vagnozzi:2018pwo,Boyle:2018rva,Parimbelli:2018yzv,Copeland:2019bho}. In this Section we discuss 
a few selected works that have a significant overlap with our study. 

Our work can be seen as a continuation of the approach put forward in Ref.~\cite{Audren:2012vy}.
This work pointed out for the first time that the theoretical error method is more efficient for the large-scale structure data analysis
than the use of the momentum cutoff $k_{\text{max}}$. 
There are three main improvements between this study and our work. 
First, Ref.~\cite{Audren:2012vy} uses a somewhat simplified theoretical model for the matter power spectrum.
It is based on the HALOFIT fitting formula for the non-linear matter power spectrum of dark matter,
and features only linear bias and redshift-space distortions (the Kaiser formula applied to the HALOFIT). 
However, it includes a correction due to spectroscopic redshift error, which has the same
function form as our redshift-space counterterms. 
Second, Ref.~\cite{Audren:2012vy} considers two theoretical errors: one
due to an imperfect description of massive neutrinos with the HALOFIT (this error is assumed to be 
correlated),
and the other one due to overall non-linear clustering uncertainties (this error is conservatively assumed to be uncorrelated). 
In contrast, the error of our neutrino treatment is taken 
into account on the same footing as the other two-loop uncertainties,
which are correlated as dictated by perturbation theory.
Third, the analysis of Ref.~\cite{Audren:2012vy} did not include the bispectrum.
However, this work performed a 
full MCMC analysis and took into account the Alcock-Paczynski effect.
Despite the mentioned methodological differences, our constraints are in good agreement
with those obtained in Ref.~\cite{Audren:2012vy} and its recent follow-ups \cite{Archidiacono:2016lnv,Brinckmann:2018owf}.
We believe that this happens because (i) the HALOFIT does capture the leading non-linear contributions,
(ii) the non-linear bias coefficients present in our analysis get partly fixed by the bispectrum information,
(iii) the counterterms present in our study have the same functional form as corrections due to spectroscopic redshift errors,
and hence produce very similar effects on parameter space after marginalization. 
Overall, the good agreement between our paper and 
Refs.~\cite{Audren:2012vy,Archidiacono:2016lnv,Brinckmann:2018owf} indicates the 
consistency of the theoretical error framework and its robustness against setup variations.

Recently, the theoretical error formalism was revisited in the context of perturbation theory in Ref.~\cite{Baldauf:2016sjb}.
This reference studied the impact of the theoretical error on the future neutrino mass and non-Gaussianity constraints 
by means of a Fisher matrix analysis of the galaxy power spectrum and bispectrum in real space.
Our work can be seen as a straightforward generalization of Ref.~\cite{Baldauf:2016sjb} to the redshift space case.
Compared to that reference, we run a full MCMC analysis and perform an exact evaluation of the one-loop
power spectrum and tree-level bispectrum of redshift-space galaxies. 
Importantly, the analysis of Ref.~\cite{Baldauf:2016sjb} was done only in real space, and hence it misses
important information from the velocity field, which enters through redshift-space distortions. 
Our study shows that taking into account this information dramatically improves the neutrino mass measurements.

Finally, let us discuss the relation between our work and a recent Euclid forecast on the cosmological  
constraints with the redshift-space bispectrum \cite{Yankelevich:2018uaz}. 
This paper performed a Fisher matrix forecast for the power spectrum and bispectrum
of Euclid galaxies in redshift space for $\L$CDM and extended dark energy models.
The main conclusion of this paper is that the redshift-space bispectrum does not significantly improve the constraints on the cosmological
parameters, both in combination with the power spectrum and with the joint power spectrum + Planck CMB likelihood.
On the one hand, our analysis agrees with Ref.~\cite{Yankelevich:2018uaz} in that the addition of the 
bispectrum to the power spectrum likelihood does not dramatically improve the constraints on the parameters of the minimal $\L$CDM. 
On other other hand, we found a notable improvement for the neutrino mass, 
which was not considered in Ref.~\cite{Yankelevich:2018uaz}. 
This result holds true for the combination of the LSS and the Planck CMB data, in which case the bispectrum also 
significantly improves the measurement of $h$ (in part by breaking the CMB degeneracy between $m_\nu$ and $h$).
Moreover, our analysis of the LSS likelihood without the CMB data yielded somewhat stronger constrains on the 
cosmological parameters compared to those presented in Ref.~\cite{Yankelevich:2018uaz}.
This improvement calls for a more detailed comparison.

Although Ref.~\cite{Yankelevich:2018uaz} uses an identical survey specification and a very similar theoretical model for tree-level the bispectrum 
(modulo IR resummation),
there are several differences.
First, Ref.~\cite{Yankelevich:2018uaz} uses a conservative momentum cutoff $k_{\text{max}}=0.15~h$/Mpc for all redshift bins, whereas the theoretical error allows us to go to much larger wavenumbers, especially at high
redshifts.
Second, Ref.~\cite{Yankelevich:2018uaz} employs the linear model for the redshift-space power spectrum,
supplemented with an exponential fingers-of-God damping.
In this approach the non-linear bias parameters are present only in the tree-level bispectrum,
whereas in our work both the one-loop power spectrum and the tree-level bispectrum share the same bias coefficients $b_2$ and $b_{\mathcal{G}_2}$.\footnote{
This is required for consistency of the theoretical description. Moreover, it has also been verified experimentally 
by measuring the bias parameters independently from the power spectrum and the bispectrum in N-body simulations (see Chapter 4.5 of \cite{Desjacques:2016bnm} for details and references).
}
Thus, these parameters are constrained from the two statistics simultaneously, 
which yields more cosmological information in the combined analysis.
Besides, the one-loop power spectrum has a stronger response to a variation of cosmological parameters as compared to linear theory, and hence its inclusion tightens the constraints by itself.
We illustrate these points in Appendix~\ref{app:kmax}.
We believe that these effects
resulted in somewhat tighter
constraints compared to those presented in Ref.~\cite{Yankelevich:2018uaz}.

It should be pointed out that Ref.~\cite{Yankelevich:2018uaz} 
features some important improvements compared to our study.
First, it properly takes into account higher multipole moments 
of the redshift space bispectrum and the binning effects,
which are ignored in our work. 
Importantly, it shows that not considering higher moments beyond the monopole bispectrum leads to a significant
loss of information.
Second, it scrutinizes the impact of 
the covariance between the power spectrum and the bispectrum\footnote{ As argued before, we do not use this covariance because formally it appears at higher orders
in perturbation theory, and for consistency requires taking into account higher order corrections to the power spectrum and bispectrum covariances. 
Note that the theoretical error formalism is designed to effectively
account for the presence of the covariance between the power spectrum and 
higher order statistics by removing the information from short scales, where this covariance
becomes important \cite{Baldauf:2016sjb}. }, 
and different prescriptions for the stochastic noise. Finally, Ref.~\cite{Yankelevich:2018uaz} analyzes 
the minimal $\L$CDM along with
extended dark energy models while in our analysis we restrict ourselves only to the base $\L$CDM with massive neutrinos.

\section{Conclusion and Outlook}
\label{sec:concl}

We have presented a forecast for the cosmological parameter and neutrino mass measurement
with a Euclid-like spectroscopic survey mock dataset.
Our analysis contains several improvements compared to previous studies. 

We use a complete analytical model for the power spectrum and bispectrum, 
which includes non-linear galaxy bias 
and redshift space distortions. 
First, we calculate explicit one-loop perturbation theory power spectra for the underlying matter 
field, which should be contrasted with the HALOFIT 
semi-analytic formula \cite{Takahashi:2012em,Bird:2011rb} adopted in some previous studies.
Second, we rigorously take into account the non-linear evolution of BAO.
Third, we use the general non-linear biasing prescription
and do not make any assumptions about the values and time-dependence of the corresponding 
bias coefficients.
Fourth, we consider the non-linear redshift-space mapping which goes beyond the linear Kaiser formula. 
We parametrize various short-scale effects (e.g. fingers-of-God and baryonic feedback) 
by means of counterterms, for which we also allow arbitrary values and time-dependence. 
This way we avoid possible biases that might be introduced by simplified
phenomenological prescriptions, e.g. the fingers-of-God exponential damping.
Another important aspect of our analysis is the bispectrum, which notably improves the constraints 
on cosmological parameters.
Our baseline theoretical model thus consists of the one-loop power spectrum and the tree-level
bispectrum and can readily be applied to data analysis.

We make use of the MCMC technique and evaluate the non-linear power 
spectrum and bispectrum for each sampled set of cosmological parameters.
We explicitly include the correlated theoretical error in our 
likelihoods. This error is based on estimates 
for higher-order loop corrections that are omitted 
in our theory calculations. 
Adding the theoretical error to the covariance matrix 
is equivalent to marginalizing over the shape of higher order non-linearities.
This makes our predictions insensitive to the choice 
of the momentum cutoff scale $k_{\text{max}}$
commonly used in the previous LSS studies.

\begin{figure}[!t]
\centering
	\includegraphics[keepaspectratio,width=0.65\linewidth]{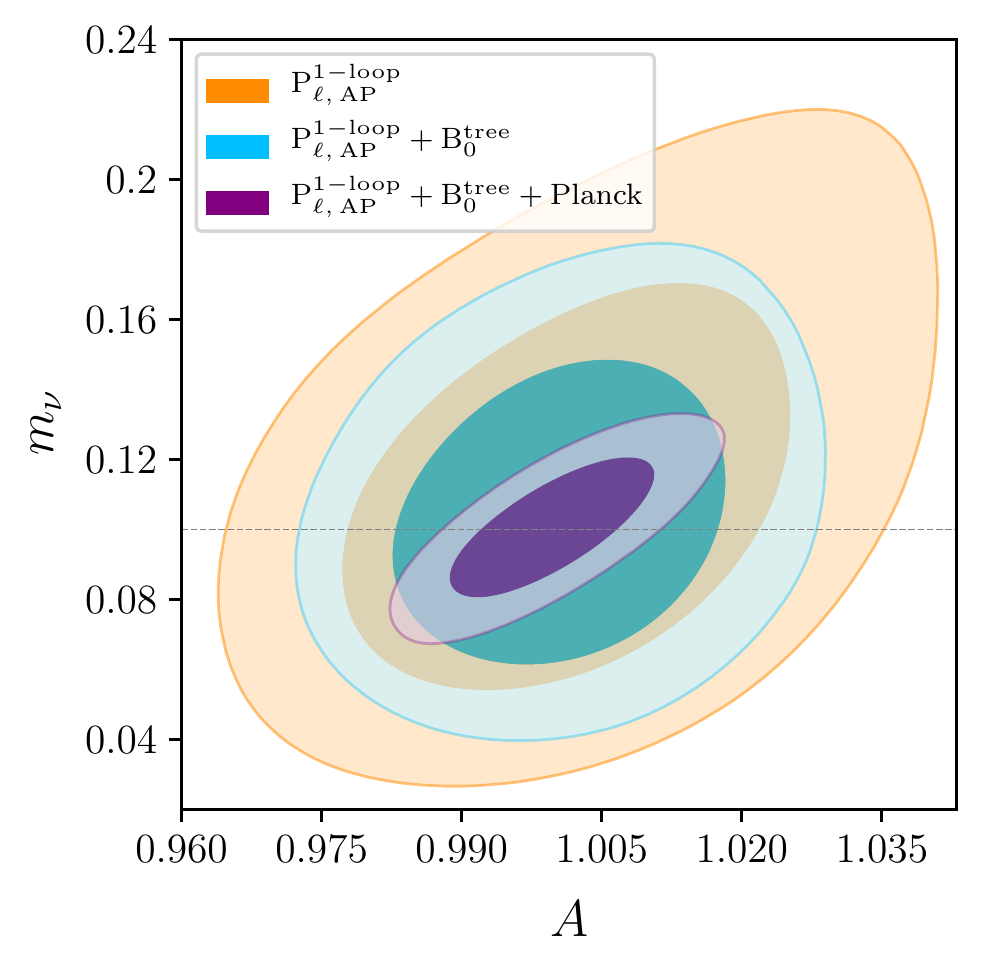}
	\caption{Posterior $1\sigma$ and $2\sigma$ contours in the $m_\nu-A(\equiv A_s/A_{s,\text{fid}})$ plane for
	the following likelihoods: 
	one-loop power spectrum only (orange), 
	one-loop power spectrum + tree-level bispectrum monopole (blue),
    one-loop power spectrum + tree-level bispectrum monopole + Planck (violet).
	$m_\nu$ is quoted in $\eV$. 
	}
	\label{fig:final} 
\end{figure}

We show that even under most conservative assumptions on the galaxy bias, redshift-space 
distortions and non-linear physics, the future galaxy clustering data alone will be able to
deliver satisfactory constrains on cosmological parameters and the sum of neutrino masses.
Specifically, by using the combination of the power spectrum and bispectrum
one will be able to probe the neutrino
mass with an errorbar of $28\meV$. 
By adding the CMB data from the Planck satellite, the error can be  
reduced down to $13\meV$. 
This forecasts the detection 
of the minimal total neutrino mass with 4.6$\sigma$ (7.7$\sigma$)
significance in the case of the direct (inverted) hierarchy.
Our results are summarized in Fig.~\ref{fig:final}.

It seems very unlikely that we will have no information on the values and time-dependence 
of the bias parameters and RSD counterterms 
by the time when the future LSS data are gathered.
The most direct way is to obtain these priors from realistic mock catalogs.
As far as the bias coefficients are concerned, their values can be estimated from 
the galaxy-galaxy lensing cross-correlation \cite{Abbott:2017wau}, 
or some semi-analytic models, e.g. the peak-background split (see \cite{Desjacques:2016bnm} and references therein).\footnote{
One might also expect that some information on counterterms and bias parameters can be extracted 
from non-perturbative observables. 
Indeed, the dark matter real-space counterterm can, in principle,
be measured from the one-point counts-in-cells statistics \cite{Ivanov:2018lcg}.    
}
Using these priors may drastically improve our
conservative limits which were obtained under very agnostic assumptions about the properties of
the Euclid galaxy sample.

There are several directions in which our study can be ameliorated.
First, one can perform a more accurate analysis of the redshift space bispectrum
that would include higher multipole moments, the Alcock-Paczynski effect,
and a more general treatment of stochastic contributions.
Second, one can extend the analysis to the case of the two-loop power spectrum and one-loop bispectrum. In that case one would have to consistently take into account 
non-Gaussian contributions to the covariance matrices and the cross-covariance between the power spectrum and the bispectrum.
Third, it would be important to see how our results can be affected by instrumental 
uncertainties of a Euclid-like survey. 
Fourth, one has to verify our assumptions on the theoretical errors with N-body simulations.
Fifth, one could make a similar analysis for other combinations 
of upcoming CMB and LSS surveys. 
Another interesting exploration venue is a forecast for non-minimal cosmological models, e.g. 
dynamical dark energy. 
We leave these tasks for future work.

\section*{Acknowledgments}

It is a pleasure to thank 
Vincent Desjacques,
Emanuele Castorina,
Cristiano Porciani,
Valery Rubakov,
Emiliano Sefusatti,
Roman Scoccimarro, and 
Matias Zaldarriaga for fruitful conversations and encouragement.
We are especially grateful to 
Deigo Blas,
Sergey Sibiryakov and
Marko Simonovi\'c for valuable comments on the draft and 
for many important suggestions 
that had a huge impact on this analysis.
We are grateful to the anonymous referee his suggestion to revisit the Planck likelihood analysis,
which improved our final results compared to the first version of the paper.
M.I. thanks the CERN Theory Department 
and MIAPP Munich
for hospitality during the
completion of this work. 
M.I. is partially supported by the Simons Foundation’s \textit{Origins of the Universe} program.
This work was partly supported through RFBR grant 17-02-01008. 
Modification of cosmological \texttt{CLASS} code has been supported by the Foundation for the Advancement of Theoretical Physics and Mathematics `BASIS'. The analysis of the Euclid sensitivity to the neutrino masses was supported by the RSF grant 17-12-01547. All numerical computations of this work were made at the MVS-10P supercomputer of the Joint Supercomputer Center of the Russian Academy of Sciences (JSCC RAS).

\appendix

\section{Explicit expressions for galaxy power spectrum multipoles}
\label{app:mult}

Let us start with the bias loop corrections to the real space galaxy power spectrum,
\be
\begin{split}
&{\cal I}_{\delta^2}(k)=2\int_\q F_2(\q,\k-\q)P_{\text{lin}}(|\k-\q|)P_{\text{lin}}(q)\,,\\
&{\cal I}_{\mathcal{G}_2}(k)=2\int_\q \sigma^2(\q,\k-\q)F_2(\q,\k-\q)P_{\text{lin}}(|\k-\q|)P_{\text{lin}}(q)\,,\\
&{\cal F}_{\mathcal{G}_2}(k)=4P_{\text{lin}}(k)\int_\q \sigma^2(\q,\k-\q)F_2(\q,\k-\q)P_{\text{lin}}(q)\,,\\
&{\cal I}_{\delta^2\delta^2}(k)=2\int_\q P_{\text{lin}}(|\k-\q|)P_{\text{lin}}(q) - 2\int_\q P^2_{\text{lin}}(q)\,,\\
&{\cal I}_{\mathcal{G}_2\mathcal{G}_2}(k)=2\int_\q \sigma^4(\q,\k-\q)P_{\text{lin}}(|\k-\q|)P_{\text{lin}}(q) \,,\\
&{\cal I}_{\delta^2\mathcal{G}_2}(k)=2\int_\q \sigma^2(\q,\k-\q)P_{\text{lin}}(|\k-\q|)P_{\text{lin}}(q) \,,
\end{split} 
\ee
where $\sigma^2(\k_1,\k_2)=(\k_1\cdot \k_2)^2/(k_1^2k_2^2)-1$, and $F_2$ is the standard SPT quadratic density kernel \cite{Bernardeau:2001qr}, 
\begin{equation}
F_2({\bf k}_1,{\bf k}_2)=\frac{5}{7}+\frac{1}{2}\frac{({\bf k}_1\cdot{\bf k}_2)}{k_1k_2}\l\frac{k_1}{k_2}+\frac{k_2}{k_1}\r+\frac{2}{7}\frac{({\bf k}_1\cdot{\bf k}_2)^2}{k_1^2k_2^2}\,.
\end{equation}
In what follows we will also use the SPT velocity kernels $G_n$.
The galaxy power spectrum in redshift space is given by
\be
\label{eq:Pg}
\begin{split}
P_{g}(k,\mu)= & Z^2_1(\k)
P_{\text{lin}}
(k)+ 2\int_{\q}Z^2_2(\q,\k-\q)
P_{\text{lin}}(|\k-\q|)
P_{\text{lin}}(q)\\
& + 6Z_1(\k)P_{\text{lin}}(k)\int_{\q}Z_3(\q,-\q,\k)P_{\text{lin}}(q)\\
& -2\tilde{c}_0 k^2 P_{\text{lin}}(k)
-2\tilde{c}_2 f \mu^2 k^2 P_{\text{lin}}(k)
-2\tilde{c}_4 f^2 \mu^4 k^2 P_{\text{lin}}(k)+P_{\text{shot}}\,,
\end{split}
\ee
where the redshift space kernels are given by 
\bseq 
\begin{align}
&Z_1(\k)  = b_1+f\mu^2\,,\\
&Z_2(\k_1,\k_2)  =\frac{b_2}{2}+b_{\mathcal{G}_2}\left(\frac{(\k_1\cdot \k_2)^2}{k_1^2k_2^2}-1\right)
+b_1 F_2(\k_1,\k_2)+f\mu^2 G_2(\k_1,\k_2)\notag\\
&\qquad\qquad\quad~~+\frac{f\mu k}{2}\left(\frac{\mu_1}{k_1}(b_1+f\mu_2^2)+
\frac{\mu_2}{k_2}(b_1+f\mu_1^2)
\right)
\,,\\
&Z_3(\k_1,\k_2,\k_3)  =2b_{\G_3}\left[\frac{(\k_1\cdot
     (\k_2+\k_3))^2}{k_1^2(\k_2+\k_3)^2}-1\right]
\big[F_2(\k_2,\k_3)-G_2(\k_2,\k_3)\big]\notag\\  
&\quad
+b_1 F_3(\k_1,\k_2,\k_3)+f\mu^2 G_3(\k_1,\k_2,\k_3)+\frac{(f\m k)^2}{2}(b_1+f \mu_1^2)\frac{\m_2}{k_2}\frac{\m_3}{k_3}\notag\\
&\quad
+f\mu k\frac{\mu_3}{k_3}\left[b_1 F_2(\k_1,\k_2) + f \mu^2_{12} G_2(\k_1,\k_2)\right]
+f\m k (b_1+f \mu^2_1)\frac{\m_{23}}{k_{23}}G_2(\k_2,\k_3)\notag\\
&\quad+b_2 F_2(\k_1,\k_2)+2b_{\mathcal{G}_2}\left[\frac{(\k_1\cdot (\k_2+\k_3))^2}{k_1^2(\k_2+\k_3)^2}-1\right]F_2(\k_2,\k_3)
+\frac{b_2f\mu k}{2}\frac{\mu_1}{k_1}\notag\\
&\quad+b_{\mathcal{G}_2}f\mu k\frac{\m_1}{k_1}\left[\frac{(\k_2\cdot \k_3)^2}{k_2^2k_3^2}-1\right]
\,,
\end{align} 
\eseq
where $\k=\k_1+\k_2+\k_3$ and
the kernel $Z_3$ must be symmetrized over its arguments.
The net expressions for the multipoles of the power spectrum 
model \eqref{eq:Pg} are obtained 
upon averaging over $\mu = (\k\cdot \z)/k$ weighed 
with the appropriate Legendre polynomials.
As for the counterterms, we use the following basis:
\be 
\begin{split}\label{Pctr_rsd}
	& P_{0,\nabla^2\delta}(k)=-k^2 \cdot \l \frac{b_1^2 f^2}{3} +\frac{2b_1 f^3}{5} + \frac{f^4}{7}\r P_{\text{lin}}(k)\,,\\
	& P_{2,\nabla^2\delta}(k)=-k^2 \cdot \l \frac{2b_1^2 f^2}{3} +\frac{8b_1 f^3}{7} + \frac{10 f^4}{21}\r P_{\text{lin}}(k)\,,\\
	& P_{4,\nabla^2\delta}(k)=-k^2 \cdot \frac{8f^2}{35} P_{\text{lin}}(k) \,,
\end{split}
\ee
in which case $c_0$ and $c_2$ correspond directly to the short-scale velocity dispersion 
$\sigma^2_v$, see \eqref{fog}.

\section{Gaussian covariance matrices for redshift space power spectrum and bispectrum}
\label{app:Cov}

The Gaussian covariance matrix for redshift-space power spectrum multipoles reads
\be
\label{eq:Crsd}
\begin{split}
C_{k_i k_j}^{(\ell_1 \ell_2)}=\frac{2}{N_k}\frac{(2\ell_1+1)}{2}\left(2\ell_2+1\right)
\int_{-1}^1d\mu \,L_{\ell_1}(\mu)L_{\ell_2}(\mu)\left[P_g(k_i,\mu)+\frac{1}{\bar{n}_g}\right]^2
\delta_{ij}.
\end{split} 
\ee
Doing the $\mu$-integral and expressing everything in terms of the multipoles one obtains
\be
\begin{split}
& C^{(00)}_{k_i k_j}=\frac{2}{N_k}\left(P_0^2+\frac{1}{5}P_2^2+\frac{1}{9}P_4^2\right)\delta_{ij}\,,\\
& C^{(02)}_{k_i k_j}=\frac{2}{N_k}\left(2P_0P_2+\frac{2}{7}P_2^2+\frac{4}{7}P_2P_4
+ \frac{100}{693}P_4^2\right)\delta_{ij}\,,\\
& C^{(04)}_{k_i k_j}=\frac{2}{N_k}\left(\frac{18}{35}P^2_2+2P_0P_4+\frac{40}{77}P_2P_4
+ \frac{162}{1001}P_4^2\right)\delta_{ij}\,,\\
& C^{(22)}_{k_i k_j}=\frac{2}{N_k}\left(
5 P_0^2+\frac{20 P_0 P_2}{7}+\frac{20 P_0 P_4}{7}+\frac{15 P_2^2}{7}+\frac{120 P_2 P_4}{77}+\frac{8945 P_4^2}{9009}\right)\delta_{ij}\,,\\
& C^{(24)}_{k_i k_j}=\frac{2}{N_k}\left(
\frac{36 P_0P_2}{7}+\frac{200 P_0 P_4}{77}+\frac{108 P_2^2}{77}+\frac{3578 P_2 P_4}{1001}+\frac{900 P_4^2}{1001}  \right)\delta_{ij}\,,\\
&C^{(44)}_{k_i k_j}=\frac{2}{N_k}\left(
9 P_0^2+\frac{360 P_0 P_2}{77}+\frac{2916 P_0 P_4}{1001}+\frac{16101 P_2^2}{5005}+\frac{3240 P_2P_4}{1001}+\frac{42849 P_4^2}{17017}
  \right)\delta_{ij}\,,
\end{split}
\ee
where we introduced the number of modes $N_k =V {k^2dk}/{(4\pi^2)}$.
Eq.~\eqref{eq:Crsd} generalizes to the isotropic 
redshift space bispectrum,
\be
C^{\ell =0}_{TT'}=\frac{\l2\pi\r^3}{V(z)}\frac{\pi s_{123}}{dk_1dk_2dk_3}\frac{\delta_{TT'}}{k_1k_2k_3}\frac{1}{4\pi}\int_0^{2\pi} d\phi \int_0^{\pi}d\omega\,\sin\omega\,
\prod_{a=1}^3\left[P_g(k_a,\mu_a)+\frac{1}{\bar n_g}\right]\,.
\ee
Choosing the coordinate system as in Ref.~\cite{Scoccimarro:1999ed}, it is trivial to obtain the final expression for the redshift space
covariance matrix. We leave it as an exercise to the reader.

\begin{figure}[h!]
	\centerline{
		\includegraphics[keepaspectratio,width=1.0\linewidth]{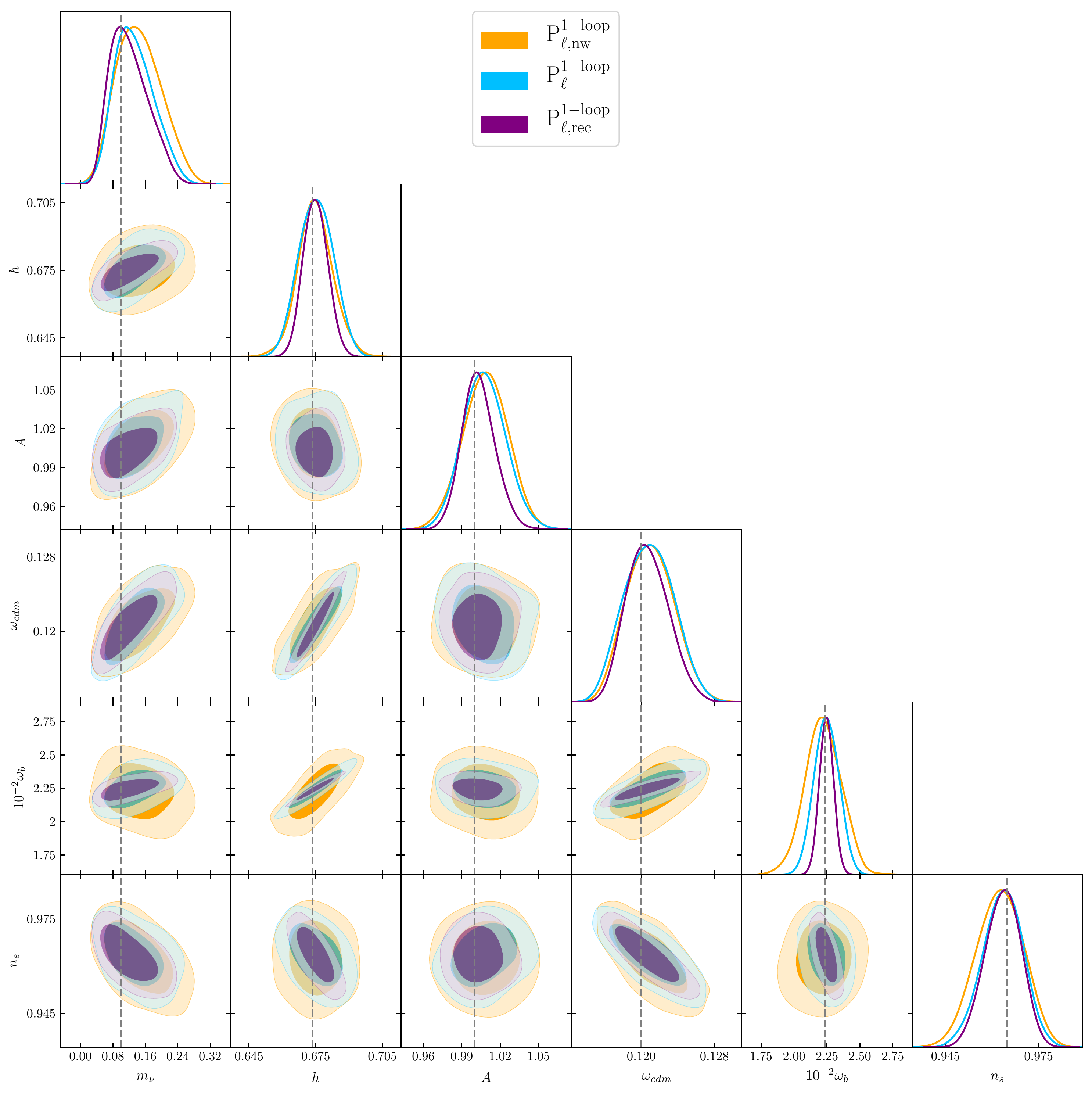}}
	\caption{$1\sigma$ and $2\sigma$ contours, see Tab. \ref{tab:constraints2} for the 1d marginalized limits.}
	\label{fig:noBAO} 
\end{figure}

\section{Information content of baryon acoustic oscillations}
\label{sec:BAO}

To quantify the information gain from the BAO we analyzed mock power 
spectrum likelihoods simulated for two extreme situations: 
the fully damped and totally undamped BAO wiggles.
Using our definition of the damping factor in Sec.~\ref{subsec:IR},
they formally correspond to the following two cases:
\bseq
\begin{align}
& \Sigma^2 \to \infty \,, \quad \text{(nw)}\,,\\
& \Sigma^2 \to 0\,, \quad \text{(rec)}\,,
\end{align}
\eseq

In the first case there are no BAO wiggles whatsoever.
This case should be contrasted with the 
BAO-only measurements (done e.g. in Ref.~\cite{Beutler:2016ixs}),
in which one throws away the broadband information and only fits the BAO wiggles. 
Our analysis does exactly the opposite: we throw away the BAO wiggles and fit the remaining shape.
A similar approach was previously used in Ref.~\cite{Hamann:2010pw}.
Note that when fitting the non-wiggly mock data we also remove the BAO 
wiggles in our theoretical templates using the same wiggly-smooth decomposition algorithm.

The second case is equivalent to using BAO reconstruction 
\cite{Eisenstein:2006nk,
Seo:2009fp,
Schmittfull:2015mja,
Obuljen:2016urm,
Zhu:2016sjc,
Schmittfull:2017uhh}, a data-driven method that aims at restoring the linear amplitude of the BAO wiggles.
This method strongly distorts the broadband part such that its cosmological information is typically discarded. 
Although the standard BAO reconstruction techniques \cite{Eisenstein:2006nk,Seo:2009fp}
do not fully recover the linear BAO wiggles \cite{Beutler:2016ixs}, 
it is, in principle, possible in more sophisticated approaches~\cite{Schmittfull:2017uhh}.
In this regard, our second case can be though of as a joint analysis of the full-shape data
and the optimally reconstructed BAO. This case should show the importance of the high-$k$ wiggles,
which are normally diluted by large-scale bulk flows.




The outcome of our analysis is presented in 
Fig.~\ref{fig:noBAO} and Table~\ref{tab:nobao}. 
There we show the results for the non-wiggly redshift-space power spectrum $P^{\text{1-loop}}_{\ell,\,\text{nw}}$, 
the reference one-loop power from our main analysis $P^{\text{1-loop}}_{\ell}$, and the power spectrum with fully reconstructed 
BAO wiggles $P^{\text{1-loop}}_{\ell,\,\text{rec}}$.

Let us first focus on the first case (nw). 
We observe the biggest change in $\omega_b$ and $n_s$.
This is explained by the argument that without the BAO wiggles $\omega_b$ 
is measured through the suppression of the power spectrum at short scales,
which can be partly compensated by the tilt. The presence of the BAO
wiggles allows one to break this degeneracy. 
Remarkably, the absence of the BAO feature does not degrade the constraints on $h$, which is responsible for the geometric distance 
information. This shows that the power spectrum shape would be a powerful tool to extract cosmological parameters even 
if the BAO were absent in the matter distribution.

Now let us consider the mock combined analysis with the reconstructed BAO~(rec).
This analysis captures information from the high-$k$ BAO wiggles, 
which benefits from smaller errors compared to the low-$k$ BAO used in our main analysis.
As a result, the parameter constrains improve quite sizable, e.g. the errorbar on $h$ reduces by $\sim 30\%$.
This points out the importance of the BAO reconstruction, which can complement cosmological parameter measurements
from the power-spectrum and bispectrum shape.


\begin{table}[h]
	\centering
	\begin{tabular}{|c|ccccc|c|}
		\hline
		Set & $10^3\,h$ & $10^2\,A$ & $10^3\,\omega_c$ & $10^4\,\omega_b$ & $10^3\,n_s$ & $m_\nu, \meV$ \\
		\hline
		$\rm P_{\ell,nw}^\lp$				&7.9&1.8&2.7 &13.8 &7.6 &$55$ \\
		$\rm P_\ell^\lp$				&7.7&1.7&2.7 &9.3 & 6.5&$48$ \\
		$\rm P_{\ell,rec}^\lp$				&5.4&1.2&2.2 &5.3 &5.6 &$45$ \\
		\hline
	\end{tabular}
	\caption{\label{tab:constraints2} 
	Marginalized $1\sigma$ error for the cosmological parameters in  $\Lambda$CDM model with one massive neutrino, see Table \ref{tab:fid}. 
We show results for the non-wiggle (smooth) one-loop redshift space power spectrum (first line), 
the reference power spectrum with realistically damped BAO wiggles (second line), 
and the power spectrum with fully undamped (reconstructed) BAO wiggles (third line).
}
	\label{tab:nobao}
\end{table}

\section{Results for the theoretical error with a strong Fingers-of-God effect}
\label{app:fog}

\begin{table}[H]
	\centering
	\begin{tabular}{|c|ccccc|c|}
		\hline
		Set & $10^3\,h$ & $10^2\,A$ & $10^3\,\omega_c$ & $10^4\,\omega_b$ & $10^3\,n_s$ & $m_\nu, \meV$ \\
		\hline
		$\rm P_{\rm \ell,\,AP}^\lp\!+\!B_0^\tree (FoG+2L)$				&9.4 &3 &3.5 &10 &10.7 &$50$ \\
		$\rm P_{\rm \ell,\,AP}^\lp\!+\!B_0^\tree (2L)$				&5.5 &1.1 &2 &6 &4.6 &$28$ \\
		\hline
	\end{tabular}
	\caption{
		Marginalized $1\sigma$ error for the cosmological parameters in $\Lambda$CDM model with a massive neutrino, see Table \ref{tab:fid} for definitions. We display results for the one-loop redshift space power spectrum and the tree-level bispectrum likelihoods with the baseline theoretical error (corresponding to dark matter two loops - 2L) and the theoretical error that includes both
		the loop contribution and a template for the strong Fingers-of-God effect (FoG).}
	\label{tab:FoG}
\end{table}

Our main analysis was performed for the theoretical error template dominated by the two-loop
corrections due to dark matter clustering. This template does not contain higher-order corrections
generated by fingers-of-God, which is motivated by several arguments. 
First, fingers-of-God were found to be small in mock catalogs of Euclid-type galaxies \cite{Orsi:2009mj,Orsi:2017ggf}. 
Second, the fingers-of-God effect is mostly produced by the velocity dispersion of the satellite galaxies. But the satellites can be removed from the catalog when computing the power spectrum \cite{Hand:2017ilm}. 
Central galaxies have much smaller velocity dispersion and its contribution is of the same order 
as the two-loop corrections due to dark matter clustering.
Third, fingers-of-God plague only the line-of-side modes, 
while the modes with the wavevectors with other directions are much less affected by them. 
Building the redshift-space wedges and throwing away the angular bins with $|\mu|\sim 1$
allows one to remove fingers-of-God while
retaining most of the information encoded in the power spectrum. In this case the residual 
theoretical error of the redshift-space wedges should indeed be dominated by the dark matter two loops.

From these arguments it is clear that there are ways to extract cosmological information
regardless of the strength of fingers-of-God. However, it is curios to see how much 
the constraints will degrade if one naively analyzes the power spectrum multipoles
in the presence of strong fingers-of-God. To this end we reanalyzed our mock data with a new 
theoretical error covariance that contains both envelops \eqref{E_P_rsd_fog} and \eqref{E_P_rsd}.
We chose the next-to-leading order contribution \eqref{E_P_rsd_fog} to be universal for all multipole moments, as dictated by the redshift-space mapping,
and assumed $\sigma_v=5\,\Mpc/h$ for all multipole moments in \eqref{E_P_rsd_fog}.

The results obtained for the power spectrum + bispectrum mock likelihoods are presented in Tab.~\ref{tab:FoG}. 
One can see that using the conservative template for the theoretical error covariance degrades constraints on all cosmological parameters roughly by a factor of 2. 
In particular, the uncertainty of neutrino mass measurement worsens to $50\,\meV$. 
However, our result shows that the effect of the strong fingers-of-God is not 
catastrophic. Even in this case the constraints on cosmological parameters from the LSS
data alone are competitive with the Planck-alone CMB measurements.

\section{Analysis with a sharp momentum cutoff}
\label{app:kmax}

\begin{table}[h]
	\centering
	\begin{tabular}{|c|ccccc|c|}
		\hline
		Set & $10^3\,h$ & $10^2\,A$ & $10^3\,\omega_c$ & $10^4\,\omega_b$ & $10^3\,n_s$ & $m_\nu, \meV$ \\
		\hline
		$\rm P_{\rm \ell,AP}^\lp\!+\!B_0^\tree~(k_{max})$				&22 &5.6 &7.8 &24.9 &20.2 &$<206$ \\
			$\rm P_{\rm \ell}^\tree\!+\!B_0^\tree~(k_{max})$				&38.6 &6.9 &14.4 &42.4 &25.5 &$<254$ \\
		$\rm P_{\rm \ell,\,AP}^\lp\!+\!B_0^\tree~(TE) $				&5.5 &1.1 &2 &6 &4.6 &$28$ \\
		\hline
	\end{tabular}
	\caption{
		Marginalized $1\sigma$ error for the cosmological parameters in $\Lambda$CDM model with one massive neutrino, see Table \ref{tab:fid}, considering one-loop/linear redshift space power spectrum and tree-level bispectrum with the sharp cutoff $k_{\text{max}}=0.15\,h\Mpc^{-1}$. For comparison, we also show the results of 
		our baseline analysis with the theoretical error, $\rm P_{\rm \ell,\,AP}^\lp\!+\!B_0^\tree~(TE) $.
		If the $1\sigma$ constraint on neutrino mass is consistent with zero, 
		we provide an upper bound of the $68\%$ confidence region.}
	\label{tab:FoG2}
\end{table}

In this Appendix we present several additional analyses that are aimed to 
(i) reveal how much the theoretical error formalism improves over the sharp cutoff approach,
and (ii) quantify the difference between the use of the linear and one-loop power spectrum in 
LSS forecasts. 

To show the difference between our baseline analysis and the $k_{\text{max}}$ results 
we construct new one-loop power spectrum and tree-level bispectrum mock likelihoods
without the theoretical error and with a sharp momentum cut at $k_{\rm max}=0.15\,h\Mpc^{-1}$.
This choice is motivated by previous 
analysis of the BOSS DR12 galaxy clustering \cite{Beutler:2016arn}
and the Euclid forecast of Ref.~\cite{Yankelevich:2018uaz}.
Note that this choice is somewhat conservative because perturbation theory
predictions are, in fact, reliable for bigger wavenumbers, and even more so at high redshifts
to be probed by Euclid-like surveys.

For the rest, we follow the methodology of our baseline analysis. 
In particular, we use the same binning for the power spectrum (see Sec.~\ref{subsec:mock} for more detail), whereas for the mock bispectrum data we use $k_\minn=0.01\,h\Mpc^{-1}$, $k_\maxx=0.15\,h\Mpc^{-1}$ splitting the corresponding momentum interval into 10 linearly spaced k-bins of width $\Delta k=0.016\,h\Mpc^{-1}$. 
Our results are presented in Tab.~\ref{tab:FoG2} (first line, $\rm P_{\rm \ell,AP}^\lp\!+\!B_0^\tree~(k_{max})$). 
One can see that constraints on all cosmological parameters degrade dramatically. 
In particular, the $1\sigma$-bound on the total neutrino mass becomes very loose, 
\mbox{$m_\nu<206\,\meV$ ($68\%$ CL)} which is comparable to the current Planck-only limit \cite{Aghanim:2018eyx}. 
This result suggests that the use of the theoretical error approach allows one to extract 
significantly more information from the mildly non-linear scales compared to an analysis with 
a conservative sharp $k_{\text{max}}$ cutoff.

Let us now discuss the impact of the power spectrum modeling on the forecasts for future LSS surveys, some of which use the linear model with the fingers-of-God damping, 
see e.g.~\cite{Yankelevich:2018uaz}.
Two important comments are in order here. 
First, the use of the tree-level expressions both for the power spectrum and bispectrum leads
to the loss of cosmological information. 
The reason is that the tree-level matter power spectrum model ignores the non-linear bias coefficients
which are present in the tree-level bispectrum.
However, the non-linear bias is a characteristic of the whole density field and hence should be treated on the same footing both in the power spectrum and bispectrum at any given order in perturbation theory. This implies that the power spectrum must necessarily be computed at the one-loop order if one analyzes the tree-level bispectrum. 
Since the tree-level bispectrum and the one-loop power spectrum share the same bias parameters $b_2$ and $b_{\mathcal{G}_2}$, 
they are better constrained in the joint analysis, which allows one to extract more cosmological information.
Second, the use of the one-loop matter power spectrum can improve the constraints by itself as compared 
to a forecast based on linear theory. 
Indeed, since $P^{\text{1-loop}}\propto P_{\lin}^2$, 
the one-loop power spectrum has a twice bigger response to a variation of cosmological parameters.
Moreover, the one-loop power spectrum introduces new redshift-dependent shape 
modifications that are absent in linear theory.
In general, the one-loop correction is smaller than the linear part, 
so this effect is not expected to be very dramatic unless one is interested in subtle phenomena like neutrino masses.

Given these reasons, we expect the effects discussed above to be relevant for our forecast.
In order to study their impact we generated new power spectrum and bispectrum mock likelihoods,
in which both these statistics are computed only at the tree level,
without the theoretical error, and with a sharp momentum cut at $k_{\rm max}=0.15\,h\Mpc^{-1}$.
To match the previous forecasts, we do not include the AP effect and IR resummation in our model.
Besides, we prefer not to include the Finger-of-God exponential damping 
in the linear power spectrum model, 
which should formally be taken into account along with the one-loop corrections omitted here.
The results are shows in Tab.~\ref{tab:FoG2} (second line, $\rm P_{\rm \ell}^\tree\!+\!B_0^\tree~(k_{max})$). 
Overall, our results are consistent with those of Ref.~\cite{Yankelevich:2018uaz}.\footnote{For some parameters our constraints are somewhat better, which can be explained by not using the fingers-of-God exponent in our analysis,
which would suppress the signal-to-noise otherwise. The other differences are the use of the full MCMC instead 
of the Fisher matrix formalism and the presence of massive neutrinos.
}
One observes that for the same choice of $k_{\text{max}}$ the use of the one-loop matter power spectrum leads to notably better results compared to the forecast based on the linear matter power 
spectrum.

\section{Analysis with a Gaussian approximation to the Planck likelihood}
\label{app:gauss_planck}

In this Appendix we present the results for the analyses which use the multivariate 
Gaussian approximation to the Planck likelihood.
The Planck posterior distribution is substantially asymmetric when the massive neutrino are included
into the fit \cite{Aghanim:2018eyx}, and hence the Gaussian approximation 
does not accurately capture
actual confidence regions \cite{Perotto:2006rj}. 
Given this reason, we focus only on the covariance matrix extracted from the Planck baseline analysis that fixed the neutrino mass to the minimal value $60$ meV. In other words, 
our Gaussian likelihood approximation does not include massive neutrinos, and consequently
does not capture their effect on the CMB data whatsoever. 
The use of this covariance matrix is, strictly speaking, inconsistent. 
However, it is still meaningful for the following two goals: 
(i) Demonstrate the robustness of our mock Planck likelihood 
used in the baseline analysis and (ii) Get an idea on the cosmological constrains which 
can be obtained from LSS if we use the CMB priors only on the cosmological parameters and not the 
neutrino mass, which is curios in light of the ``lensing tension'' present in the Planck CMB data \cite{Aghanim:2018eyx}.


\begin{table}[h!]
	\centering
	\begin{tabular}{|c|ccccc|c|}
		\hline
		Set & $10^3\,h$ & $10^2\,A$ & $10^3\,\omega_{cdm}$ & $10^4\,\omega_b$ & $10^3\,n_s$ & $m_\nu, \meV$ \\
		\hline
		Planck	Gaussian (Planck G)& 5.4& 1.4& 1.2& 1.5& 4.2& $-$\\ \hline
		$\rm P^\lp\!+\!Planck~G$			& 1.3& 1.4& 0.3& 1.2& 3.1& $ <226$ \\
		$\rm P^\lp\!+\!B^\tree\!+\!Planck ~G$	&0.8 & 1.3&0.2& 1.1& 3& $48$\\
		\hline
		$\rm P_{\rm \ell,\,AP}^\lp\!+\!Planck ~G$			& 1.8&1 & 0.4& 1.1&2.9 &$24$ \\
		$\rm P_{\rm \ell,\,AP}^\lp\!+\!B_0^\tree\!+\!Planck ~G$			&0.8 &0.9 &0.2 &1.1 &1.9 & $19$\\
		\hline
		$\rm P_{\rm \ell,\,AP}^\lp\!+\!B_0^\lp\!+\!Planck ~G$	&0.8 &0.7 &0.2 &1 &1.7 & $17$\\
		\hline
	\end{tabular}
	\caption{ Marginalized $1\sigma$ errors for the cosmological parameters in $\Lambda$CDM with one massive neutrino (see Table \ref{tab:fid}) for different combinations of likelihoods. For comparison, we also show current Planck 
		limits on the parameters of the base $\Lambda$CDM with \textit{a fixed minimal neutrino mass}, hence the corresponding error is omitted in the topmost row. If $1\sigma$ posterior contour of neutrino mass overlaps with zero, we show an upper bound of corresponding $68\%$ CL limit.}
	\label{tab:constraints_Gauss}
\end{table}

All in all, we downloaded Markov chains that sampled the 
cosmological parameters
of the \texttt{base\_plikHM\_TTTEEE\_lowl\_lowE\_lensing} 
likelihood from the Planck Legacy Archive\footnote{\href{http://pla.esac.esa.int/pla/\#cosmology}{
\textcolor{blue}{http://pla.esac.esa.int/pla/\#cosmology}}
}
and computed the covariance matrix for a subset of cosmological parameters $h,\omega_{cdm},\omega_b,n_s,A,\tau$. Then we used the Gaussian likelihood with this covariance matrix to 
derive constraints for the same combinations of likelihoods 
as in our baseline analysis. The results are displayed in Table~\ref{tab:constraints_Gauss}.
The first observation is that the errorbars on the cosmological parameters are 
somewhat different in the full mock and the Gaussian Planck likelihoods without the LSS data.
However, the difference reduces when we add the LSS likelihoods.
On the one hand, 
we observe a very good agreement between 
the CMB+LSS runs with
the Gaussian approximation
and the realistic mock Planck likelihood
for the parameters of the minimal $\L$CDM. 
On the other hand, the Gaussian likelihood worsens significantly
the constraints on the neutrino masses,
which is a result of neglecting the correlations between $m_\nu$ and other cosmological 
parameters present in the CMB data, e.g. $m_\nu$ and $h$.
Our results represent an important consistency check proving that the information gain indeed
comes from breaking of the degeneracies between the LSS and CMB.

\bibliographystyle{JHEP}
\bibliography{short}

\end{document}